%% file: 01_RevisionReStat_wp.tex
\documentclass[english,12pt]{article}
\usepackage[T1]{fontenc}
\usepackage[utf8]{inputenc}
\usepackage{geometry}
\usepackage{amsmath,appendix}
\usepackage{amsthm}
\usepackage{comment}
\usepackage{tabularx}
\usepackage{amssymb}
\usepackage{placeins}
\usepackage{bbm, dsfont}
\usepackage{setspace}
\usepackage{multirow}
\usepackage{pdflscape,longtable}
\usepackage{threeparttable}
\usepackage{booktabs}
\usepackage{setspace}

\usepackage[sort&compress]{natbib}
\usepackage{graphicx}
\usepackage{stackrel}
\usepackage{mathpazo}
\usepackage[citecolor= black,
			colorlinks=true,
		    linkcolor = black,
            urlcolor  = black,
            pdfpagelabels]{hyperref}

\usepackage{tikz}
\usetikzlibrary{er,positioning}
\usetikzlibrary{calc,trees,positioning,arrows,chains,shapes.geometric,%
    decorations.pathreplacing,decorations.pathmorphing,shapes,%
    matrix,shapes.symbols}

\tikzstyle{bag} = [align=center]
\tikzset{
  decoration={border},
  tuborg/.style={decorate},
  tubnode/.style={midway, right=2pt}
}

\makeatletter
\theoremstyle{plain}

\theoremstyle{plain}

\usepackage{babel}

\makeatother

\title{Coronavirus Perceptions and Economic Anxiety\thanks{\scriptsize Thiemo Fetzer, University of Warwick, CEPR, CESifo, Thiemo Fetzer@warwick.ac.uk; Lukas Hensel, University of Oxford, lukas.hensel@bsg.ox.ac.uk; Johannes Hermle, University of California, Berkeley and IZA, j.hermle@berkeley.edu;  Christopher Roth, University of Warwick, briq, CESifo, Cage, CEPR, Christopher.Roth@warwick.ac.uk. We thank the editor and two anonymous referees for their insightful comments and suggestions. We also thank Eric Avis, Joshua Dean, Stefano DellaVigna, Jonathan de Quidt, Armin Falk, James Fenske, Marta Golin, Yuriy Gorodnichenko, Johannes Haushofer, Matt Lowe, Andrew Oswald, Daniel Sgroi, Leah Shiferaw, Thomas Ferguson, and Dmitry Taubinsky for very useful comments as well as Anna Lane and Ivan Yotzov for excellent research assistance. Financial support from a grant by the The Institute of New Economic Thinking (Grant INO20-01) is gratefully acknowledged. Lukas Hensel gratefully acknowledges financial  support from the Wellspring Philanthropic Fund. Ethical approval was received by the Blavatnik School of Government’s Departmental Research Ethics Committee (BSG\_C1A-20-16) of the University of Oxford and the Humanities and Social Sciences Research Ethics Committee at the University of Warwick (protocol HSSREC 76/19-20).}}
\author{\begin{tabular}{cp{1cm}c}
Thiemo Fetzer &&Lukas Hensel \\[0.3cm]
  Johannes Hermle&& Christopher Roth \\
  &&\\
  \end{tabular}}
\date{First version: March 6, 2020\\[0.25cm]
This version: \today\\[0.25cm]Forthcoming, Review of Economics and Statistics
}

\begin{document}
\maketitle

\abstract{ \noindent We provide one of the first systematic assessments of the development and determinants of economic anxiety at the onset of the coronavirus pandemic. Using a global dataset on internet searches and two representative surveys from the US, we document a substantial increase in economic anxiety during and after the arrival of the coronavirus. We also document a large dispersion in beliefs about the pandemic risk factors of the coronavirus, and demonstrate that these beliefs causally affect individuals’ economic anxieties. Finally, we show that individuals’ mental models of infectious disease spread understate non-linear growth and shape the extent of economic anxiety.}\\

\vspace*{1cm}



\vspace*{2cm}

\pagebreak

\section{Introduction}
\doublespacing

\noindent The worldwide spread of the novel coronavirus (SARS-CoV-2) \citep{zhu2020novel,wu2020nowcasting,li2020early} has led to a substantial disruption of global economic activity. This article provides one of the first systematic assessments of the rapid emergence and causal determinants of economic anxiety at the onset of the coronavirus pandemic, when there was large uncertainty about the extent of its economic impact. We focus on how perceptions of pandemic risk factors shape economic anxieties. Understanding the development and causes of economic anxiety in the wake of the coronavirus pandemic is essential from both a scientific and practical perspective, particularly given recent empirical evidence demonstrating that perceptions and expectations about the macroeconomic environment substantially shape households' economic decisions \citep{dacunto2016effect,bailey2016social,bailey2017beliefsleverage,coibion2019does,kuchler2015personal}.


Predicting the development of economic anxiety and assessing its underlying mechanisms in the context of a pandemic is difficult when relying on historical accounts. Unlike regular economic downturns which begin with a moderate but accelerating decline in economic activity, the arrival and rapid global spread of the coronavirus pose a rare, sudden shock \citep{mackowiak2018lack}. Several aspects of human belief and expectation formation render the environment of the coronavirus pandemic distinct from that experienced during a conventional economic downturn. In particular, individuals have difficulty forming beliefs about the future in the wake of infrequent major events \citep{gallagher2014learning,rabin2002inference}. Moreover, when updating their beliefs, individuals place a disproportionate weight on the most recent events \citep{malmendier2011depression}, especially when these events are particularly salient \citep{tversky1973availability,bordalo2013salience}. As a consequence, belief formation may differ substantially in the unprecedented environment of the coronavirus pandemic as compared to more conventional economic shocks. Thus, relative to relying on historical accounts, employing contemporaneous data provides a more promising approach to assess the evolution of contemporaneous economic anxiety.

In this article, we collect contemporaneous data to systematically investigate the development and determinants of economic anxiety at the onset of the coronavirus pandemic. We study the underlying psychological mechanisms that shape economic anxiety in the environment of a pandemic by assessing the role of beliefs and information about pandemic risk factors as well as individuals' subjective mental models of infectious disease spread.

To set the stage for our analysis, we document a rapid increase in economic anxiety during and after the coronavirus has reached a country. Employing global data during the period of massive global spreading in January and February 2020, we show that Google search intensity for topics indicative of economic anxiety surged substantially after the virus has reached a country. To measure economic anxieties directly and in real-time after the arrival of the coronavirus, we conducted two survey experiments with representative samples of the US population on March 5, 2020 and March 16, 2020. In this 11 day period, the United States saw massive within-country spreading with a 26-fold increase in the number of confirmed cases from 176 to 4576. Moreover, public communication of the crisis' severity had shifted dramatically once the WHO declared it a pandemic on March 11. The data indicate a substantial increase in economic anxiety after the arrival of the coronavirus in the United States. 


The rapid surge in economic anxiety sets the stage to study the underlying informational and psychological mechanisms that shape economic anxiety in the wake of a pandemic. First, we study individuals' beliefs about the mortality and contagiousness of the coronavirus - two key characteristics relevant for assessing pandemic risks and predicting the severity of the coronavirus crisis. We elicited these beliefs in our March 5 survey before any lockdown measures had been put in place and the crisis had not yet been declared a pandemic. We document substantial dispersion in beliefs about both mortality and contagiousness. Moreover, the median participant overestimates both the mortality and contagiousness of the virus relative to the upper bound of estimates currently available in the medical literature. We show that beliefs about mortality and contagiousness are associated with participants' economic worries about the aggregate economy and their personal economic situation.

To further understand the precise causal relationships between coronavirus perceptions and economic anxiety, we embedded two experiments in our March 5 survey that varied the framing of coronavirus mortality as well as a treatment that studied the role of information about contagiousness. These real-time experiments allow us to shed light on the influence of information and its framing in an environment marked by large uncertainty about the future extent of the crisis. 

The first component of our experiment focuses on the framing of mortality risk. Participants were either truthfully informed, based on official estimates at the time of the survey, that the death rate from the coronavirus is ``20 times higher than for the flu'' (high mortality treatment) or ``5 times lower than for SARS'' (low mortality treatment). The wording was chosen to mirror the way such information is commonly communicated in the media.\footnote{For instance, the New York Times and The Telegraph compared the coronavirus to the flu and SARS (\url{https://www.nytimes.com/2020/02/29/health/coronavirus-flu.html}; \url{https://www.telegraph.co.uk/news/2020/03/06/coronavirus-vs-sars-flu-mers-death-toll/}, last accessed April 30$^{th}$ 2020).} We find that participants in the high mortality treatment report significantly higher concerns, both in a statistical and economic sense, about the aggregate economy and their personal economic situation. These results highlight the influence of the framing of news on public perceptions and economic expectations in times of high uncertainty \citep{prat2013political,chong2007framing}.

To investigate the effect of information regarding contagiousness, participants in a treatment group were, based on scientific estimates at the time of the experiment \citep{li2020early,wu2020nowcasting}, informed that \textit{``approximately 2 non-infected people will catch the coronavirus from a person who has the coronavirus'.'} Given that 81\% of respondents overestimate this statistic, the information treatment should decrease the perceptions of the contagiousness of the virus. We find that treated respondents report significantly lower worries about their personal economic situation. These causal results underscore the role that information plays in shaping economic anxiety in an environment characterized by large uncertainty and highlight the importance of both factual and targeted communication during health crises \citep{razum2003sars,person2004fear}. 

Second, besides taking information into account, forward-looking individuals also rely on their subjective mental models of the world to make predictions about the future \citep{andre2019subjective}. To understand the role of these mental models in shaping crisis beliefs and economic anxiety, we elicited participants' predictions of the growth of a fictitious disease in our March 16 survey. Consistent with exponential growth bias \citep{stango2009exponential,wagenaar1975misperception,levy2016exponential}, we document that the majority of individuals underestimate the non-linear nature of infectious disease spread. Furthermore, we show that mental models of infectious disease spread are substantially associated with participants' beliefs about the severity of the current coronavirus crisis: respondents who show a better understanding of non-linear disease spread anticipate a higher severity of the crisis and display higher worries about the aggregate US economy.


We contribute to the literature by documenting the development and underlying determinants of economic anxiety in the wake of a global pandemic. In particular, we provide novel causal evidence on the impact of information about pandemic risk factors on the formation of economic anxiety. Furthermore, we demonstrate the role subjective mental models of infectious disease spread play in shaping heterogeneity in economic anxiety. Our paper is most closely related to concurrent work by \cite{binder2020coronavirus} who conducted a survey using a sample from Amazon Mechanical Turk in early March. \cite{binder2020coronavirus} documents cross-sectionally that greater concerns about the coronavirus are associated with higher inflation expectations and more pessimistic unemployment expectations. She also studies how information provision about the Fed's interest rate cut in response to the coronavirus affects inflation and unemployment expectations. Our paper complements \cite{binder2020coronavirus} by documenting how the spread of coronavirus affects economic anxieties over time, and by providing both descriptive and causal evidence on how perceptions of the pandemic risk factors affect economic anxiety. We also relate to subsequent work studying the impact of the coronavirus on the economy \citep{coibion2020labor,hanspal2020income,adams2020inequality,Bartik2020}.



More generally, our work is related to a growing literature investigating the formation of economic sentiment and expectations about the macroeconomy among households and firms \citep{coibion2019monetary, fuster2010natural,fuster2012natural,coibion2012can,coibion2015information,coibion2015phillips,coibion2018firms,binder2018stuck,malmendier2011depression}. Relative to prior work, our evidence is unique in assessing economic sentiment and its drivers before and during a historic public health crisis in real time. We particularly relate to the literature studying the role of information in shaping economic sentiment and behavior \citep{coibion2020inflation,armona2016home,dacunto2016effect,roth2019expectations, binder2018experiment,bailey2016social,bailey2017beliefsleverage}. We also relate to the literature studying the role of cognitive processes in forming economic sentiment and macroeconomic expectations \citep{d2019iq,andre2019subjective}. We add to this literature by highlighting the importance of subjective mental models about infectious disease spread for shaping economic anxiety in the wake of a rare, unexpected, and unfamiliar public health shock.
 
Finally, we also contribute to the broad literature on the perception of health risks \citep{Fortson2011mortrisk,Oster2013testing,carbone2005smoking,kerwin2018scared,heimer2019yolo}. While existing evidence has primarily focused on individuals' beliefs about risks to their own health \citep{winter2014obese,kan2004obesity,liu1995risk}, we contribute to this literature by providing new evidence on the perception of factors relevant to pandemic in addition to individual risks. 



\section{Emergence of Economic Anxiety}\label{sec:development}

We begin by documenting the emergence of economic anxieties at the onset of the coronavirus pandemic. First, we focus on the period of the initial global spread of the coronavirus during January and February 2020. Leveraging global data on Google searches indicative of economic anxieties, we study the evolution of economic anxiety during the arrival of the coronavirus in a country. Next, we use survey data to study the development of economic anxiety within the US after the arrival of the coronavirus.

\subsection{Observational Evidence from Internet Searches during Global Spread} 

\paragraph{Data and Empirical Specification:}

We leverage data on internet search intensity from Google Trends. These data have been used in the past to detect influenza epidemics \citep{ginsberg2009} and to nowcast economic activity \citep{choi2012predicting}.\footnote{Moreover, as shown by prior studies, such internet searches serve as a measure of economic sentiment among households and thus as a predictor of future economic demand and activity \citep{vosen2011forecasting,choi2012predicting}. To qualify this claim, in Online Appendix Table \ref{table:recessionindicators} we use quarterly data from 2015 to 2019 and show that real GDP growth and real growth in consumption and imports are significantly lower, in both a statistical and economic sense, in the quarters following increases in "Recession" topic searches.} The Google Trends platform provides an interface to query search data, providing for each query a measure of search intensity scaled from 0 to 100, with 100 representing the highest proportion among the queried terms within a selected region and time frame. Google Trends queries can be constructed based on individual search terms or search topics which encompass groups of related individual search terms. We employ queries by search topics, an approach that has the advantage of capturing a broader set of search terms and not requiring any translations across languages.

To study the development of economic anxiety, we extracted Google search activity for the topics "Recession" and "Stock Market Crash" for a total of 194 countries and territories listed in Online Appendix Table \ref{table:countrylist}. We also leverage data on the search topics "Survivalism" and "Conspiracy Theory" which capture panic reactions among the public. We collected these data for January and February 2020 to study the developments during the initial global spread of the coronavirus when there was still significant uncertainty over whether a pandemic would emerge.\footnote{Online Appendix Figure \ref{fig:searchapp} shows the time series for the four topics of interest at the global level.} To make effect sizes interpretable, we normalize the search intensity at the country level by the mean search intensity prior to the arrival of the coronavirus in each country.\footnote{Specifically, for the normalization we use the mean search intensity between December 1, 2019 and the date of arrival of the coronavirus in a given country. This normalization makes the coefficient estimates interpretable as percentage changes relative to pre-coronavirus levels without having to resort to the mean of the dependent variable for interpretation. Results are not affected by this normalization, see Online Appendix Table \ref{table:main_DD_nonormalisation}. }


To study the impact on search activity we exploit the precise timing of coronavirus arrival in a country. The underlying coronavirus case data are from \cite{dong2020}. Econometrically, we perform the following difference-in-differences regression using daily data:

\begin{align}\label{eq:didmodel}
y_{c,t} = \alpha_c + day_t + \beta \times C_{c,t} + \epsilon_{c,t} 
\end{align}

where $y_{c,t}$ measures the search intensity in country $c$ on day $t$ for a specific topic. $C_{c,t}$ is a dummy variable indicating either having had at least one confirmed case or having had at least one human-to-human transmission of the coronavirus in country $c$ at time $t$. 
The regressions control for country fixed effects $\alpha_c$, absorbing fixed and time-invariant different levels of search intensities across countries. The time fixed effects $day_t$ absorb a level shifter for each day, capturing the global trend. We cluster standard errors at the country level. Intuitively, this analysis captures the impact of the \text{local} arrival of the coronavirus conditional on the global trend. 

\begin{figure}[!htbp]
\caption{Impact of Coronavirus Arrival on Internet Searches: Global Evidence}
\label{fig:search}
\centering
\begin{tabular}{c}
\includegraphics[height=6cm]{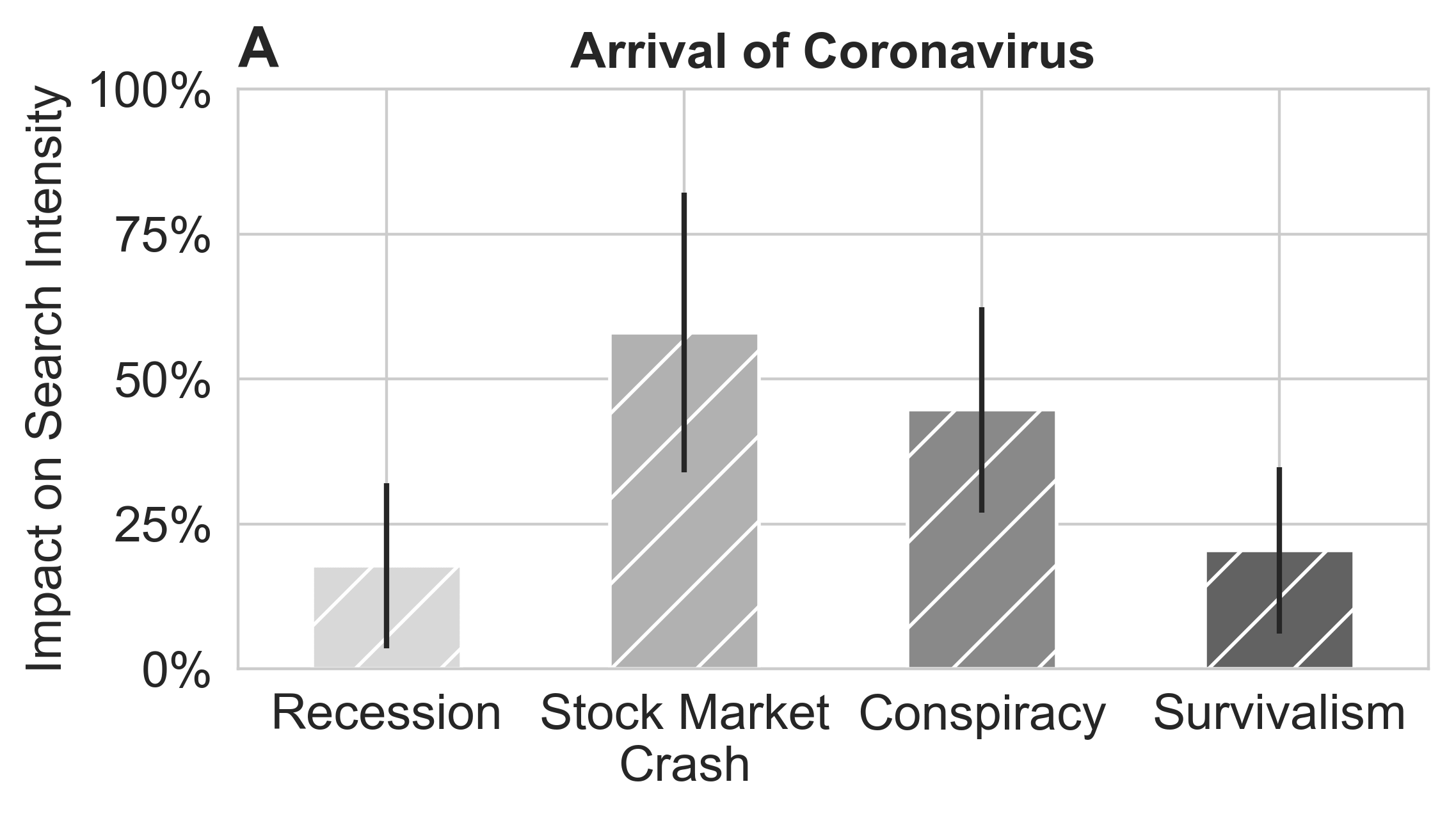}\\
\includegraphics[height=6cm]{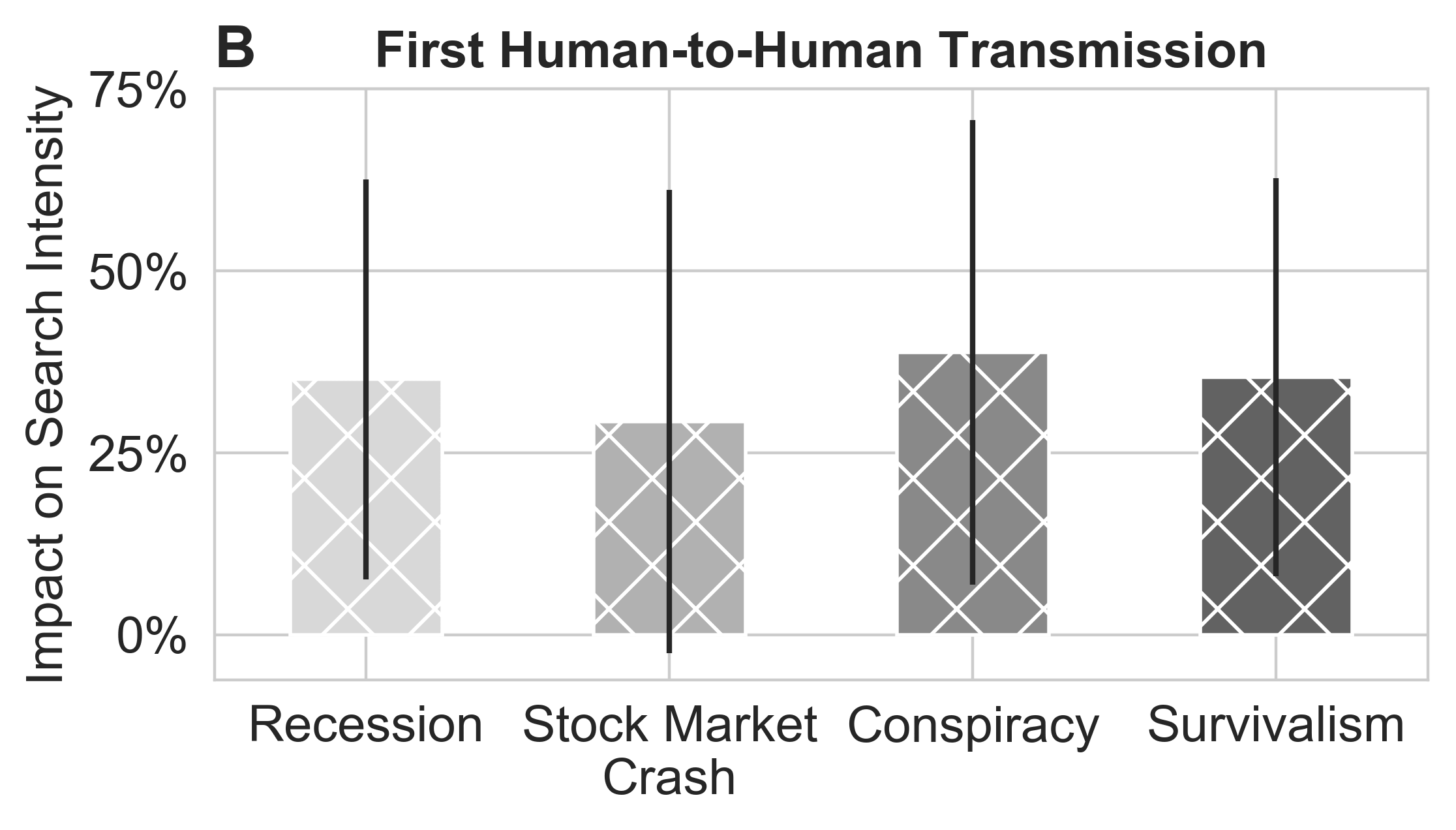}
\end{tabular}
\parbox{15cm}{\footnotesize \textit{Notes:} Figure \ref{fig:search} shows the impact of the arrival of the coronavirus (Panel A) and first human-to-human transmission (Panel B) in a country on Google search intensity for the topics "Recession", "Stock Market Crash", "Conspiracy Theory", and "Survivalism" obtained from difference-in-differences regressions conditional on country and day fixed effects. The dependent variable measures Google search intensity by topic indicated in column header, normalized by the average search intensity in a country prior to the coronavirus arrival. The Google searches are collected for the time span between January 1st and February 29th, 2020. In all panels, error bands indicate 95\% confidence intervals obtained from standard errors clustered at the country level.}
\end{figure}

\paragraph{Results:}
The data indicate that the arrival of the coronavirus in a country substantially increased search intensity for topics related to economic recessions by 17.8 ($s.e.=7.3$) percent relative to the pre-coronavirus search patterns (Figure \ref{fig:search}, Panel A and Online Appendix Table \ref{table:main_DD}). Similarly, search intensity for topics related to stock market crash rose by 58 ($s.e. = 12.4$) percent. In addition, an increase of 20.4 ($s.e.=7.3$) and 44.7 ($s.e.=9.1$) percent can be observed for topics related to survivalism and conspiracy theories respectively.\footnote{Google searchers for prayers also increased sharply during the coronavirus crisis \citep{bentzen2020crisis}. } Additionally, the response of search intensity to the first human-to-human transmission of the coronavirus in a country corroborates these results (Figure \ref{fig:search}, Panel B and Online Appendix Table \ref{table:main_DD}). In a placebo test, we find no impact of the arrival of the coronavirus on a series of unrelated Google searches such as 'Dog', 'Horse', 'Insect', 'Rain', or 'Rainbow' (Online Appendix Table \ref{table:placebo_DD}).\footnote{Results are further robust to dropping each country in turn or to dropping all countries pertaining to any of the 17 subregions globally in turn (see Online Appendix Figures \ref{fig:robdropping} and \ref{fig:robdroppingsubregions}).}  In sum, this evidence indicates that the arrival of the novel coronavirus leads to a spike in economic anxieties.

\subsection{Micro-evidence after Arrival of Coronavirus in the United States}\label{sec:descriptives}

\begin{figure}[!h]\centering
\caption{Evolution of Beliefs about Severity of Crisis and Economic Worries between Early and Mid-March: Evidence from the United States}
\label{fig:comp_time}
\begin{tabular}{cc}
\includegraphics[height=6cm]{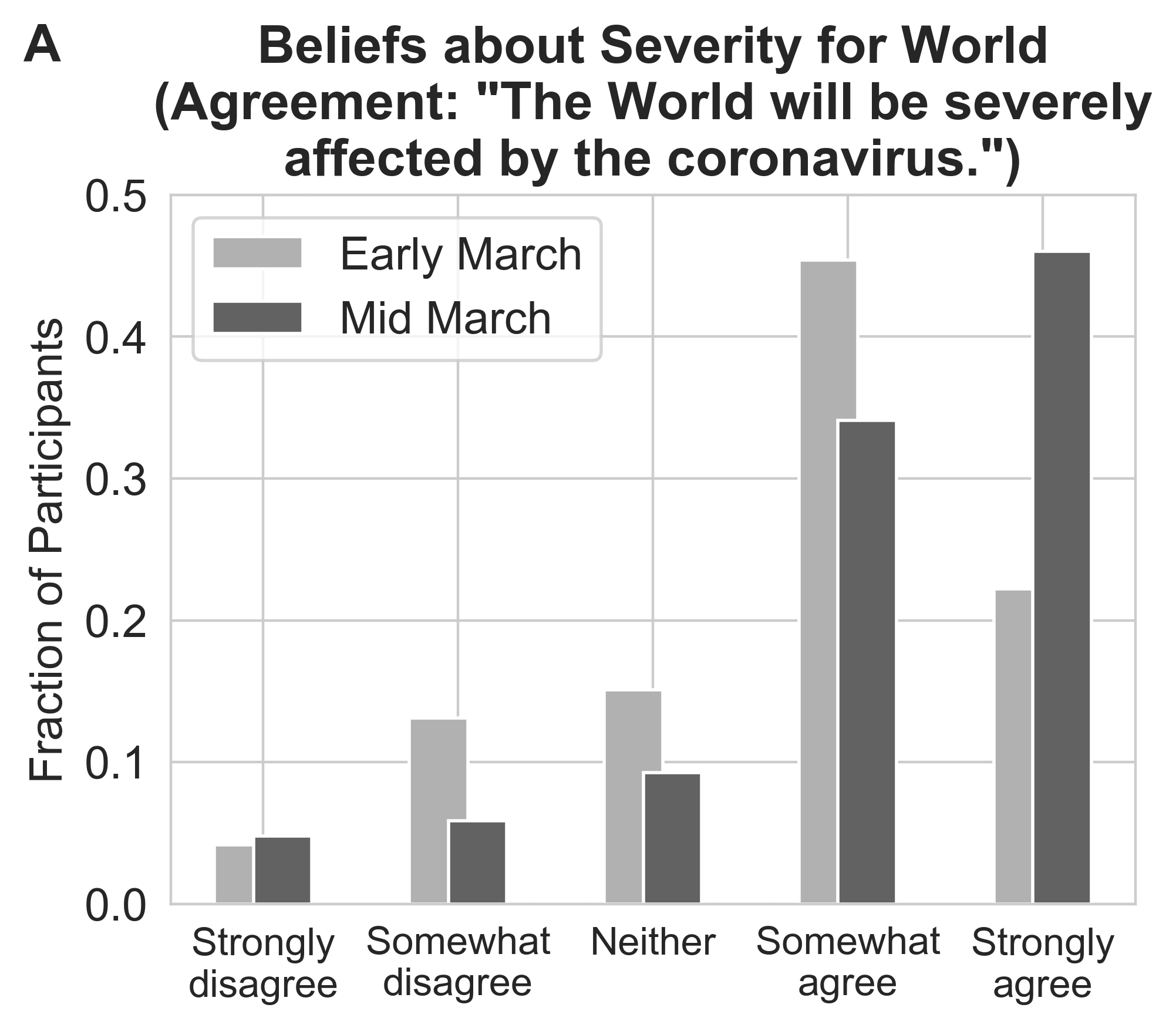} &\includegraphics[height=6cm]{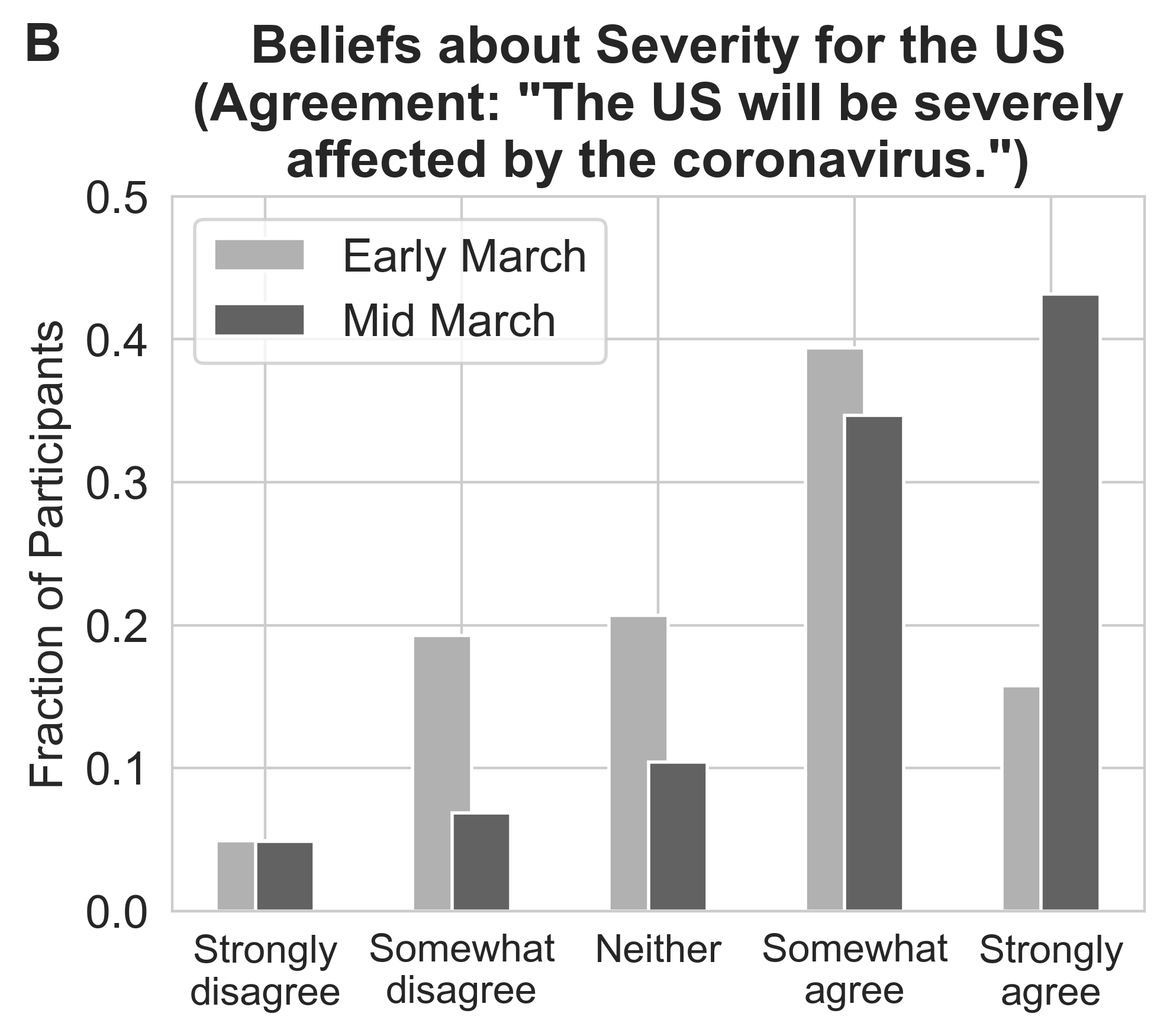}\\
\includegraphics[height=6cm]{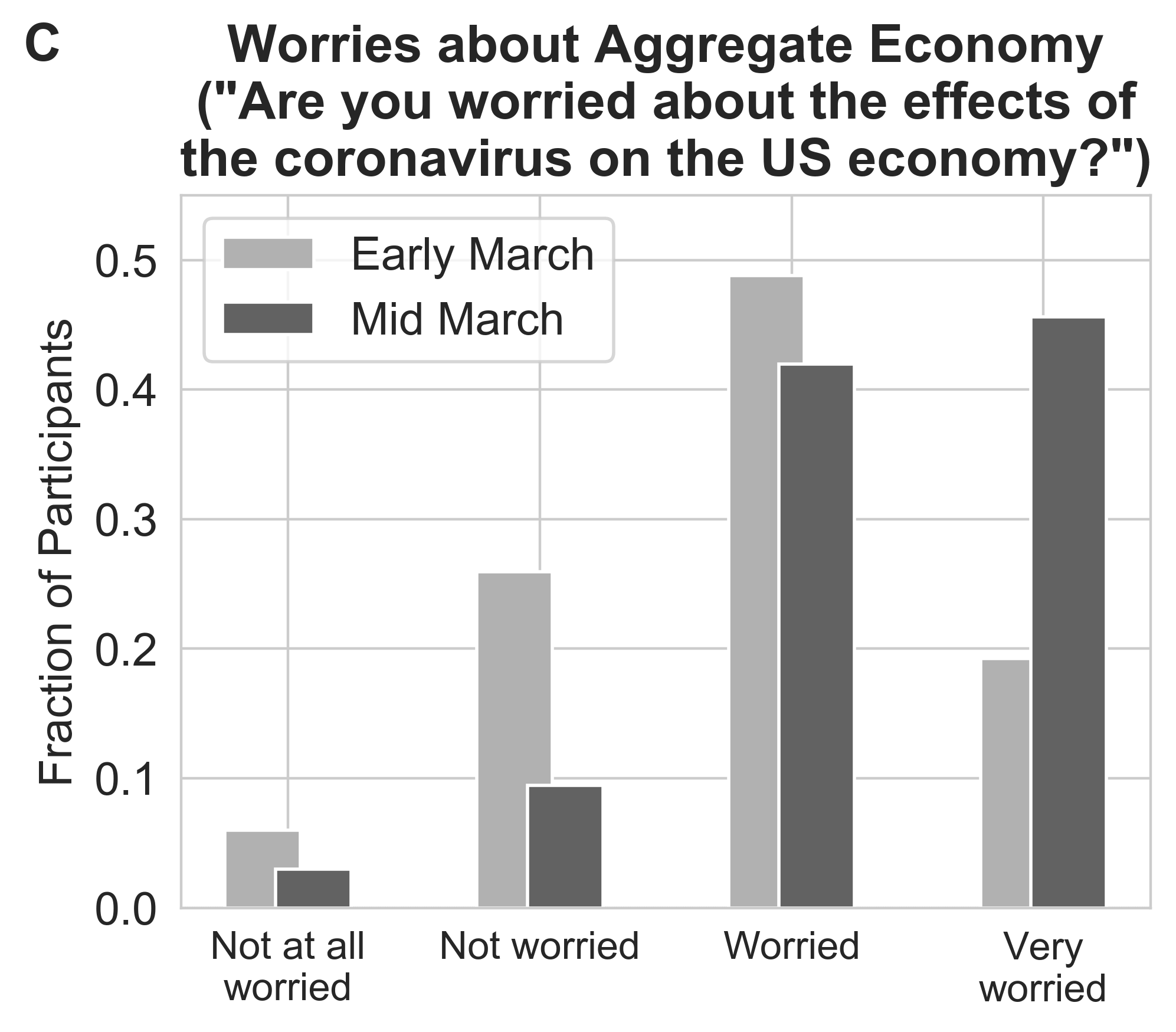} &\includegraphics[height=6cm]{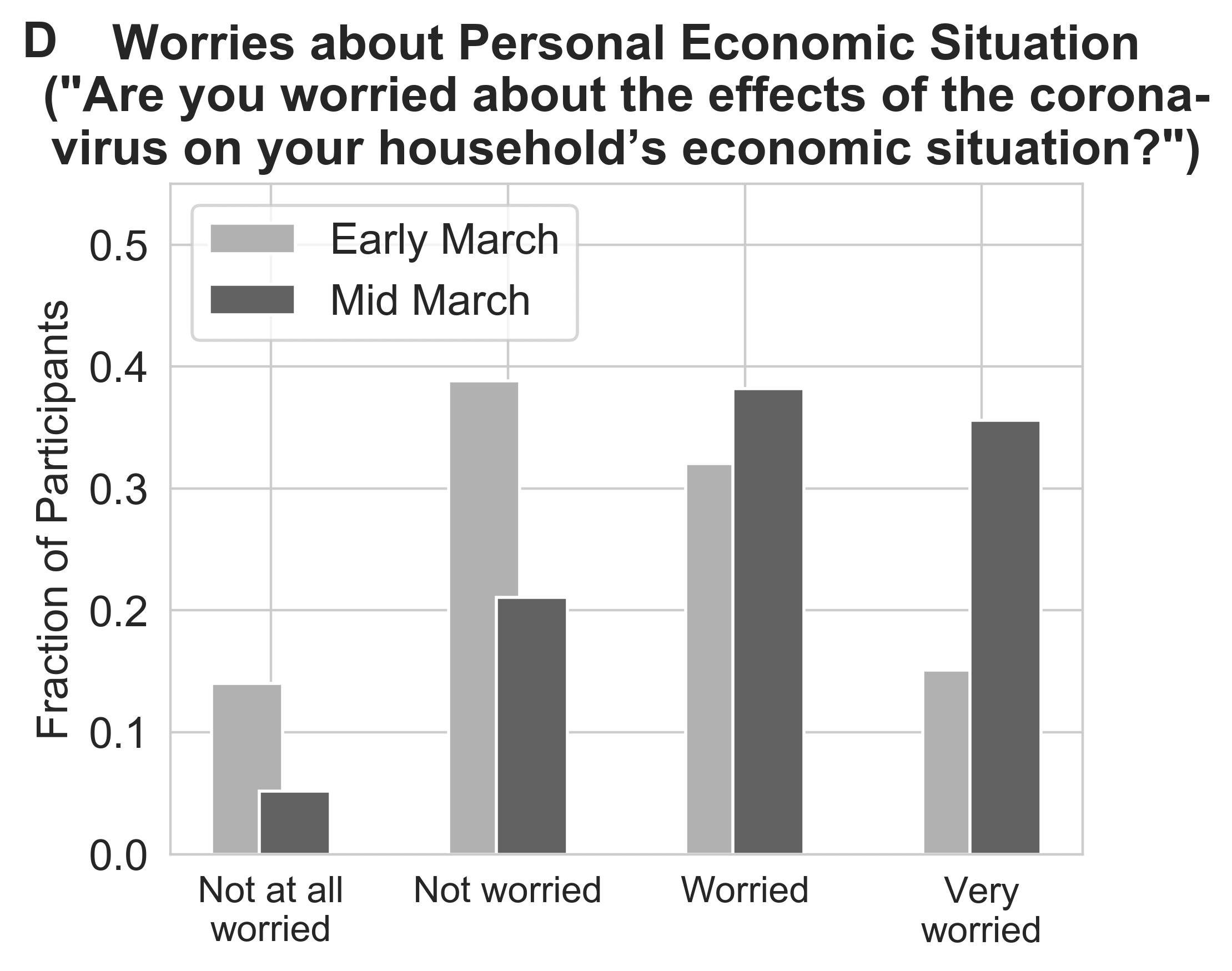}\\
\end{tabular}
\parbox{15cm}{\footnotesize \textit{Notes:} Figure \ref{fig:comp_time} compares the levels of economic anxiety in the US in surveys conducted on March 5 and March 16. Panel A shows beliefs about the severity of the coronavirus crisis for the world. Panel B shows beliefs about the severity of the coronavirus crisis for the US. Panel C shows worries about the US economy. Panel D shows worries about one's personal economic situation. Lighter columns indicate data collected on March 5, 2020, while darker columns indicate data collected on March 16, 2020.} 
\end{figure}

Does economic anxiety increase further as the novel coronavirus spreads within a country after the first domestic case occurs? We provide real-time evidence on this question using two surveys that measure economic anxiety in the United States. The surveys were administered to broadly representative samples of the US population on March 5 ($n=915$) and March 16, 2020 ($n=1,006$).\footnote{Our sample is representative of the US population in terms of income, region, gender, age, and education (see Online Appendix Tables \ref{table:sum_exp} and \ref{table:sum_expgrowth}). We collaborated with an online panel provider (Luc.id)  which is widely used in the social sciences.} Within this 11-day time span the number of confirmed cases within the United States jumped by a factor of approximately 26, from 176 to 4576. Hence, this time frame captures a period of substantial within-country spread.\footnote{In Online Appendix A we present cross-sectional results from the US in mid-February, during a time in which the US reported only 13 cases across the whole country. Respondents from states with any coronavirus cases exhibit significantly more pessimistic expectations (Online Appendix Table \ref{table:opinionpoll}).}
In both surveys, we investigate participants' beliefs about the severity of the crisis for the world and the US as well as their worries about the aggregate economy and their personal economic situation. For the precise wording of the questions and response scales, see Figure \ref{fig:comp_time}.

\paragraph{Results:} 

The evolution of our survey measures over time between March 5 and 16 is visualized in Figure \ref{fig:comp_time}.  We document a substantial increase in participants' beliefs about the severity of the crisis for the world (Figure \ref{fig:comp_time}, Panel A) and the US (Figure \ref{fig:comp_time}, Panel B) as well as in their worries about the aggregate US economy (Figure \ref{fig:comp_time}, Panel C) and their personal economic situation (Figure \ref{fig:comp_time}, Panel D). Quantitatively, these increases are sizable. For instance, the fraction of respondents who were worried about the impact on their personal economic situation increased from 47\% to 74\% (p < 0.001) (see also Online Appendix Table \ref{table:comparison_time}).\footnote{In our March 5 survey, we elicit economic anxieties after the relative mortality framing described in Section 3.2. The descriptive patterns of an increase in economic anxiety from March 5 to March 16, however, hold in both cases: when we focus either on respondents exposed to the high relative mortality framing or respondents exposed to the low relative mortality framing.}

In addition, in Online Appendix Figure \ref{fig:subgroupevol} we investigate heterogeneity across several subgroups, dividing the sample by gender (Panel A), age (Panel B), and political affiliation (Panel C). We do not find any differences between women and men (Online Appendix Figure \ref{fig:subgroupevol}, Panel A).
Similarly, old and young individuals do not differ strongly except that young people show substantially higher worries regarding their personal economic situation, potentially due to their higher unemployment risk (Online Appendix Figure \ref{fig:subgroupevol}, Panel B). Finally, we observe stark partisan differences (Online Appendix Figure \ref{fig:subgroupevol}, Panel C). Individuals who identify as Democrat hold substantially higher beliefs about the severity of the crisis and show higher economic concerns. However, independent of the specific demographics, we observe that beliefs about the severity of the crisis as well as economic worries increased for all subgroups between March 5 and March 16.


In sum, the data indicate that over 11 days individuals' perceptions of the severity of the crisis strongly intensified and their economic worries substantially increased. This finding is in line with results obtained using other data sources. Within the same time frame, aggregate Google search intensity for the "Recession" topic increased by a factor of 10 in the US and by a factor of 5.5 on the global level (Online Appendix Figures \ref{fig:trend_late}, Panels A and B).\footnote{The evolution of the search patterns for the topics "Stock Market Crash", "Conspiracy Theory", and "Survivalism" was qualitatively similar (Online Appendix Figures \ref{fig:trend_late}, Panels C-H).} We also confirm our findings using other nationally representative opinion polls conducted between March 5 to 8 and March 16 to 19 2020 (Online Appendix Figure \ref{fig:polldata}).

\pagebreak


To further quantify the effects of the within-country spread of the coronavirus, we analyze the local arrival in a difference-in-differences analysis at the state level (described in section \ref{sec:opinionpolling} of the Online Appendix), exploiting the fact that some states saw their first confirmed case in the time between our two surveys. The time fixed effects we include allow us to control for aggregate developments, such as stock market movements. Online Appendix Table  \ref{table:survey_DiD} shows that having at least one case is associated with significantly more pessimistic beliefs about the severity of the impact on the world (0.23 standard deviations, $s.e.=0.10$) and on the US (0.26 standard deviations, $s.e.=0.09$). It is also associated with higher worries about the US economy by 0.22 standard deviations ($s.e.=0.11$).



\section{Perceptions of Pandemic Risk Factors}\label{sec:experimentsperceptions}

The rapid increase in economic anxieties during and after arrival of the coronavirus sets the stage to test for the underlying determinants. In this section, we examine the role of perceptions of pandemic risk factors along two dimensions. First, we conducted two experiments that allow us to causally assess the impact of individuals' perceptions of coronavirus mortality and contagiousness on economic anxiety. Importantly, through experimental variation we are able to isolate the direct effect of perceptions from other environmental variables that affect all participants symmetrically, such as stock market conditions. Second, we study respondents' mental models of infectious disease spread and the role of these mental models in shaping economic anxiety.

\subsection{Descriptives: Perceptions of Coronavirus Mortality and Contagiousness}

What beliefs did people hold about pandemic risk factors at the onset of the coronavirus crisis? At the time of our first survey on March 5, there was still substantial uncertainty and public disagreement about how severely the US economy would be affected by the coronavirus. In our survey, we measured participants' beliefs about two key characteristics that are relevant for the pandemic threat of the coronavirus: mortality and contagiousness (R0), i.e. the expected number of infections directly caused by one infected person.

We elicited participants' beliefs about the mortality of the coronavirus using the following question: \textit{``Out of 100 people who are infected with the coronavirus, how many do you think will die as a result of catching the virus?''}. Beliefs about the contagiousness (R0) of the virus were elicited using the following question: \textit{``Think of a person who has the coronavirus. How many non-infected people do you think will catch the virus from this person?''}. 

 Our data indicate substantial heterogeneity in participants' beliefs about these characteristics of the coronavirus (see Panels A and B of Figure \ref{fig:exp_res}). On average, participants' beliefs about both the mortality from the coronavirus as well as its contagiousness were substantially higher than official and scientific estimates. The median participant estimated a mortality of 5\% (mean of 14\%) compared to an estimate of 3.4\% provided by the World Health Organization (WHO) around the time of the surveys. Similarly, the median participant estimated a contagiousness (R0) of 10 (mean of 43) relative to scientific estimates at the time of the survey in the range of R0 $\approx 2$ \citep{li2020early,wu2020nowcasting}.

These coronavirus beliefs are substantially positively associated with economic anxiety (Figure \ref{fig:exp_res}, Panel C and Online Appendix Table \ref{table:experiment_correlations}, Panel A). For instance, holding mortality and contagiousness beliefs that are higher than the official WHO and scientific estimates \citep{li2020early,wu2020nowcasting} is associated with higher worries about one's personal economic situation by a magnitude of $0.48$ ($s.e.=0.063$) and $0.41$ ($s.e.= 0.082$) standard deviations, respectively. The quantitative results remain virtually unchanged when controlling for demographic and socioeconomic controls (Online Appendix Table \ref{table:experiment_correlations}, Panel B) and persist when using a continuous measure of perceptions rather than binary variables (see Online Appendix Table \ref{table:experiment_correlations_cont}). Following a complementary analysis, Online Appendix Figure \ref{fig:nonpar} visualizes non-parametric relationships, underscoring the substantial positive association with economic worries along the belief distribution.\footnote{The positive association is particularly pronounced for individuals who hold lower beliefs about coronavirus mortality and contagiousness, potentially because increases at low levels induce larger perceived marginal effects of the crisis on economic prospects.}

\subsection{Experimental Treatments}

To understand whether beliefs about the mortality and contagiousness of the coronavirus causally affect economic anxiety, we administered one framing treatment as well as an information treatment. The structure of the experiments was as follows: in the first component of the experiments, a random subset of respondents was assigned to receive the ``high relative mortality'' treatment, while the remaining respondents were assigned to receive the ``low relative mortality treatment''. Subsequently, we elicited participants' economic worries. In the second component of the experiments, we randomly assigned some respondents to get truthful information about the contagiousness and then re-elicited participants' economic worries.

\paragraph{Framing of relative mortality:}
Our first experimental variation focuses on the framing of mortality risk. In the experiment, participants were either truthfully informed, based on the same scientific estimate of coronavirus mortality at the time of the survey, that the death rate from the coronavirus is \textit{``20 times higher than for the flu''} (high mortality treatment) or \textit{``5 times lower than for SARS''} (low mortality treatment). The wording was chosen to mirror how information is commonly communicated in the media. We study how these different framings of mortality of the coronavirus affect participants' expectations about the severity of the effects of the coronavirus in general, and participants' worries about the effects on the aggregate economy and their personal economic situation. Econometrically, we estimate treatment effects using the following specification: 

 \begin{equation}
   y_i = \beta_0 + \beta_1 \textnormal{highrelativemortality}_{i}+\varepsilon_i
 \end{equation}
 where $y_i$ is the z-scored outcome of interest for individual $i$ and $\textnormal{highrelativemortality}_{i}$ is a dummy variable indicating whether individual $i$ was exposed to the high mortality framing.\footnote{Randomization achieved excellent balance (see Online Appendix Table \ref{table:balance}).} In additional robustness tests, we also test for the robustness of the results when including demographic and socioeconomic controls, including gender, age bin dummies, log income, log income squared, dummies for having a high school degree and having some college education, dummies for being unemployed, currently working, a student and for self-identifying as Democrat or Republican.

Relative to the low mortality treatment, the high mortality treatment causally leads participants' to hold higher beliefs about the crisis' severity for the world and the US: respondents in the high mortality treatment display $0.28$ ($s.e.=0.066$) and $0.23$ ($s.e.=0.066$) standard deviations higher beliefs about the crisis' severity for the world and the US, respectively (Online Appendix Table \ref{table:experiment_mainpart1}). 

\begin{figure}[!h]
\caption{Beliefs About Coronavirus and the Effect of Information on Economic Worries}
\label{fig:exp_res}
\centering
\begin{tabular}{cc}
\includegraphics[height=4cm]{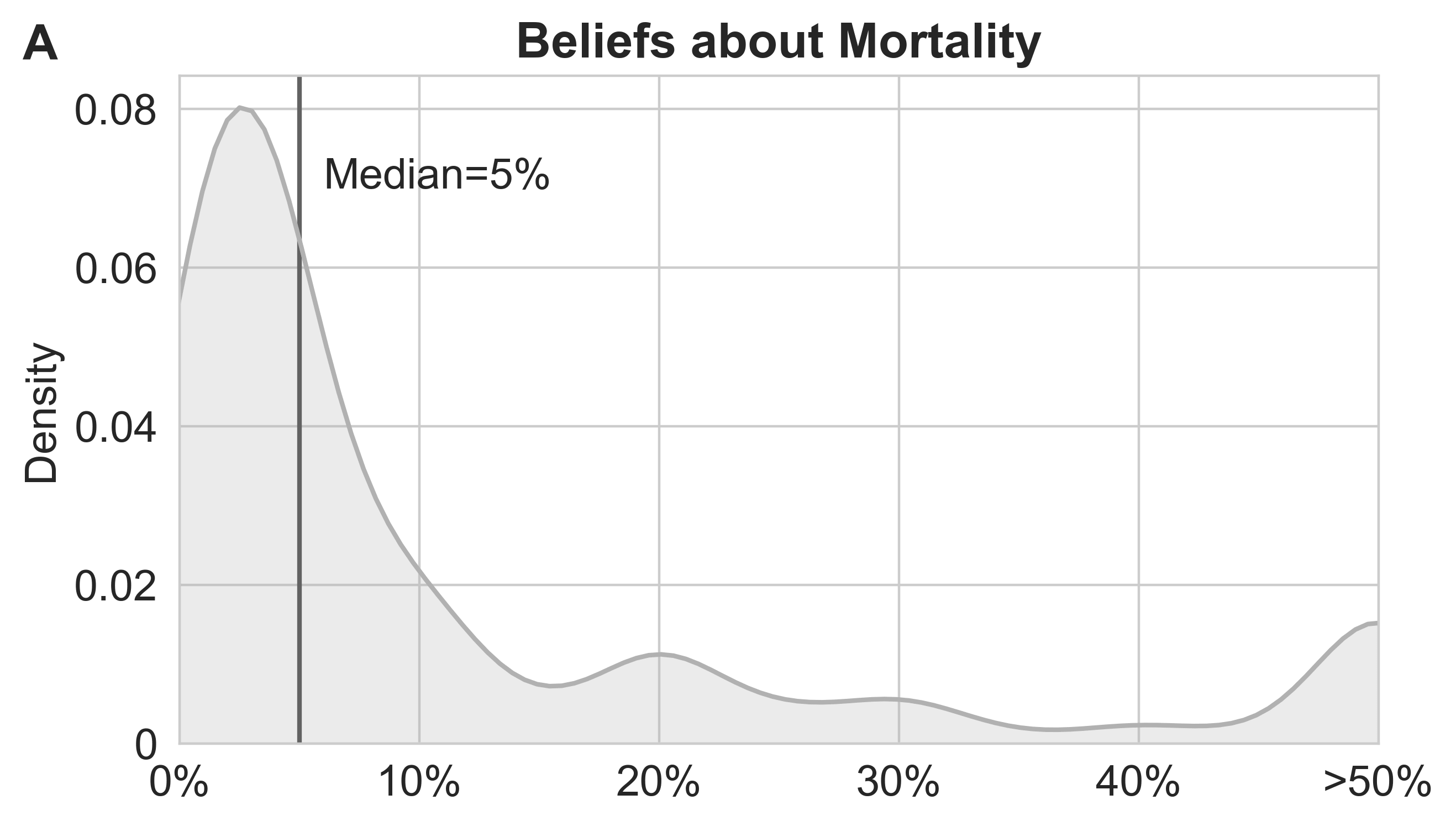} &\includegraphics[height=4cm]{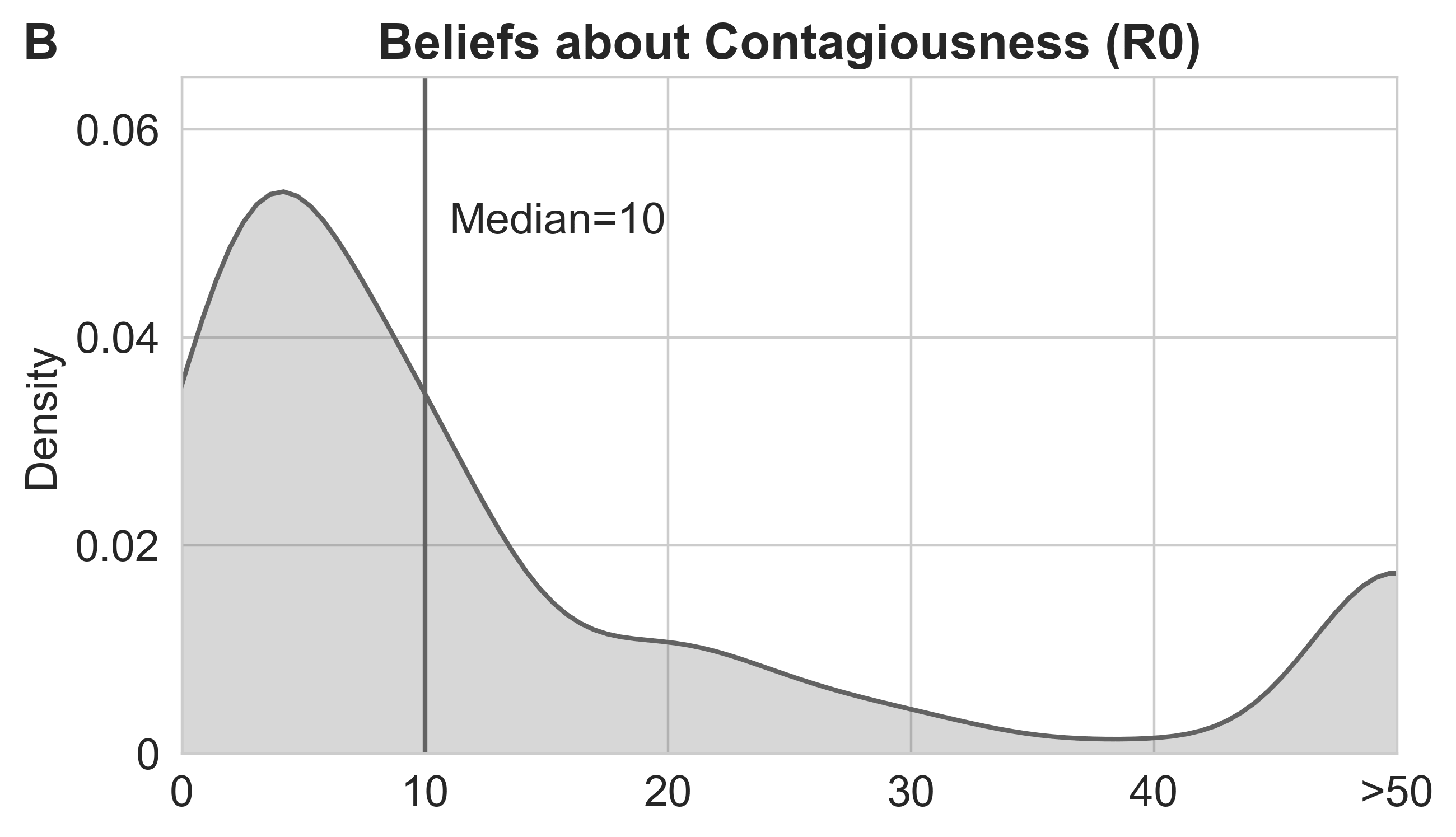}\\
\includegraphics[height=5cm]{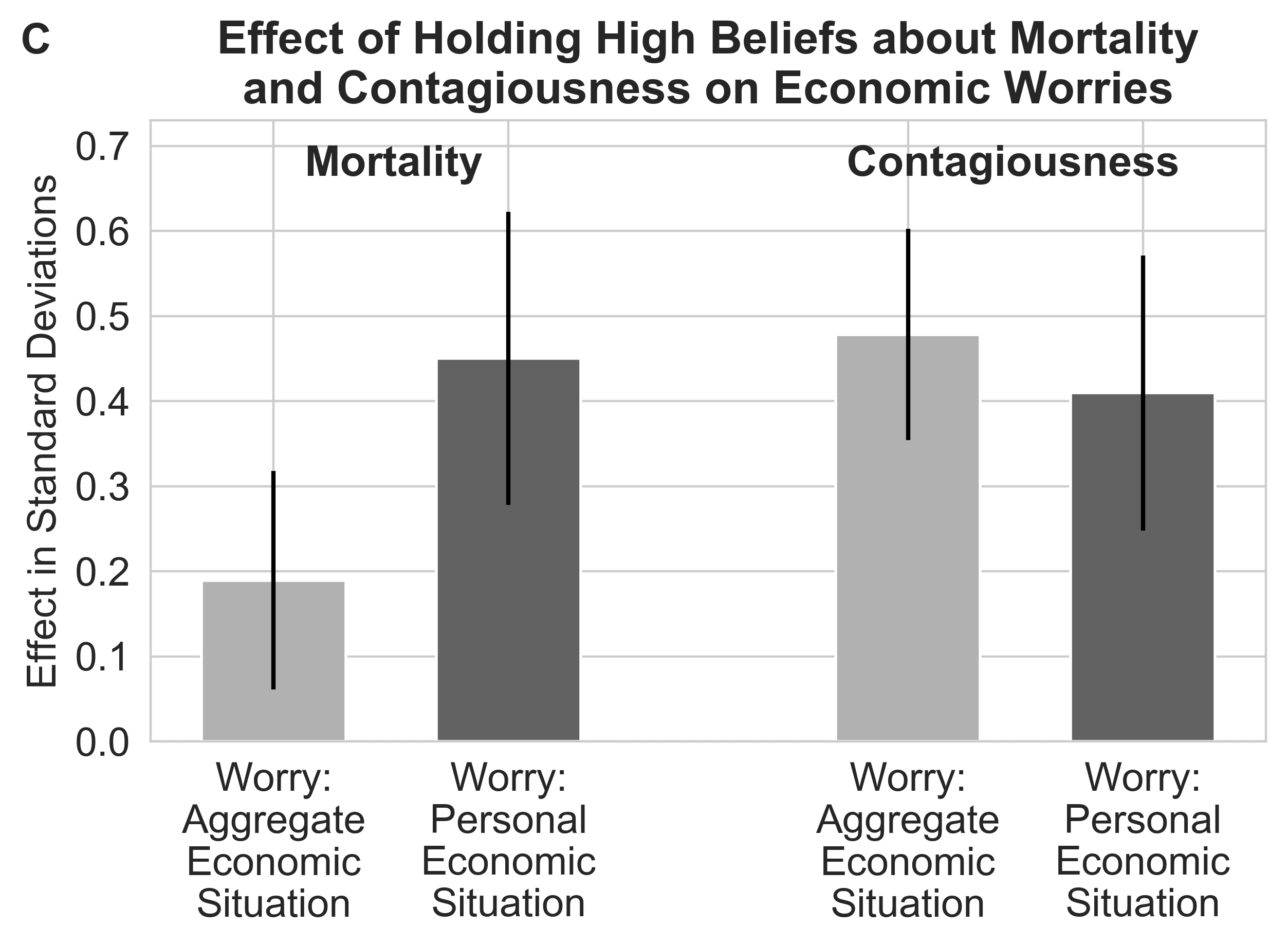}  &\multicolumn{1}{c}{\includegraphics[height=5cm]{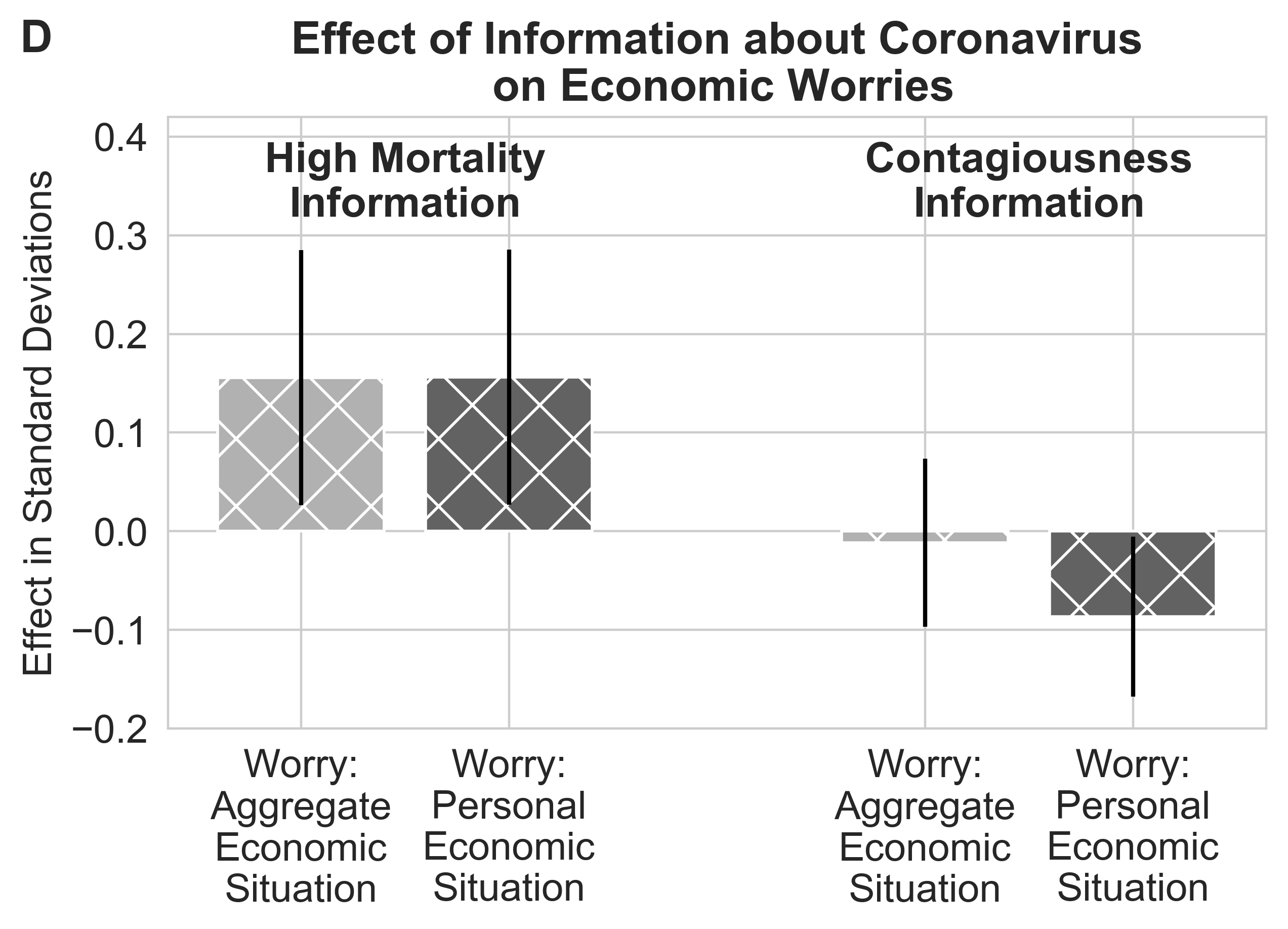}} 
\end{tabular}
\parbox{15cm}{\footnotesize \textit{Notes:} Figure \ref{fig:exp_res} displays perceptions of the novel coronavirus and the experimental results. The data were collected on March 5. Panel A and B show the distribution of beliefs about mortality and contagiousness (R0) of the coronavirus. Panel C shows the effect of overestimating mortality and contagiousness relative to official numbers on worries about the aggregate US economy and respondents' personal economic situation. Panel D shows the experimental results on the effect of information about the coronavirus on economic worries. The two leftmost bars in Panel D show the effect of information suggesting high relative mortality as opposed to low relative mortality on worries about the aggregate US economy and one's personal economic situation. The two rightmost bars in Panel D show the effect of information about contagiousness on worries about the aggregate US economy and one's own personal economic situation. In all panels, error bars indicate 95\% confidence intervals.}

\end{figure}

These treatment differences also persist for participants' economic worries (Figure \ref{fig:exp_res}, Panel D and Online Appendix Table \ref{table:experiment_main}, Panel A): relative to the low mortality treatment, respondents in the high mortality treatment increase their worries about the effects of the coronavirus on the US economy by 0.16 ($s.e.=0.066$) standard deviations and about their personal economic circumstances by 0.16 ($s.e.=0.066$) standard deviations. The quantitative effect sizes correspond to 106\% and 102\% of the Republican-Democrat gap, respectively, and are virtually unchanged when controlling for demographic and socioeconomic controls  (Online Appendix Table \ref{table:experiment_main}, Panel B). These results highlight the influence of framing of news stories on public perceptions \citep{prat2013political,chong2007framing}.

\paragraph{Information about contagiousness:}

Besides mortality, contagiousness is a key characteristic that influences the pandemic risk of an infectious disease. The higher disease contagiousness, the larger is the risk of widespread and fast infection of the population which can lead to an overload of the health care system \citep{massonnaud2020covid} and costly disruption of economic activity \citep{Abba2016spread}. To understand the causal effect of beliefs about contagiousness on economic anxiety, in the second part of the experiment we administered an additional information treatment.

After eliciting participants' beliefs about the contagiousness (R0) of the coronavirus, the participants were randomly assigned to be either in a ``contagion information group'' or a control group, which received no information. Based on scientific estimates \citep{li2020early,wu2020nowcasting}, respondents in the contagion information group were informed that \textit{``approximately 2 non-infected people will catch the coronavirus from a person who has the coronavirus''}. Given that 81.4\% of respondents overestimate this statistic, the information treatment should decrease the perceptions of the contagiousness. 

To test for the effect on economic anxieties, we re-elicited participants' worries about the effects of the coronavirus on the US economy and their household's economic situation as before. To analyze this treatment, we use an ANCOVA specification of the following form:

 \begin{equation}
   y_i = \delta_0 + \delta_1 \textnormal{contagiousnessinfo}_i+ \delta_2 y_{i,-1}+\varepsilon_i
 \end{equation}
where $\textnormal{contagiousnessinfo}_i$ is a dummy variable indicating whether individuals were provided the treatment information.\footnote{Randomization achieved excellent balance (see Online Appendix Table \ref{table:balance}).} $y_{i,-1}$ is the outcome of interest measured in the same survey prior to the second experiment.\footnote{We do not control for $y_{i,-1}$ in the first specification because we did not collect any outcome data prior to the relative mortality treatment.} 

Respondents in the contagiousness information treatment show $0.09$ ($s.e.=0.041$) standard deviations lower worries about the effects of the coronavirus on their personal economic situation and a small decrease in their worries about the aggregate US economy (0.01 sd, $s.e.=0.043$) (Figure \ref{fig:exp_res}, Panel D and Online Appendix Table \ref{table:experiment_main}, Panel A). The quantitative effect sizes correspond to 57\% and 7\% of the Republican-Democrat gap, respectively, and are virtually unchanged when controlling for demographic and socioeconomic controls (Online Appendix Table \ref{table:experiment_main}, Panel B) or when controlling for a treatment indicator for the relative mortality treatment (Online Appendix Table \ref{table:experiment_main_crossrcheck}).\footnote{As Online Appendix Table \ref{table:experiment_main_int} indicates, there are no significant interaction effects between the treatments.}

In sum, the experimental evidence indicates that perceptions and information regarding coronavirus mortality and contagiousness are significant causal determinants that shape individuals' expectations about the aggregate economy and their personal economic situation at a time of high uncertainty.

\pagebreak

\subsection{Mental Models of Infectious Disease Spread}

The public health impacts associated with a pandemic vary as a disease spreads through space and time. We already documented that economic anxieties evolve dynamically with this spread. However, so far we have not analyzed how the anticipation of such developments shapes economic anxieties. Besides information about risk factors, forward-looking individuals rely on mental models of the world to make predictions about the future, and in the context of a pandemic, the future extent of disease spread. To analyze this question, we investigate participants' subjective mental models of infectious disease spread to understand the role of cognitive processes and their limitations in shaping economic anxiety in response to the outbreak of the coronavirus pandemic.

As humans are organized in networks, disease spread typically follows a non-linear (e.g. logistic or quasi-exponential) function, at least in the beginning of an outbreak \citep{keeling2011modeling,kermack1927contribution}. Hence, a small number of cases can rapidly evolve into a widespread pandemic. Such a trajectory can be vastly underestimated if individuals do not take into account the non-linear nature of disease spread but rather adopt a mental model of linear growth. 

To systematically investigate this question, we asked participants in our March 16 survey to predict the spread of a fictitious infectious disease under simplifying assumptions. We elicited participants' predictions about the spread of a fictitious disease rather than asking participants for their estimates of the future number of coronavirus cases for three reasons. First, investigating the role of cognitive processes requires the elicitation of individuals' abstract mental models rather than their predictions for the specific case of the coronavirus pandemic. Second, predictions about the future severity of the coronavirus pandemic will be crucially shaped by individuals' expectations about the extent of endogenous containment measures as well as societal reactions which are independent of the general nature of infectious disease spread. Third, no reliable benchmark is available for the future spread of the coronavirus, making it infeasible to assess the ex-ante accuracy of estimates. 
 
Participants were instructed to assume that on a day 1, one person has a fictitious disease and that each day a newly infected person infects two healthy people before stopping being contagious. To provide some guidance, participants were further informed that on day 2, 3 people will be infected as the person who had the disease on day 1 spread it to two other people on day 2. Participants were then asked to predict the total number of people infected with the fictitious disease on day 5, 10, and 20.
 
\begin{figure}[!h]\centering
\caption{Mental Models about the Spread of Infectious Diseases and Economic Worries}
\label{fig:nonlin}
\begin{tabular}{cc}
\includegraphics[height=6cm]{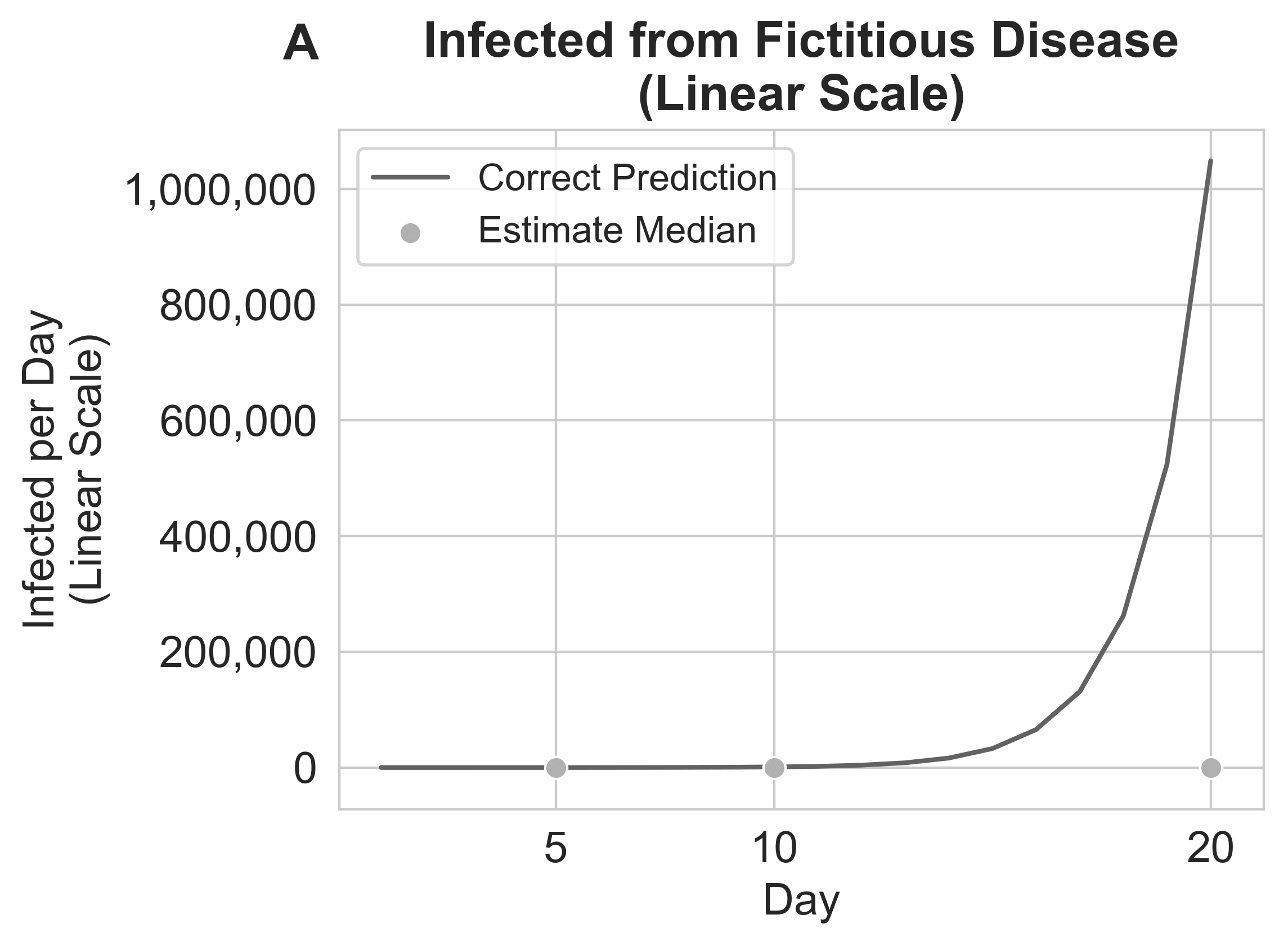} &\includegraphics[height=6cm]{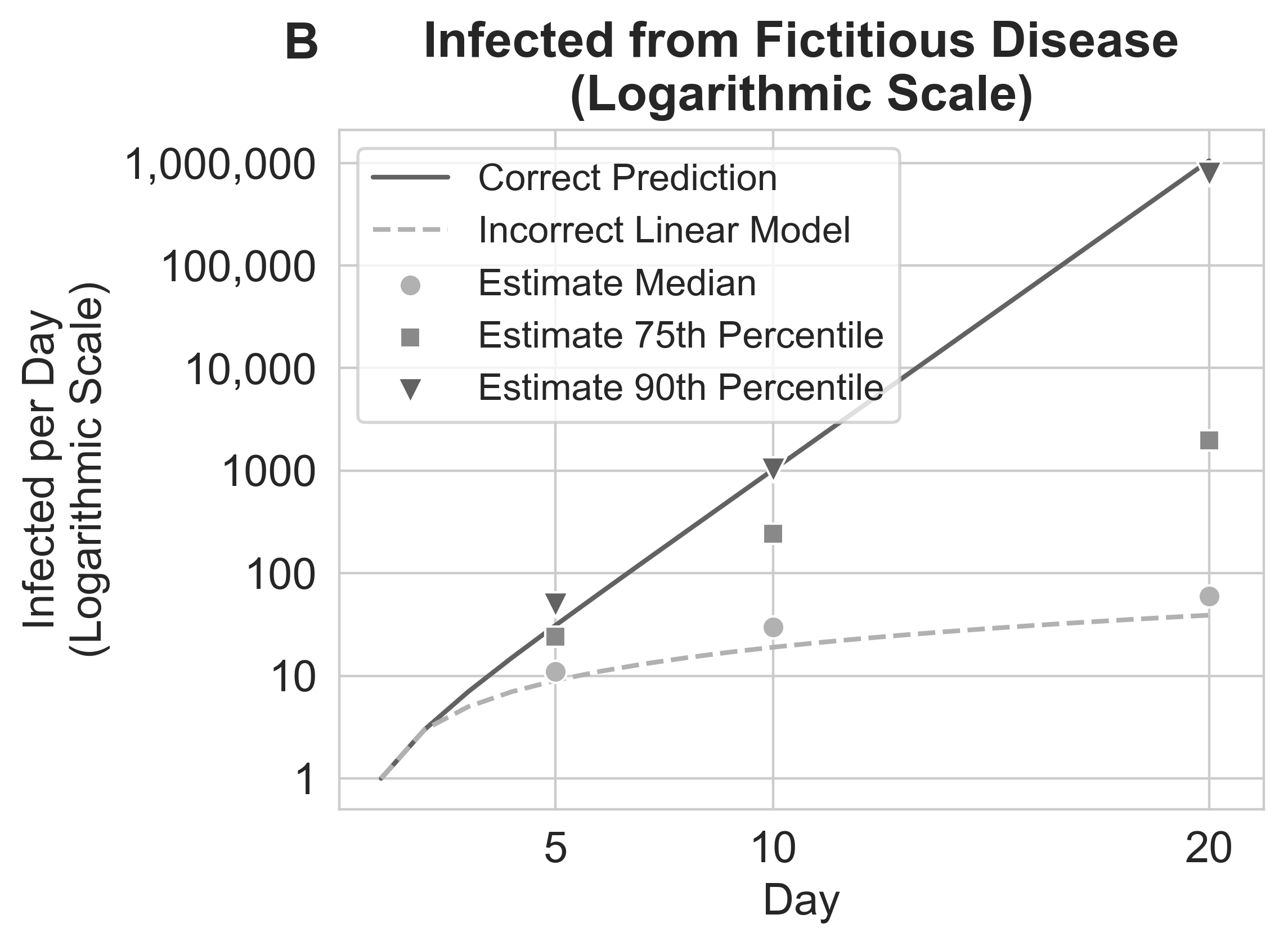}\\
\multicolumn{2}{c}{\includegraphics[height=6cm]{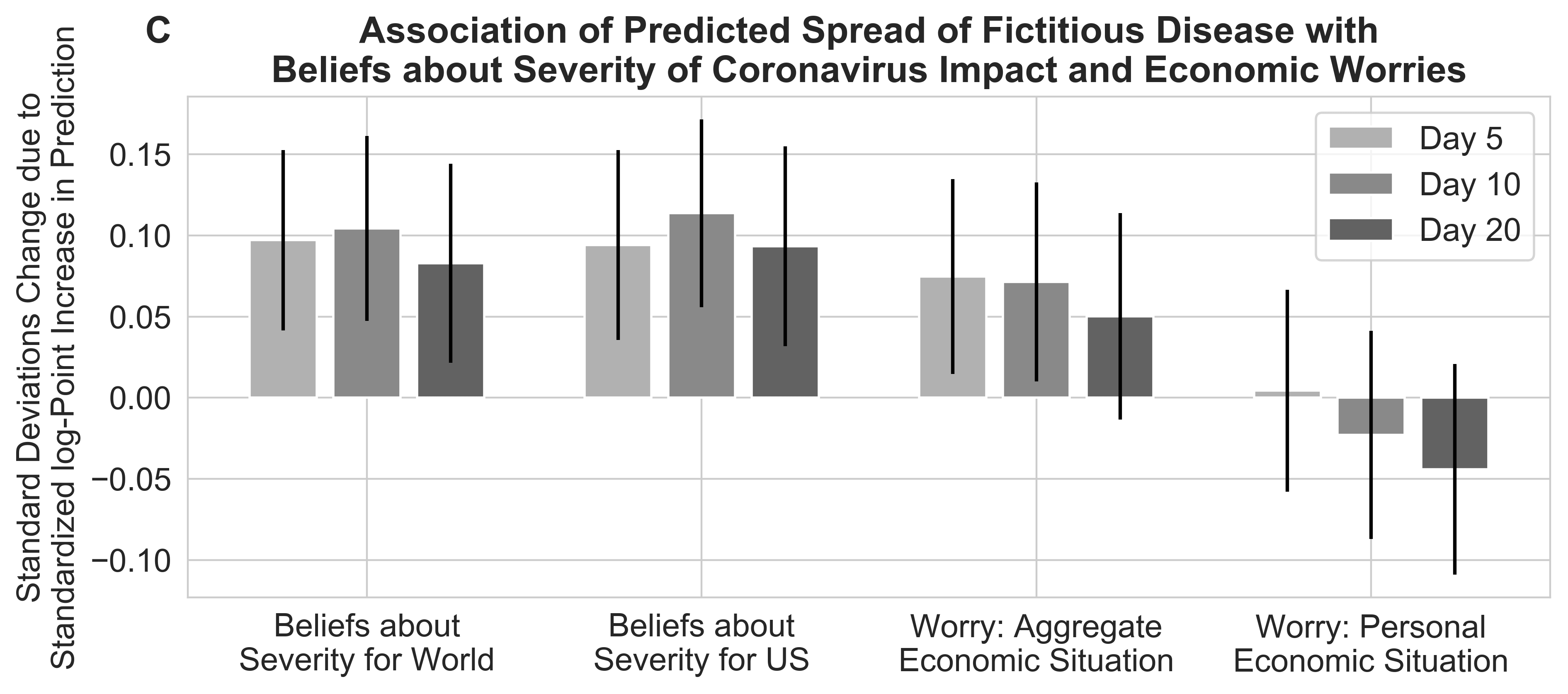}}
\end{tabular}
\parbox{15cm}{\footnotesize \textit{Notes:} Figure \ref{fig:nonlin} visualizes mental models of infectious disease spread and their association with economic anxieties. The data were collected on March 16. Panel A shows participants' median belief about the spread of a fictitious disease on a linear scale. Panel B shows participants' median, 75th percentile, and 90th percentile belief about the spread of a fictitious disease on a logarithmic scale. Participants were instructed to predict the number of cases of a fictitious disease on day 5, 10, and 20. Participants were informed that on day 1, one person has the disease and that each day a newly infected person infects two healthy people and then stops being contagious. In both panels, the solid line indicates the correct prediction. In panel B the dashed line indicates an incorrect linear model with a growth rate of 2 per day. Panel C displays the association of predicted spread of the fictitious disease with participants' beliefs about the severity of the impact of the coronavirus pandemic on the world and the US, as well as worries about the aggregate US economy and their personal economic situation. In all panels, error bars indicate 95\% confidence intervals.}
\end{figure}

Figure \ref{fig:nonlin}, Panel A shows the median participant's estimates and the correct prediction values. The results indicate that the average individual highly underestimates the spread of the fictitious disease. In contrast to correct prediction values of 31 on day 5, 1023 on day 10, and 1,048,575 on day 20, the median participant estimates a case number of 16 on day 5, 30 on day 10, and 60 on day 20. Inconsistent with non-linear growth, the predictions of the median participant can be well approximated by a subjective linear growth model (as exemplified by the green line in Panel B of Figure \ref{fig:nonlin} for a linear growth rate of 2 per day). A linear mental model, however, is not uniformly present for the entire population. In particular, the 90th percentile prediction in our sample very well captures the correct quasi-exponential growth, indicating heterogeneity in individuals' mental models of infectious disease spread.\footnote{To investigate the sources of heterogeneity, we explore the correlates of mental models in Online Appendix Table \ref{table:pred_expgrowth}. Across several specifications, we find that being older than 65 as well as having higher levels of education and income are positively associated with a more accurate mental model of infectious disease spread.}

To understand how contemporaneous economic anxiety is associated with individuals' mental models of the spread of infectious diseases, we correlate economic anxieties described in Section \ref{sec:descriptives} with participants' predicted number of people infected with the fictitious disease on day 5, 10, and 20 (Figure \ref{fig:nonlin}, Panel C and Online Appendix Table \ref{table:correlations_expgrowth}). To address outliers in participants' predictions, we use a z-scored transformation of the logarithm of the predicted number of infected people. 


The data show statistically significant positive associations between participants' predictions and their beliefs about the crisis' severity for the world and the US as well as their worries about the aggregate US economy. For example, a one standard deviation increase in the estimate of infectious disease spread after 10 days is associated with an increase of 0.11 ($s.e.=0.03$) in participants' beliefs about the severity for the US. The quantitative effect sizes are virtually unchanged when controlling for demographic and socioeconomic controls, including importantly education level. Interestingly, there is no significant association between participants' worries about their personal economic situation and the predicted number of infected people. This finding squares with previous evidence that individuals do not fully extrapolate their individual risk from aggregate societal risk \citep{weinstein1989optimistic,bord2000sense}.

To complement this analysis, we classify each individual's mental model by implementing a k-means clustering algorithm using 3 clusters on the log space of predicted cases on day 5, 10, and 20. Panel A in Online Appendix Figure \ref{fig:mentalclass} indicates that the obtained 3 types can be summarized as a linear, exponential, or an intermediate non-linear model. Panel B reveals that around 64.8\% of participants exhibit a roughly linear model, while 15.7\% display a roughly exponential model, and 19.5\% an intermediate non-linear mental model. Finally, Panel C confirms the previous results that holding a more accurate (non-linear) mental model of infectious disease spread is associated with higher beliefs about the crisis' severity as well as higher worries regarding the aggregate economy. 

In sum, the results indicate that individuals who exhibit a more accurate mental model of non-linear growth of infectious disease spread are more worried about the aggregate effects of the coronavirus pandemic, potentially as they foresee a greater potential for a widespread contagion of the population.

\section{Conclusion}\label{sec:conclusion}
Combining global data from internet searches and two online experiments with representative samples of the US, this article documents a rapid emergence of economic anxiety at the onset of a major pandemic, and studies perceptions of pandemic risk factors as correlational and causal determinants.

Our results point to a critical role of subjective beliefs about pandemic risks as well as mental models of infectious disease spread in shaping public perception of the severity of the contemporaneous health crisis and economic anxiety. For the present case of the coronavirus, we find substantial heterogeneity in beliefs about mortality and contagiousness, two key characteristics relevant for pandemic risk. In real-time experiments, we show that information provision regarding these characteristics causally shapes economic anxiety among the population. Our experiment also shows that framing of information about the coronavirus matters for the inferences that people make. Specifically, our experiment highlights that even if journalists base their comparisons on the same mortality statistics, the choice of comparison matters. These results speak to an important debate on how media coverage and public communication of disease outbreaks affect people’s beliefs \citep{bursztyn2020misinformation}.

Moreover, consistent with exponential growth bias \citep{stango2009exponential,levy2016exponential}, for the majority of the population subjective mental models understate the non-linear nature of infectious disease spread. The heterogeneity in individuals' mental models crucially shapes their perception of the severity of a major global pandemic and affects their worries about the impact on the aggregate economy.

\pagebreak

\bibliographystyle{aer}
 
\bibliography{coronabib}


\newpage
\appendix

\begin{center}
\textbf{\LARGE Online Appendix}
\bigskip

\textbf{\large Coronavirus Perceptions and Economic Anxiety}

{ Thiemo Fetzer, Lukas Hensel, Johannes Hermle, Christopher Roth}
\end{center}

\bigskip

\setcounter{table}{0}
\renewcommand{\thetable}{A.\arabic{table}}
\setcounter{figure}{0} 
\renewcommand{\thefigure}{A.\arabic{figure}}

\begin{figure}[!htbp]
\caption{Time Series of Global Internet Searches for January and February 2020}
\label{fig:searchapp}
\centering
\begin{tabular}{c}
\includegraphics[height=10cm]{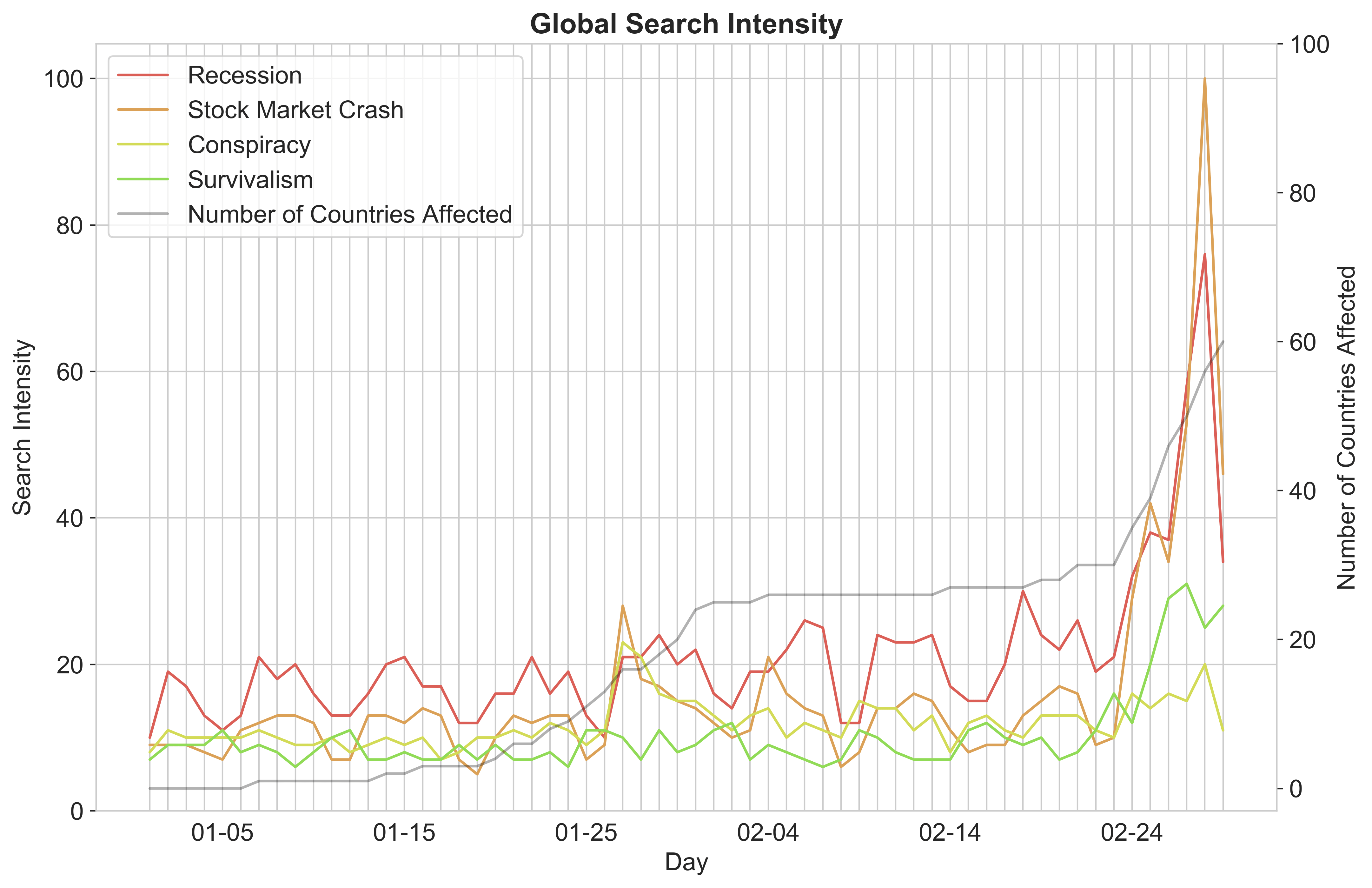} 
\end{tabular}
\parbox{15cm}{\footnotesize \textit{Notes:} Online Appendix Figure \ref{fig:searchapp} shows the time series of the search intensity for the Google topics "Recession", "Stock Market Crash", "Conspiracy Theory", and "Survivalism" from January 1st to February 29th, 2020 as well as the number of countries with a confirmed coronavirus case.}
\end{figure}

\begin{landscape}
\begin{figure}[h]
\caption{Robustness of results to dropping individual countries \label{fig:robdropping}}
\begin{center}$
\begin{array}{ll}
  \includegraphics[scale=.6]{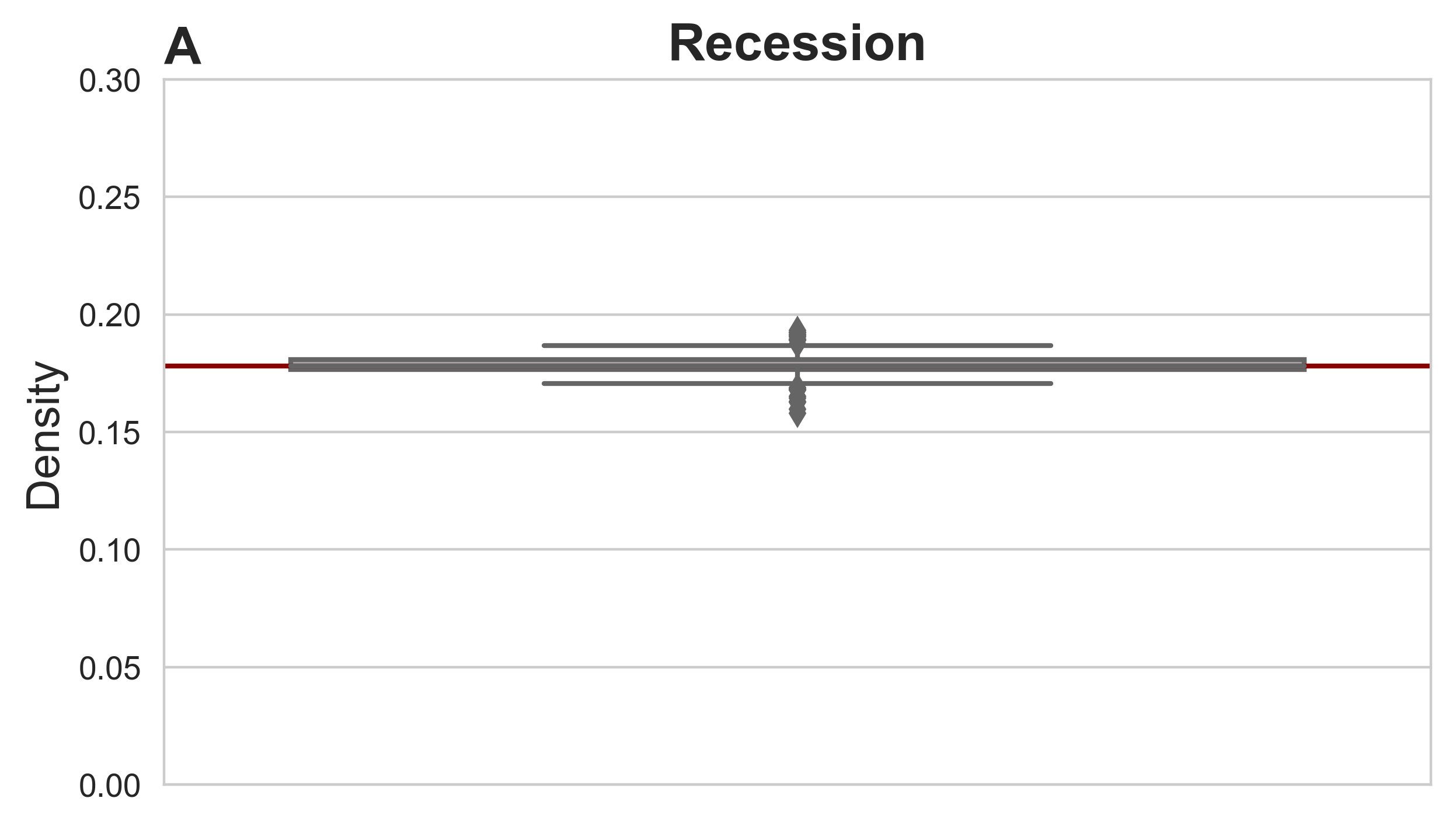}  &   \includegraphics[scale=.6]{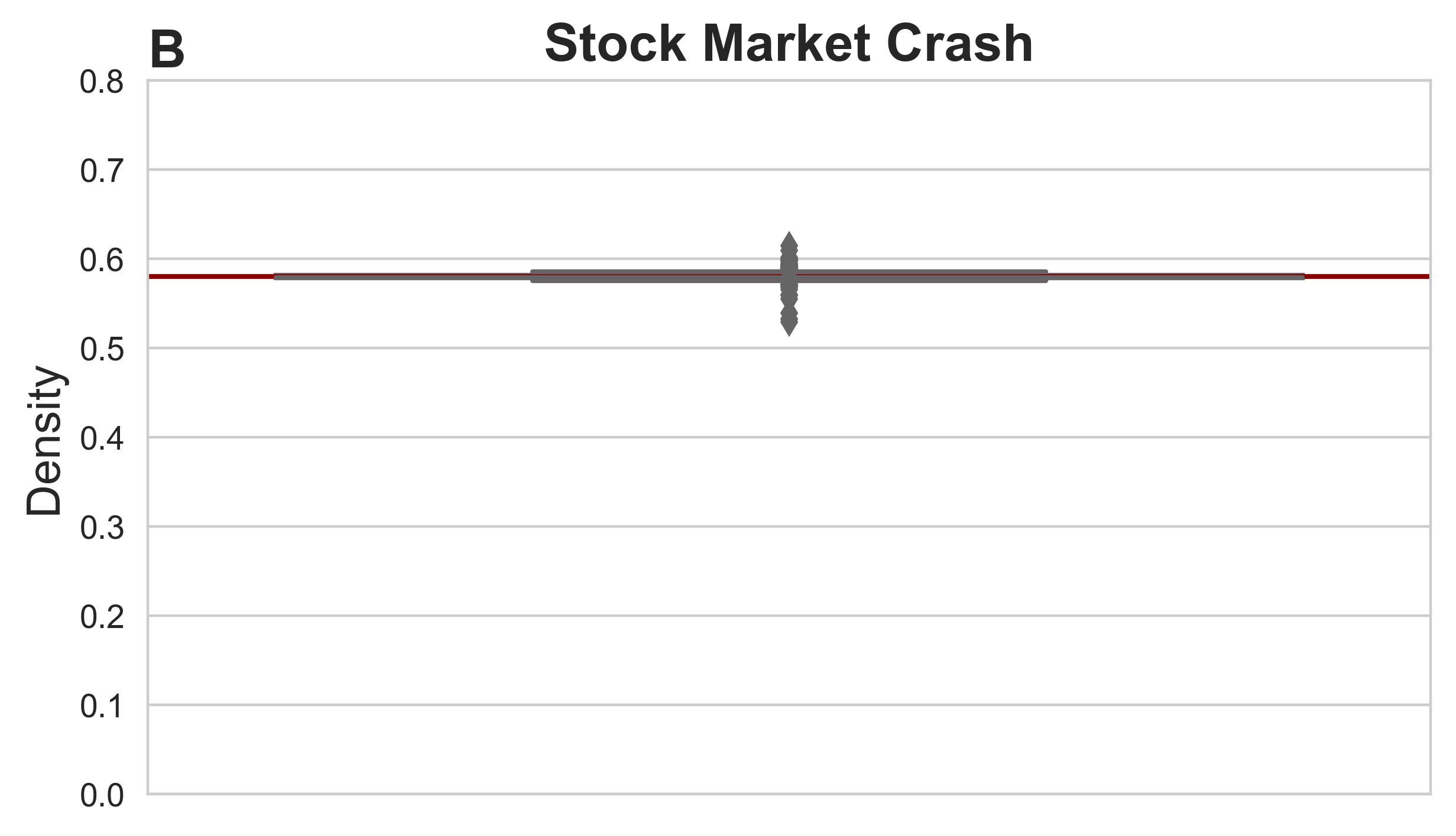} \\
  \includegraphics[scale=.6]{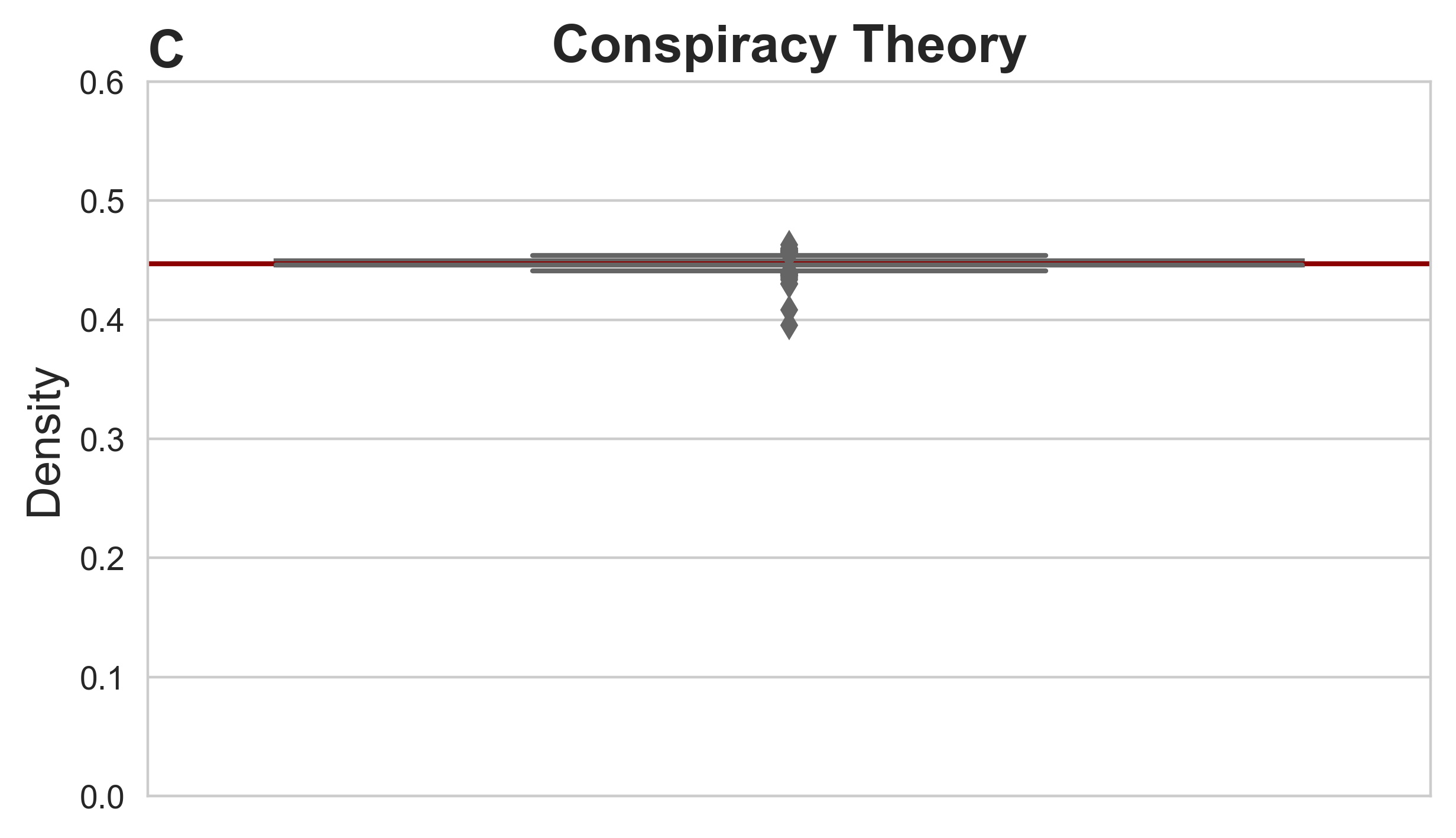}  &   \includegraphics[scale=.6]{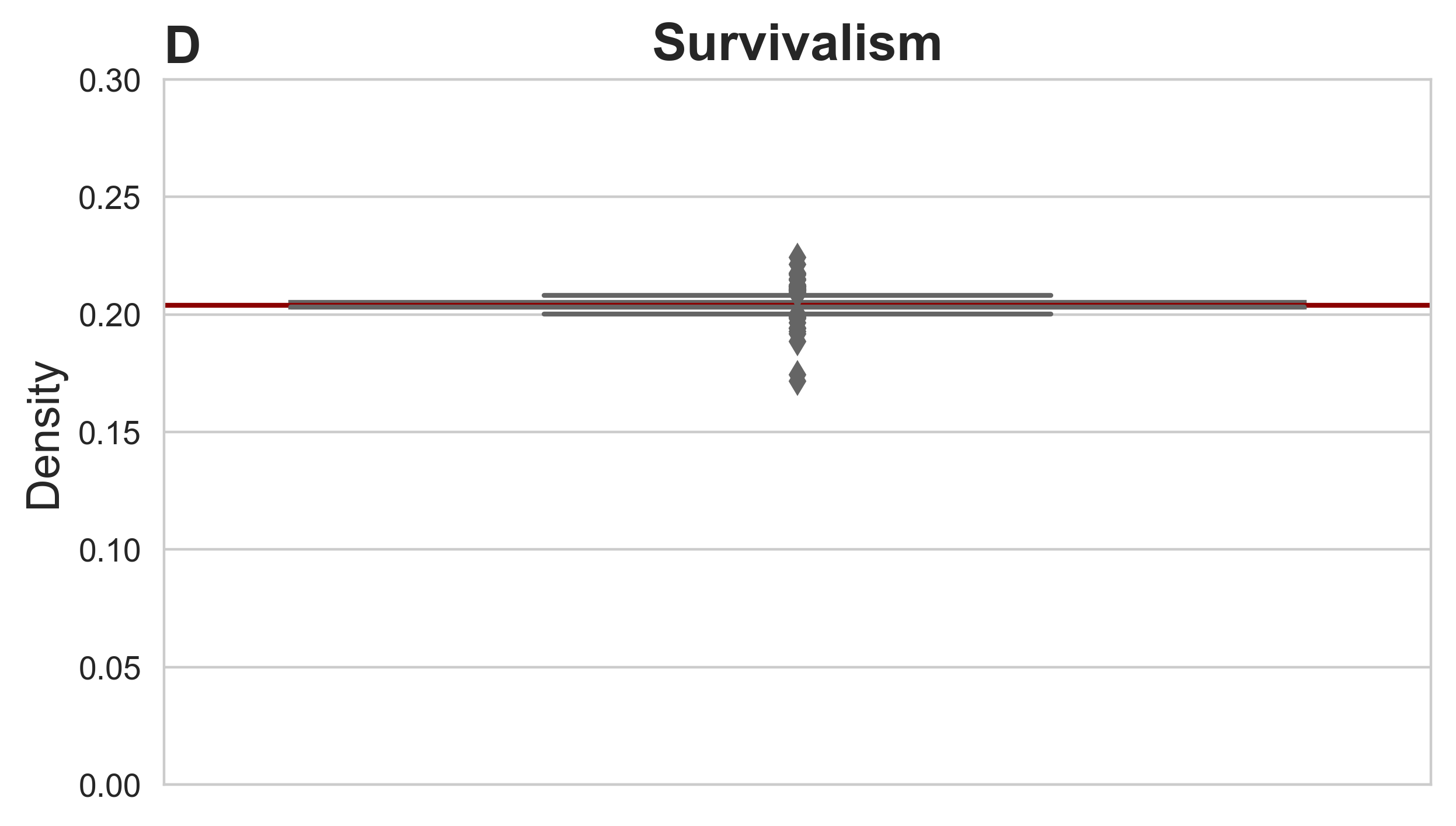} \\
  \end{array}$
\end{center}
\scriptsize{\textbf{Notes:} Online Appendix Figure \ref{fig:robdropping} shows boxplots of the point estimates obtained from estimating equation \ref{eq:didmodel} when dropping each country in turn. The coefficient obtained from estimation on the full sample is indicated by the horizontal red line.}
\end{figure}
\end{landscape}

\begin{landscape}
\begin{figure}[h]
\caption{Robustness of results to dropping individual sub-regions \label{fig:robdroppingsubregions}}
\begin{center}$
\begin{array}{ll}
  \includegraphics[scale=.6]{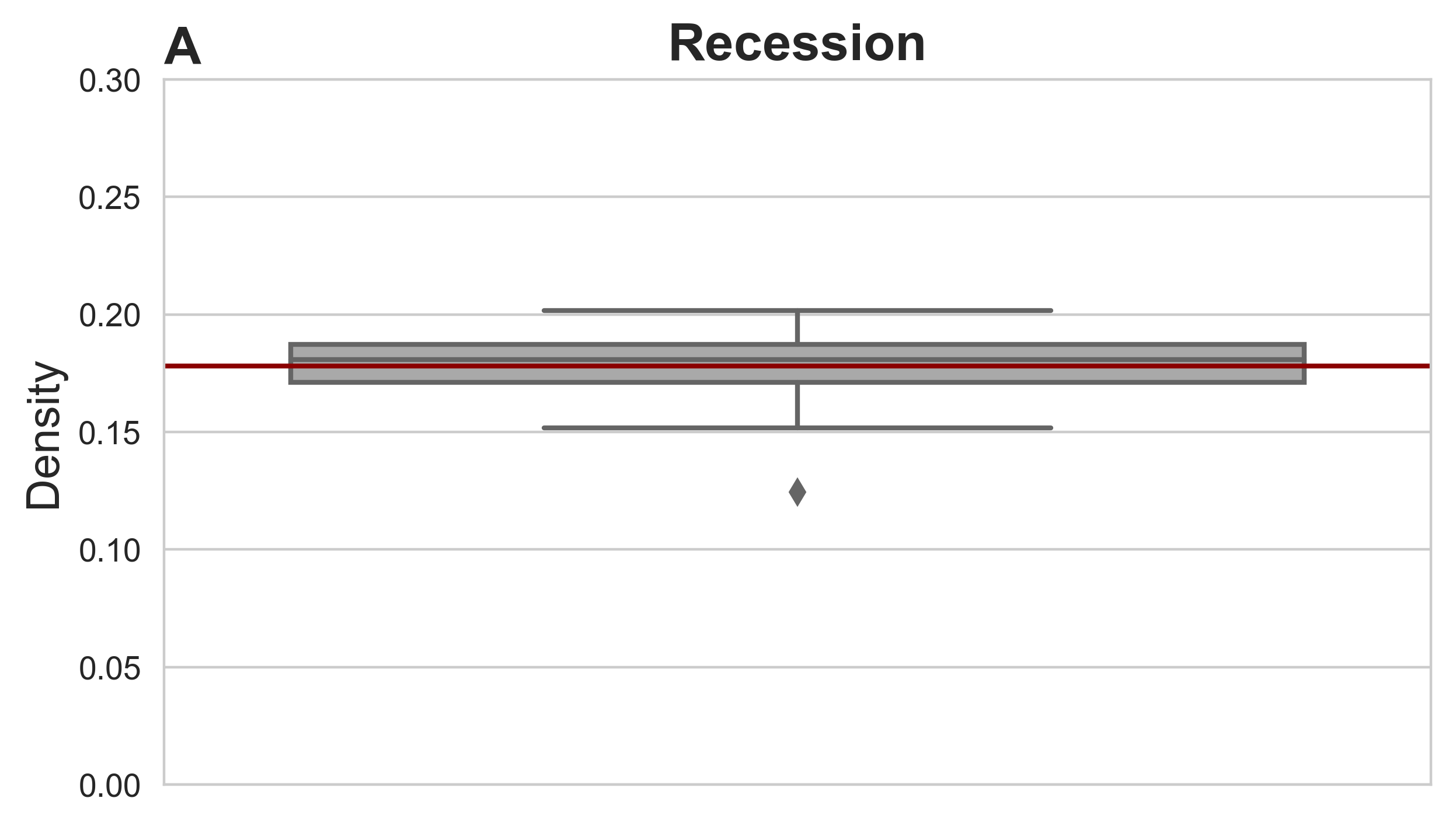}  &   \includegraphics[scale=.6]{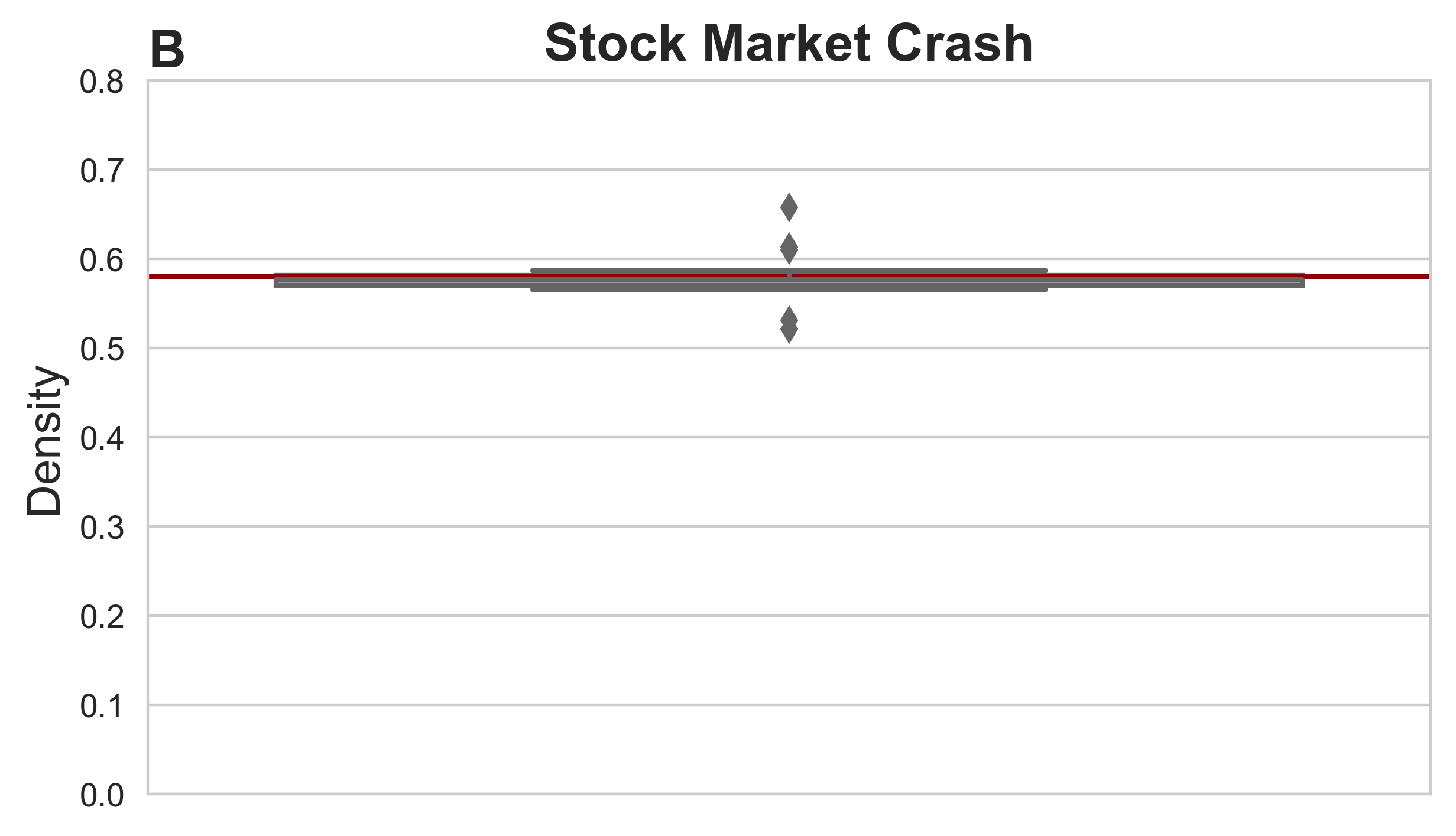} \\
  \includegraphics[scale=.6]{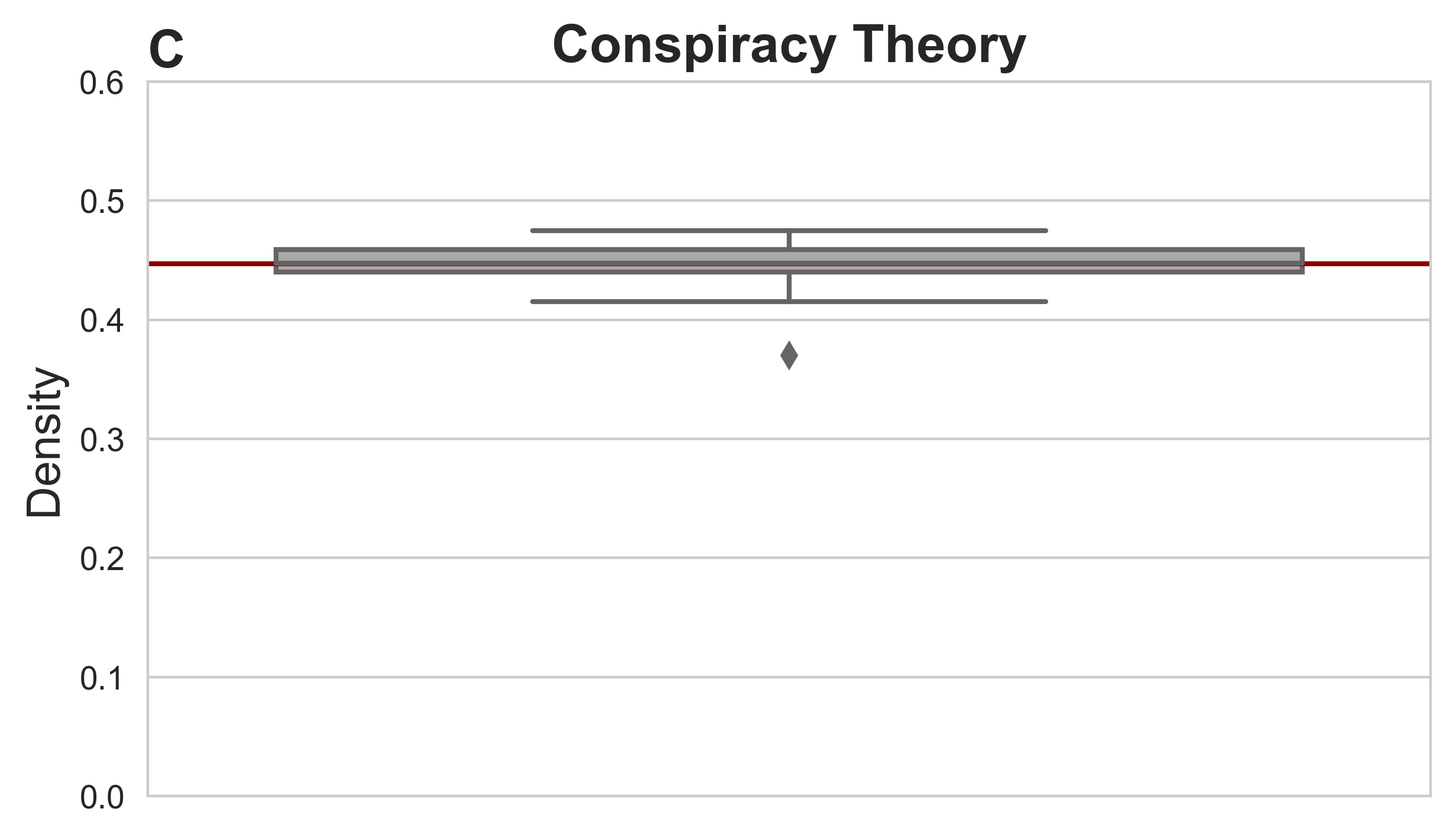}  &   \includegraphics[scale=.6]{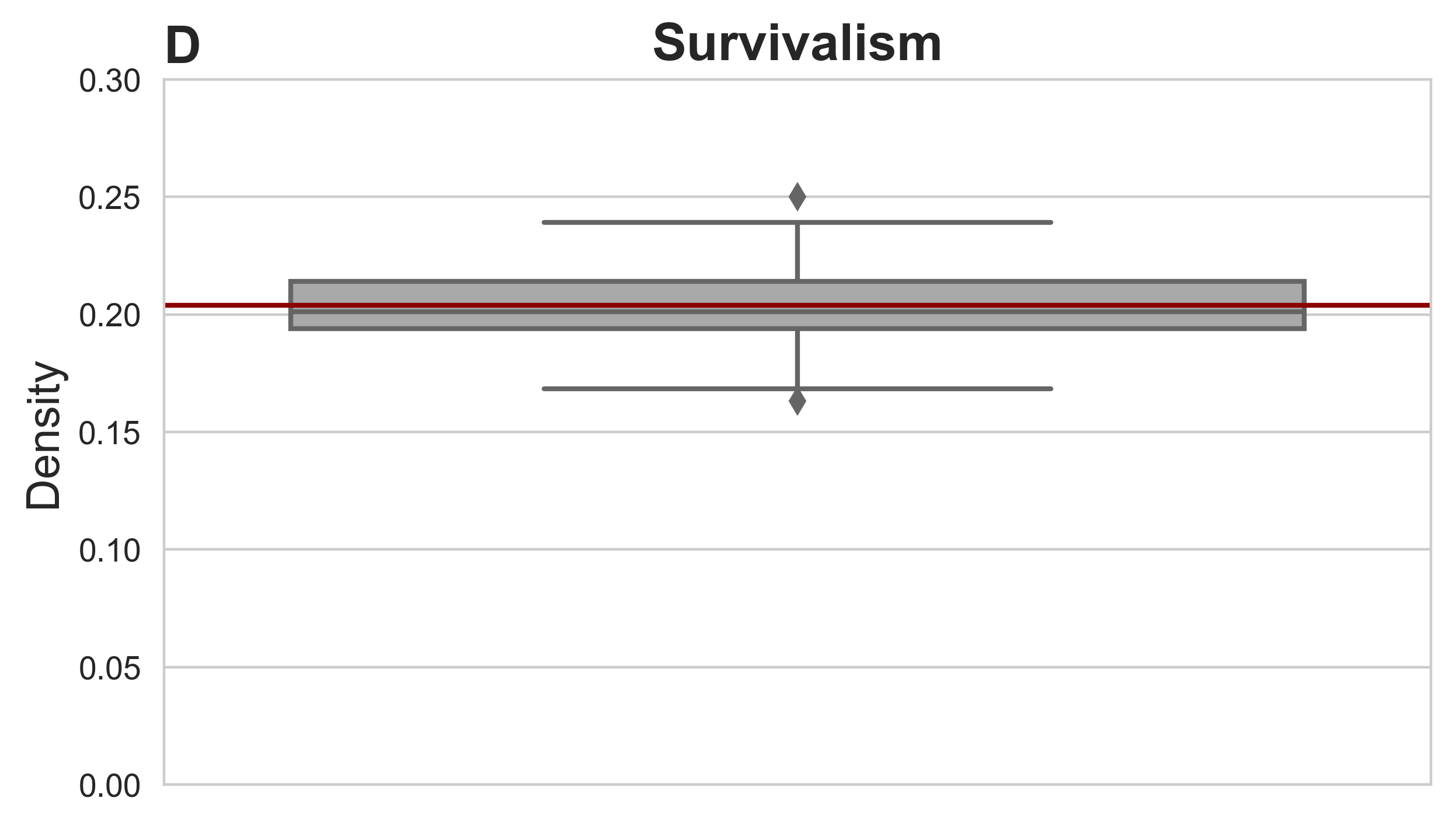} \\
  \end{array}$
\end{center}
\scriptsize{\textbf{Notes:} Online Appendix Figure \ref{fig:robdroppingsubregions} shows boxplots of the point estimates obtained from estimating equation \ref{eq:didmodel} when dropping countries belonging to each of the 17 different sub-regions in turn. The coefficient obtained from estimation on the full sample is indicated by the horizontal red line.}
\end{figure}
\end{landscape}

\begin{figure}[!htbp]
\caption{Evolution of Beliefs about Severity of Crisis and Economic Worries by Subgroups}
\label{fig:subgroupevol}
\centering
\begin{tabular}{cccc}
\\[0.1cm]
\multicolumn{4}{c}{\hspace{0cm} A\hspace{0.5cm} By Gender}\\
\includegraphics[height=5cm]{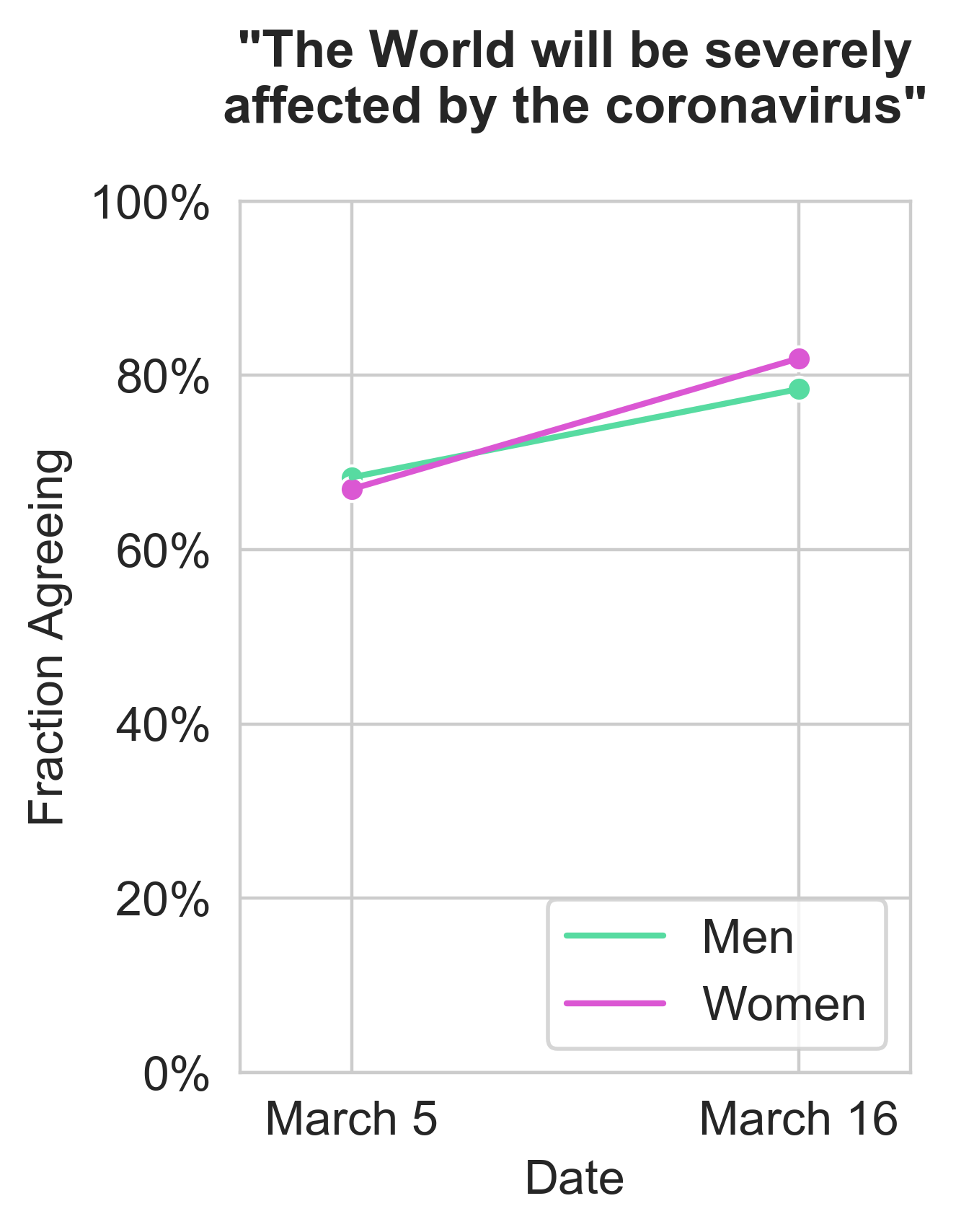} &
\includegraphics[height=5cm]{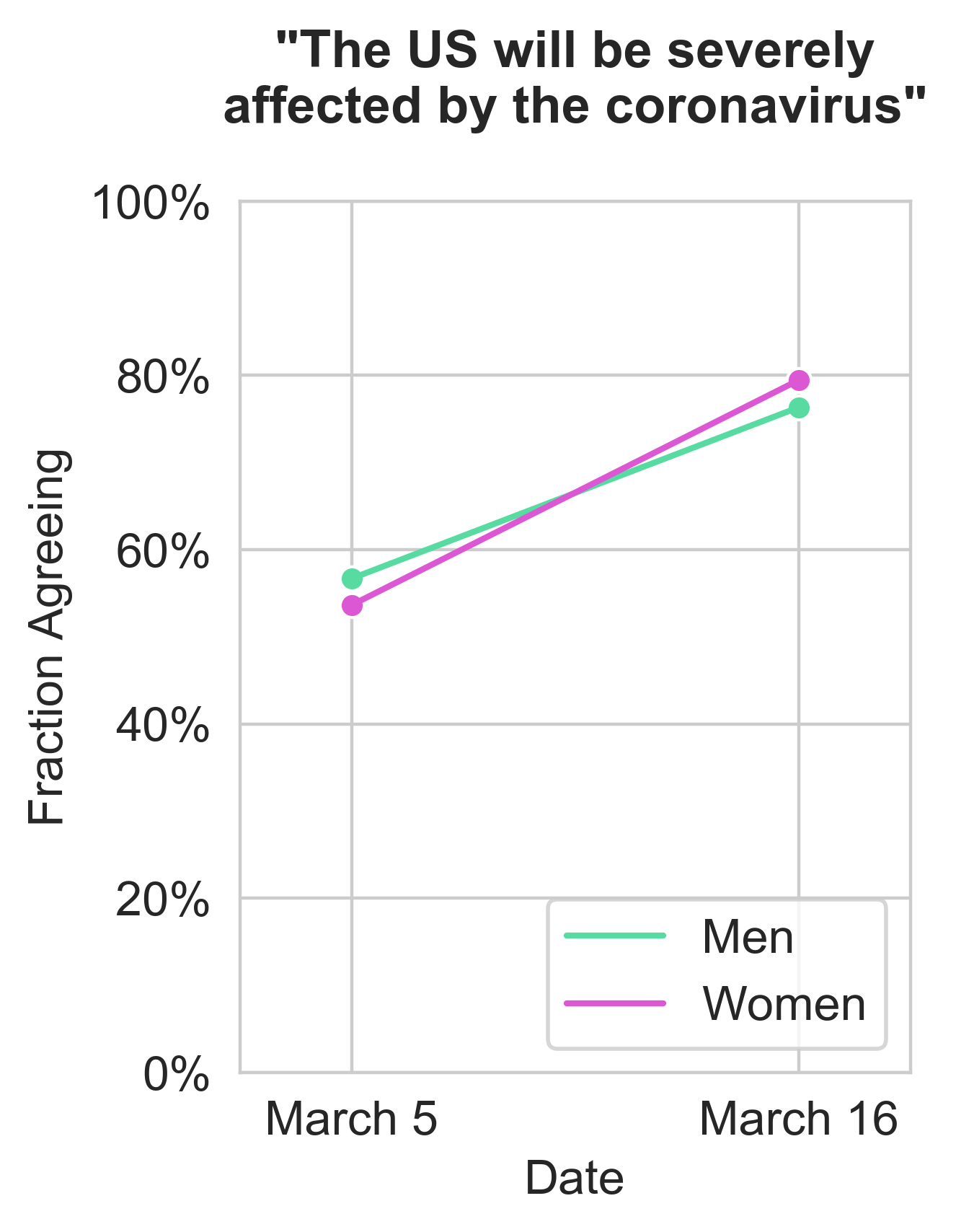} &
\includegraphics[height=5cm]{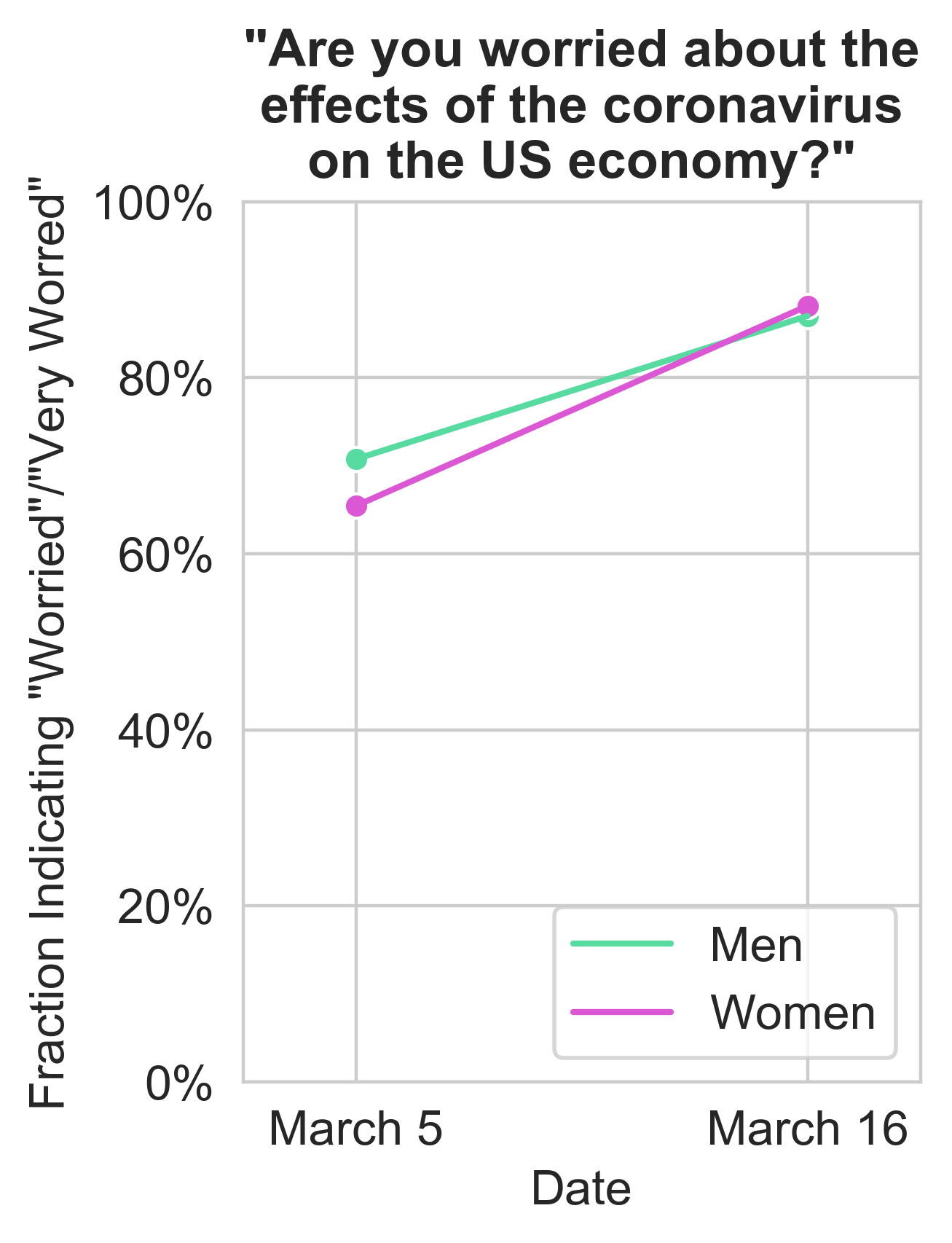} &
\includegraphics[height=5cm]{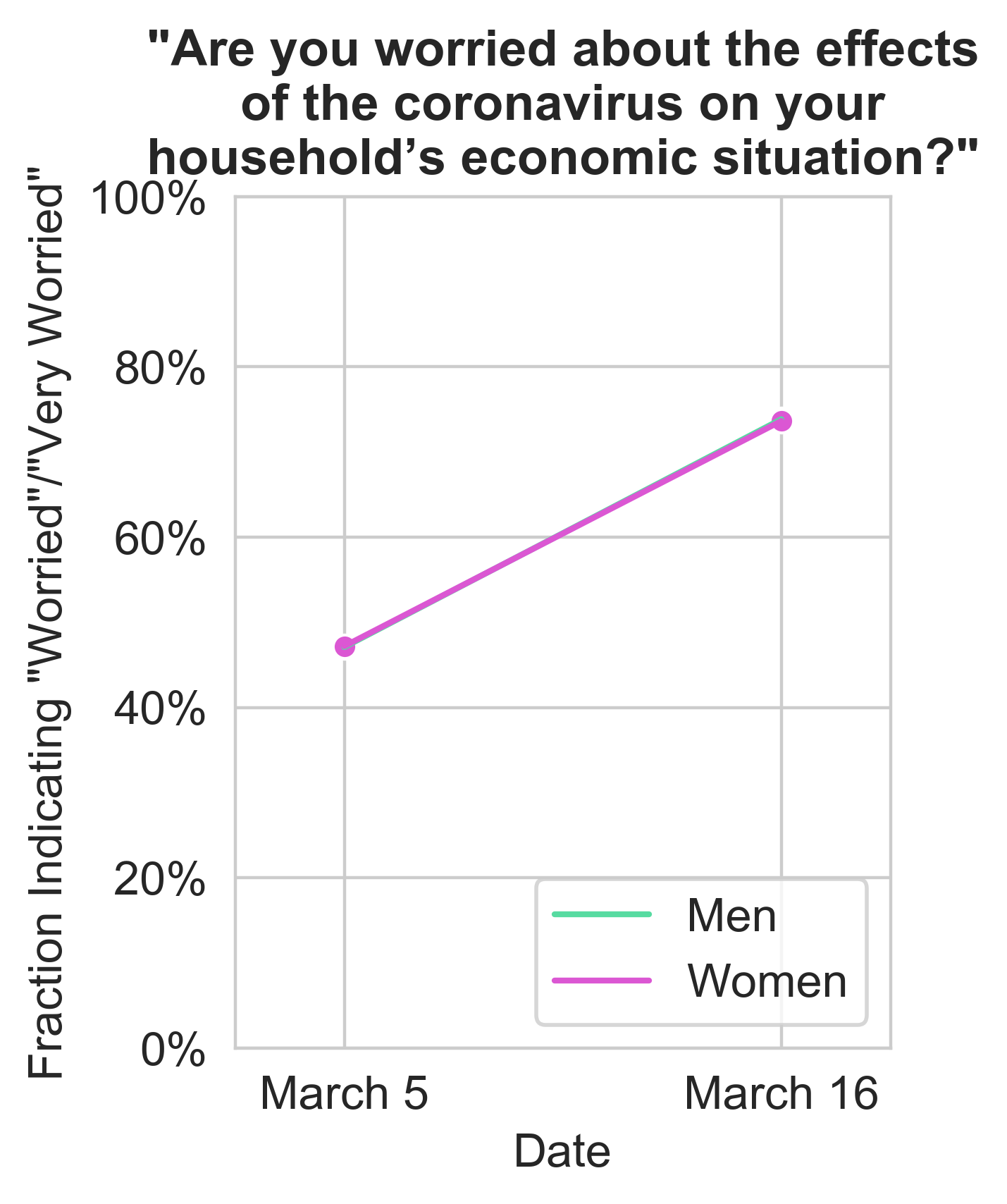} \\
\\[0.1cm]
\multicolumn{4}{c}{\hspace{0cm} B\hspace{0.5cm} By Age}\\
\includegraphics[height=5cm]{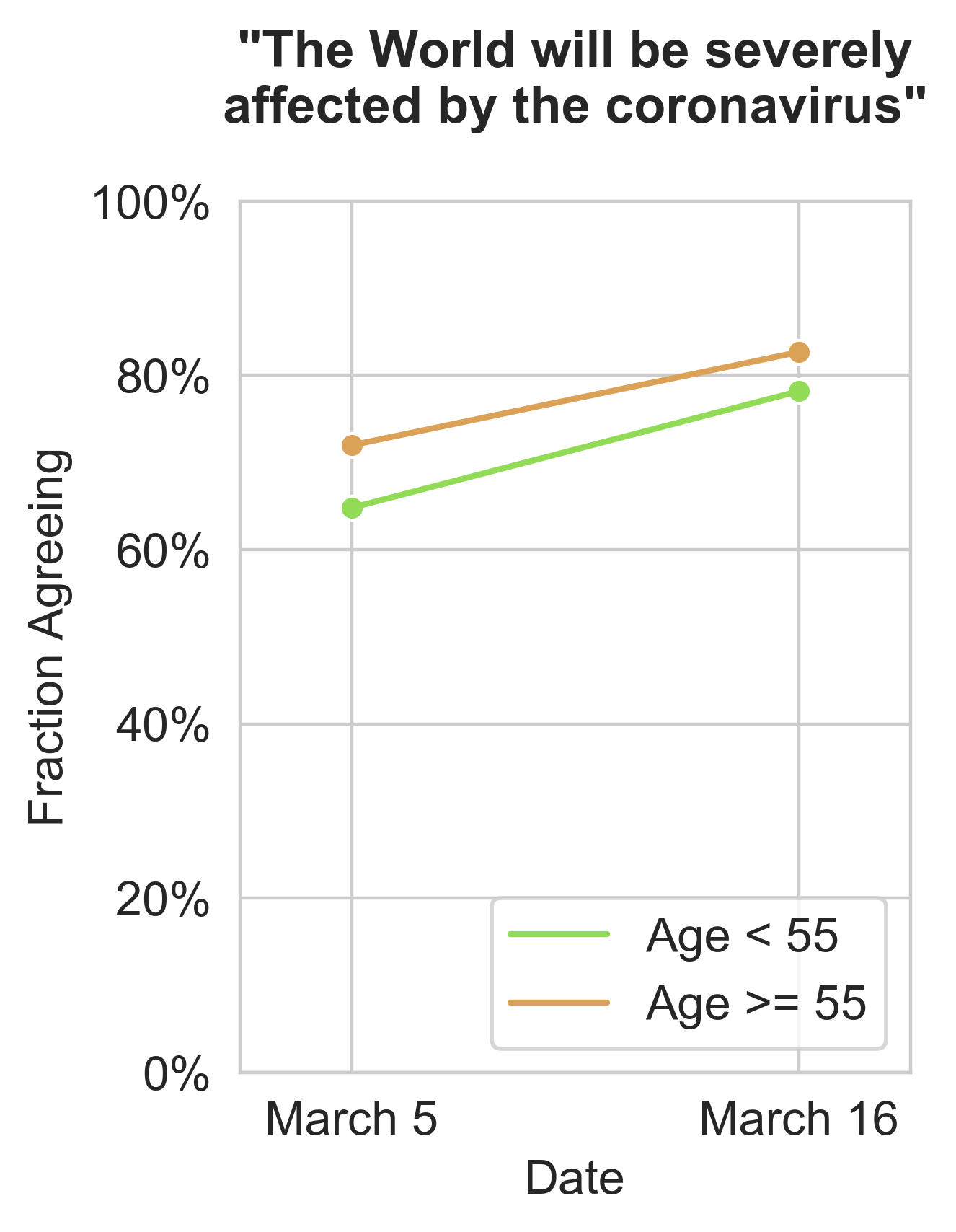} &
\includegraphics[height=5cm]{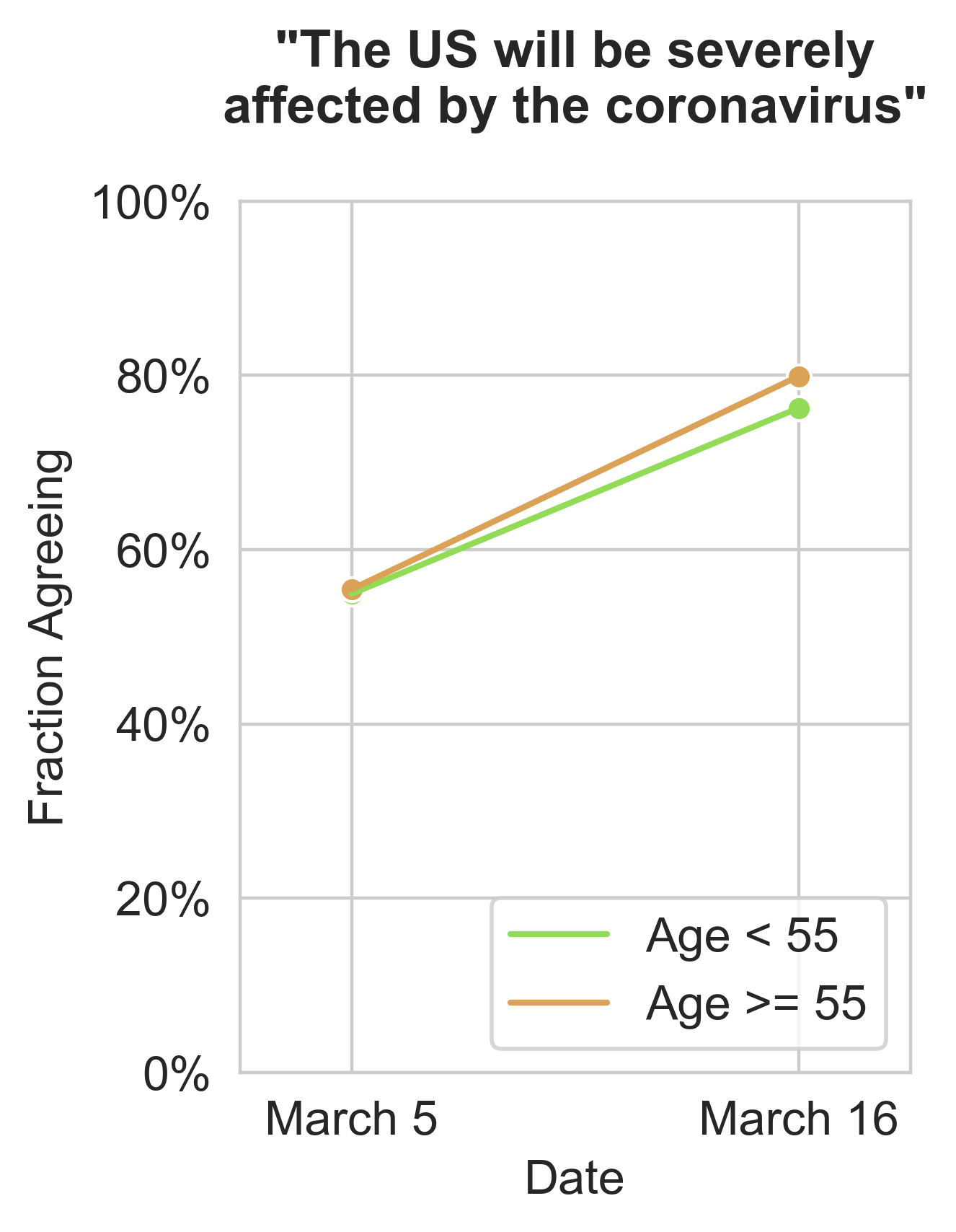} &
\includegraphics[height=5cm]{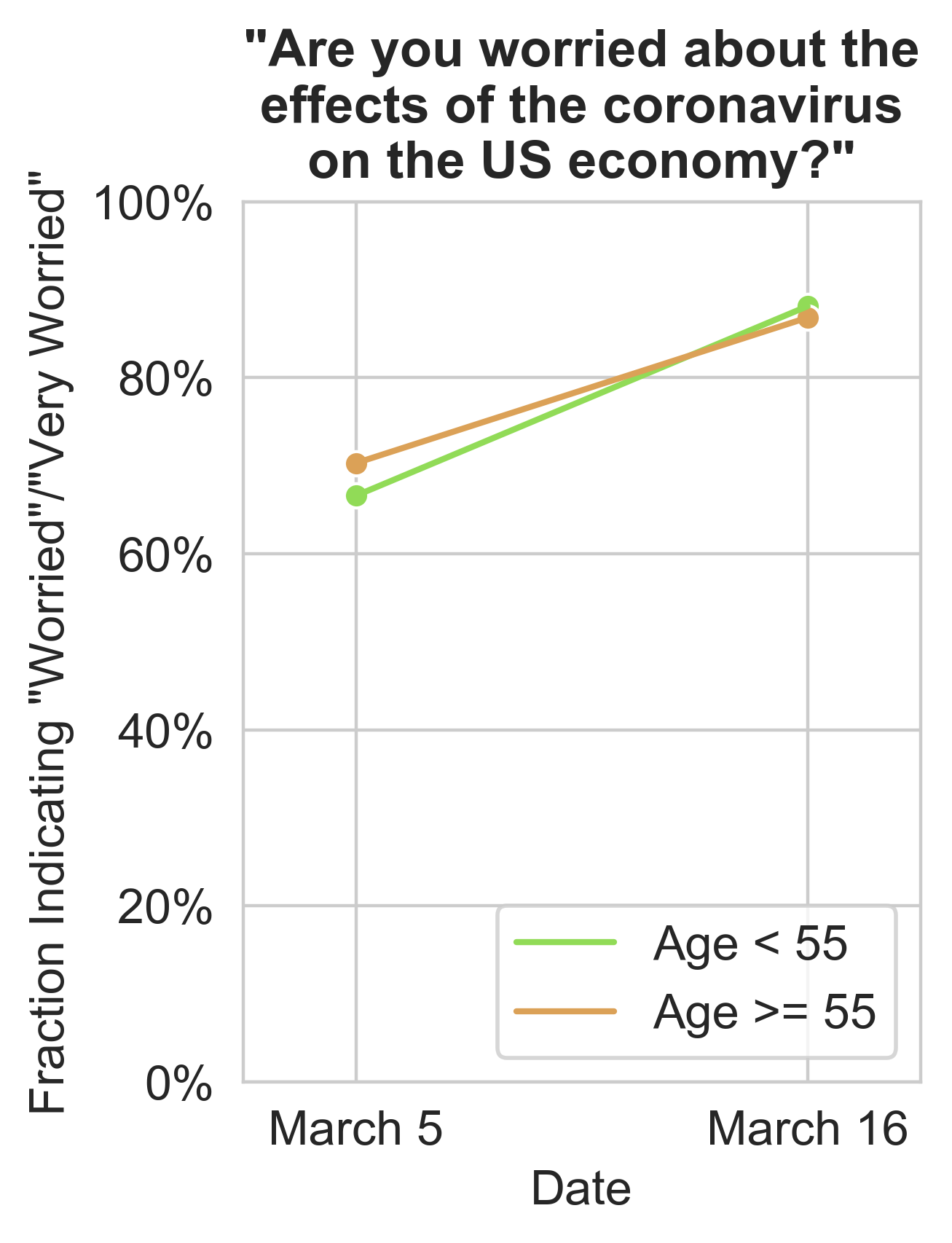} &
\includegraphics[height=5cm]{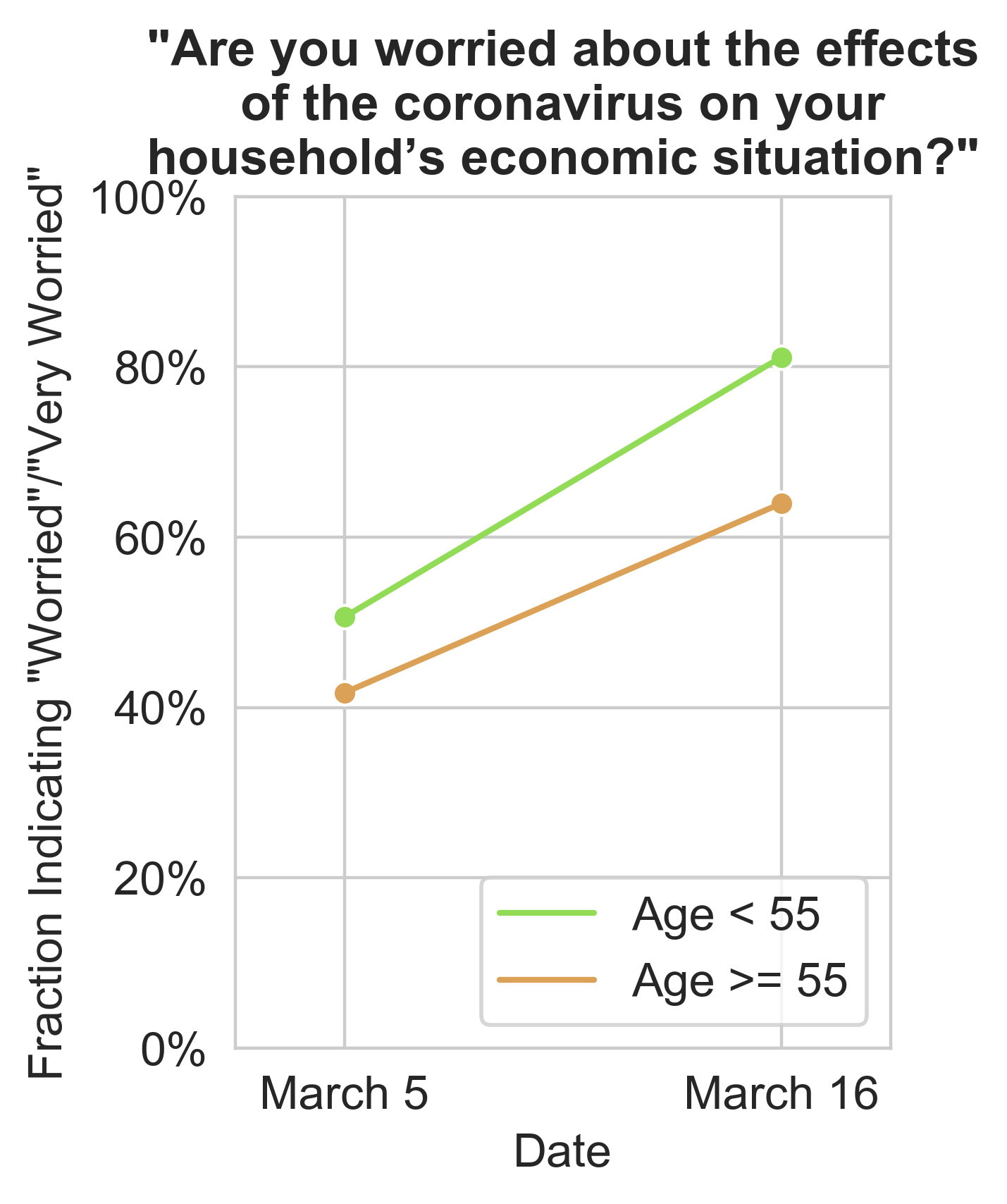} \\
\\[0.1cm]
\multicolumn{4}{c}{\hspace{0cm} C\hspace{0.5cm} By Political Affiliation}\\
\includegraphics[height=5cm]{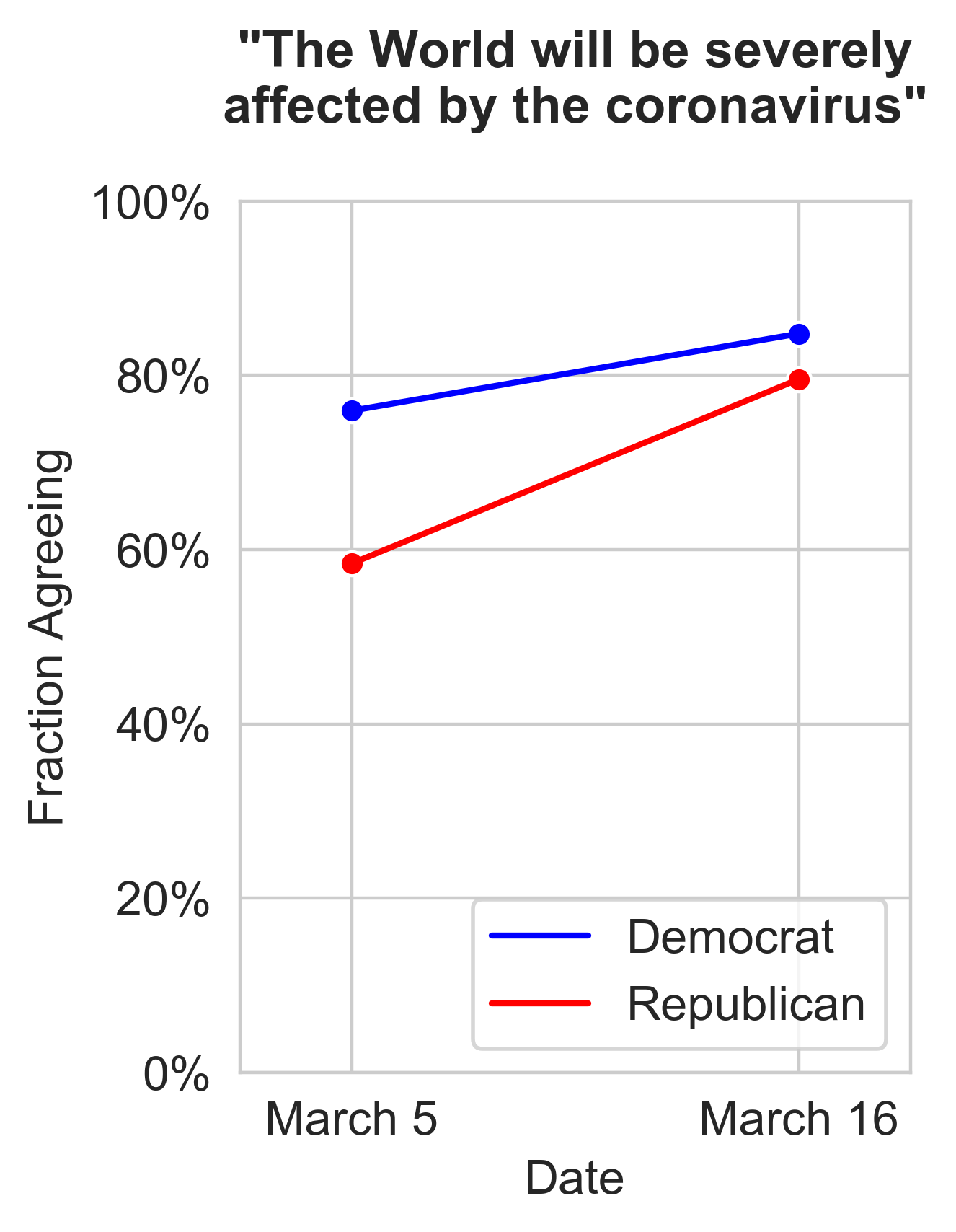} &
\includegraphics[height=5cm]{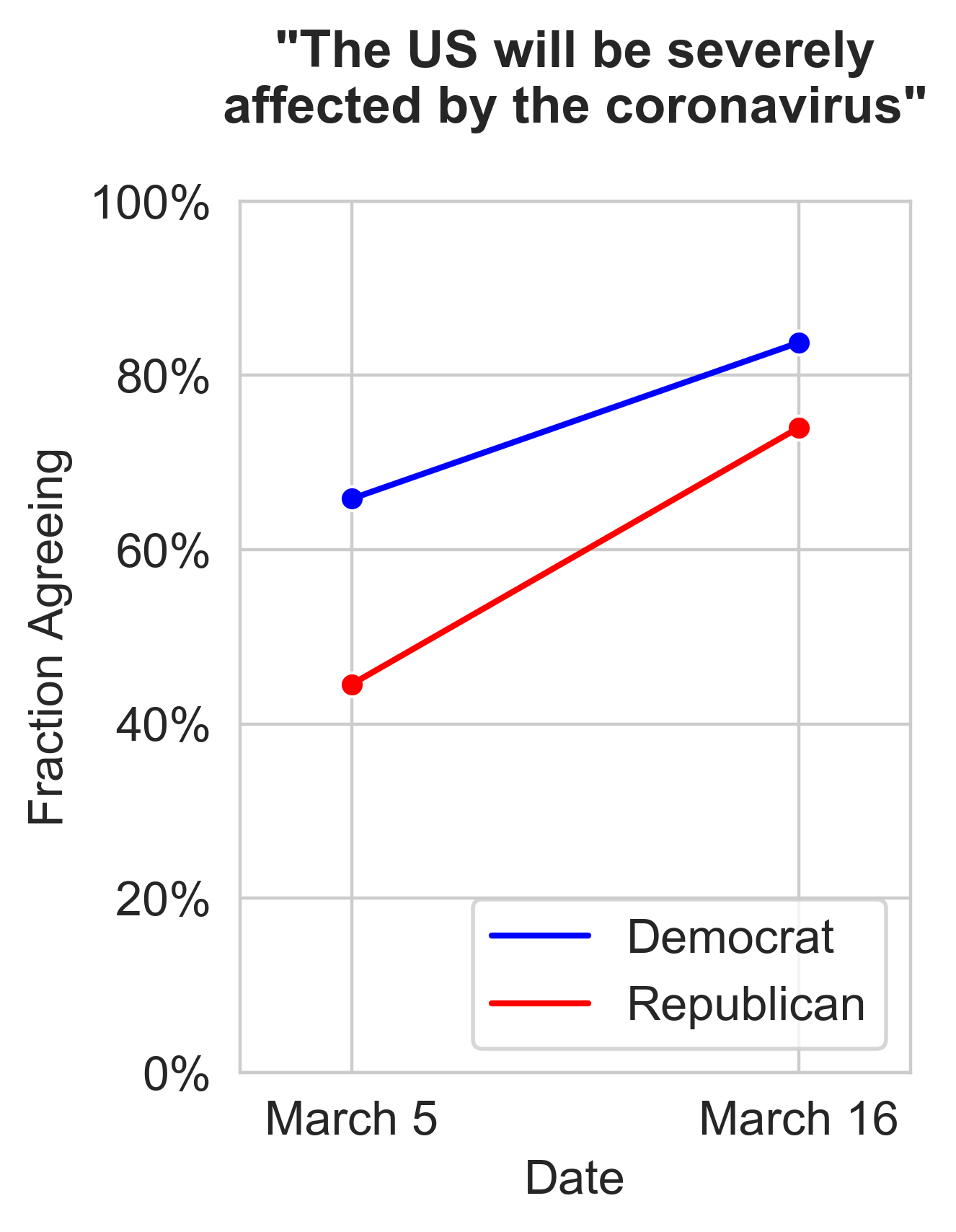} &
\includegraphics[height=5cm]{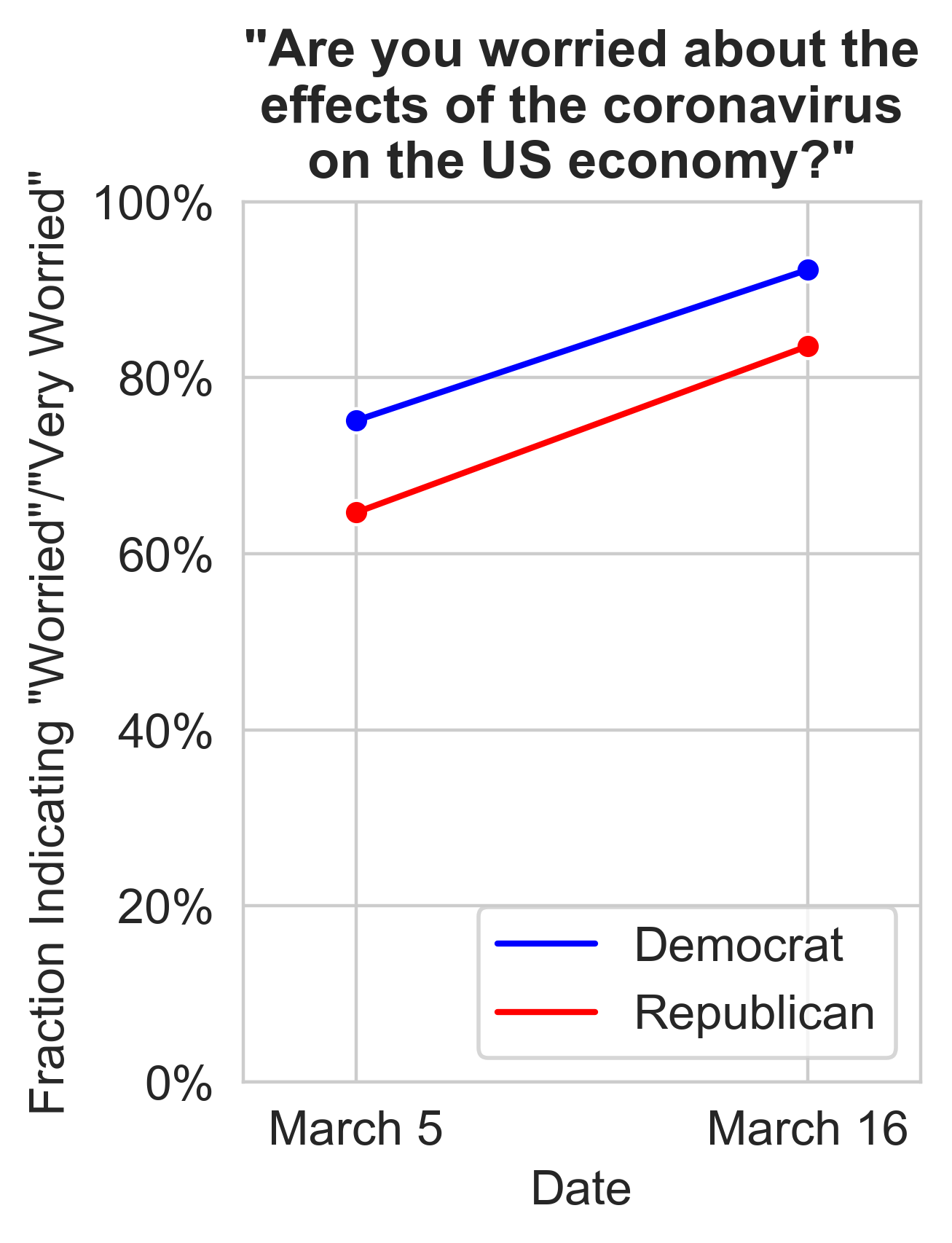} &
\includegraphics[height=5cm]{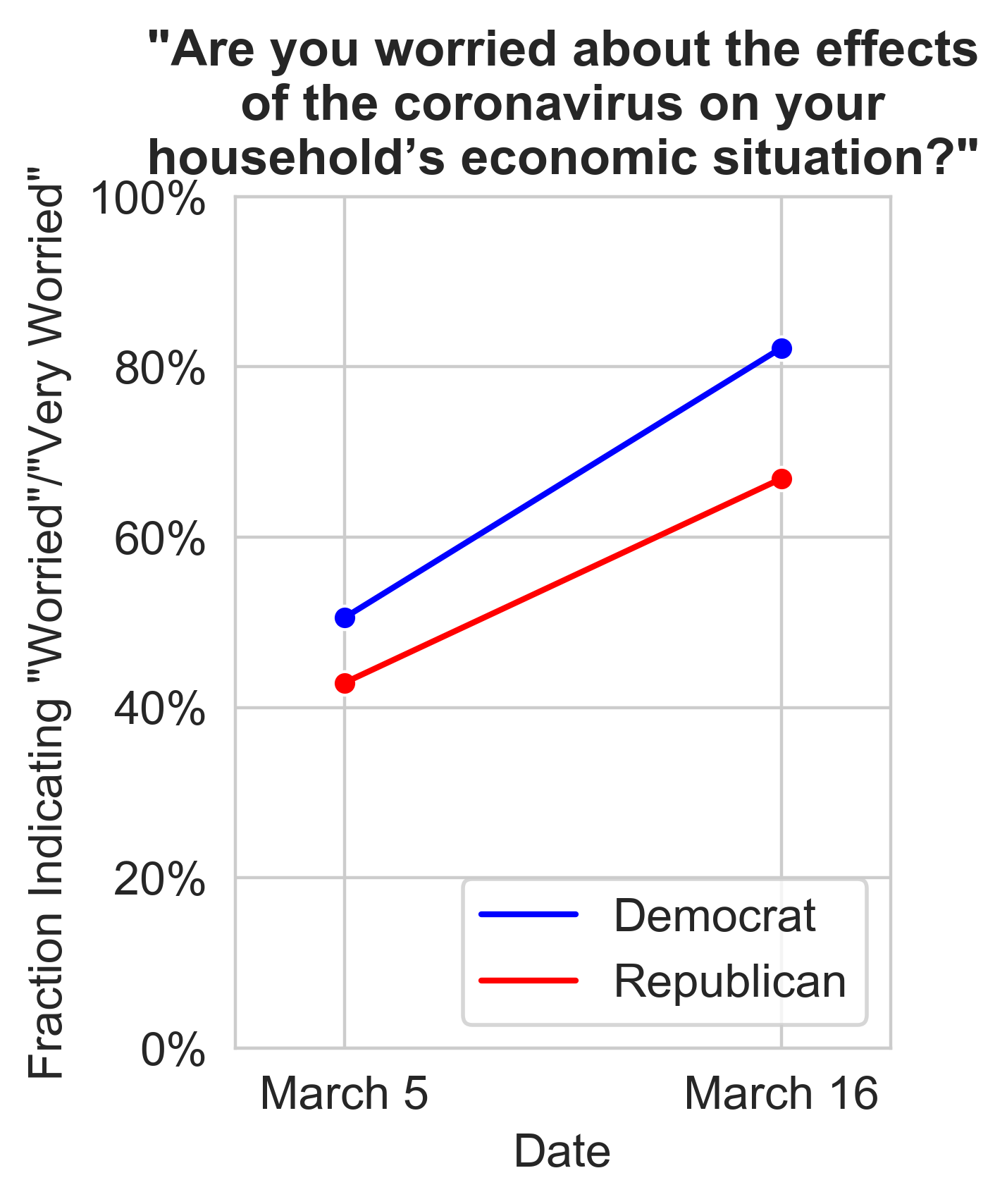} 
\end{tabular}
\parbox{15cm}{\scriptsize \textit{Notes:} Online Appendix Figure \ref{fig:subgroupevol} shows for different subgroups the evolution of beliefs about the  severity of the crisis for the world (leftmost panels) and the United States (second leftmost panels) as well as worries about the aggregate economy (second rightmost panels) and worries about respondents' personal economic situation (rightmost panels). The data were collected on March 5 and March 16, 2020.}
\end{figure}

\begin{figure}[!h]\centering
\caption{Time Series Google Search Intensity for the United States and Worldwide from 02-19-2020 to 03-16-2020.}
\label{fig:trend_late}
\begin{tabular}{cc}
\includegraphics[height=4cm]{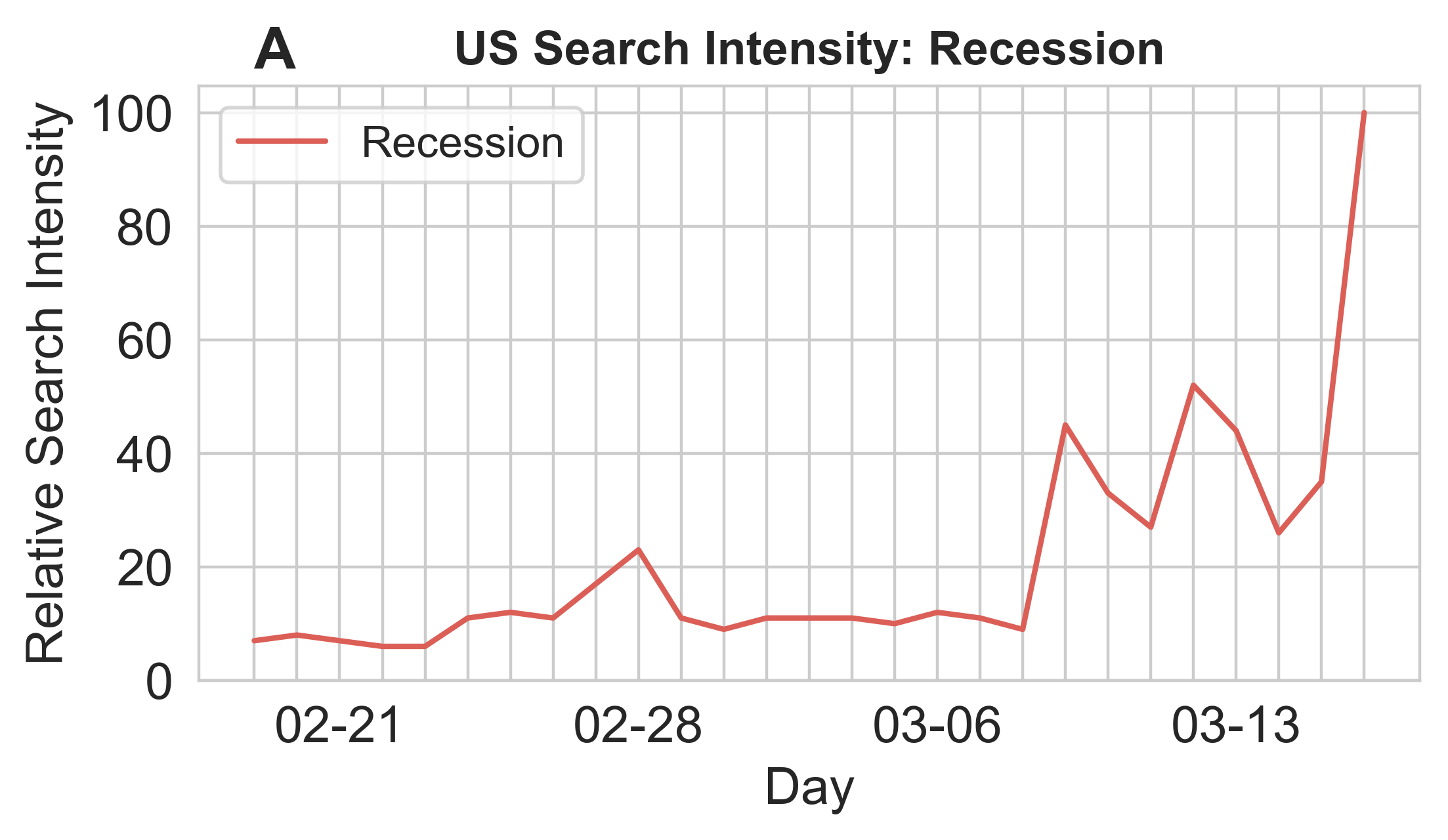} & \includegraphics[height=4cm]{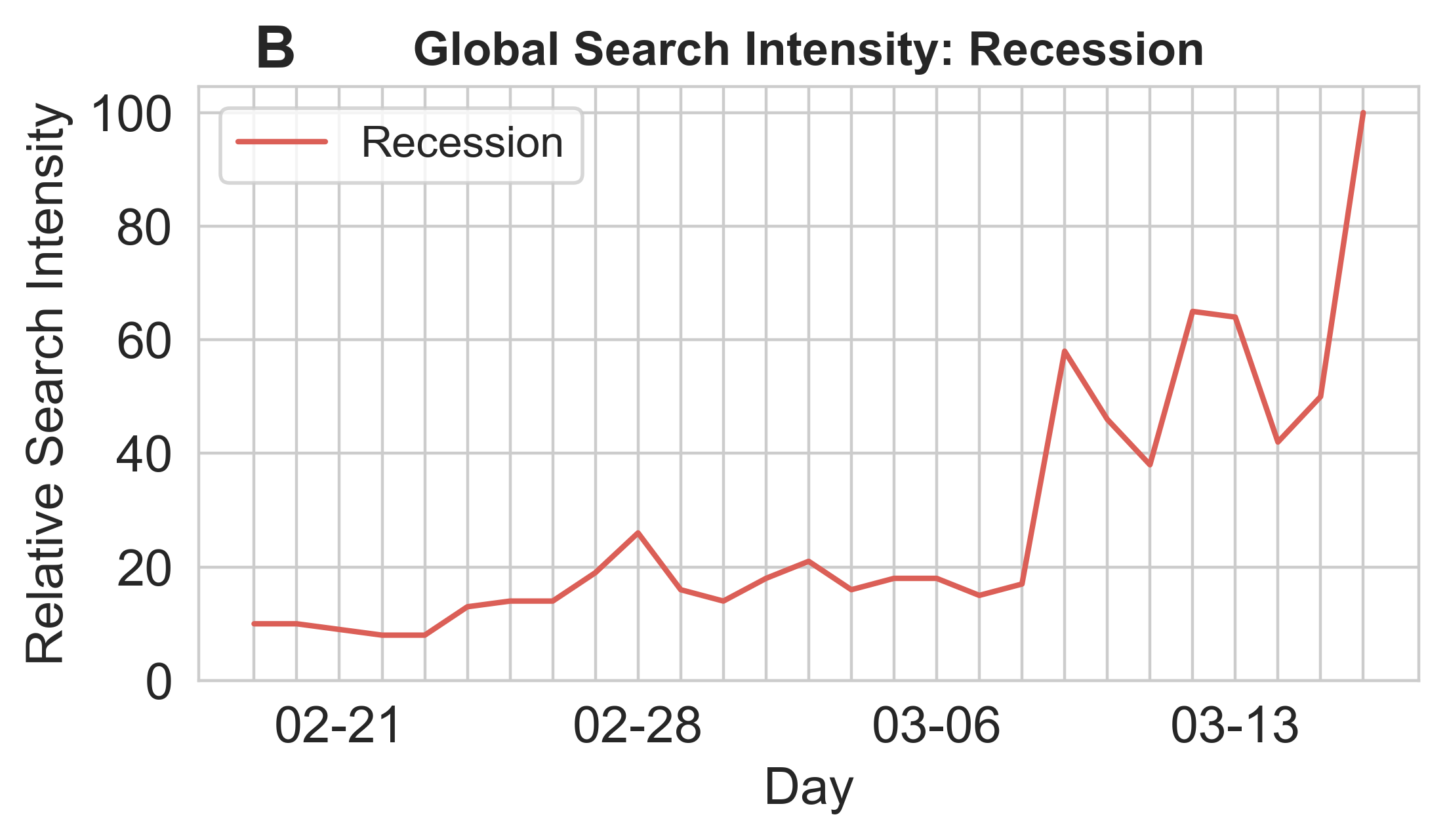}\\
\includegraphics[height=4cm]{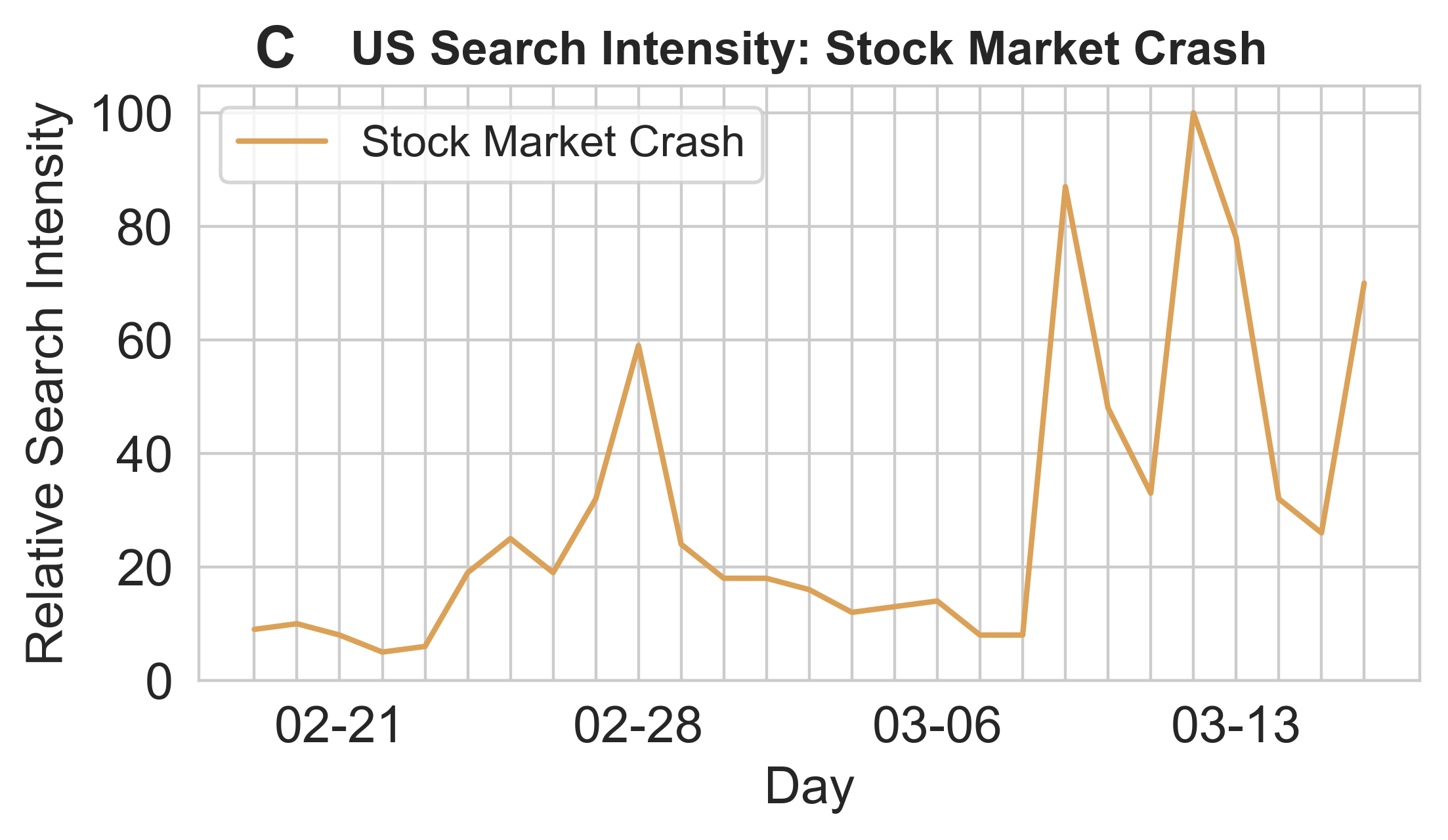} & \includegraphics[height=4cm]{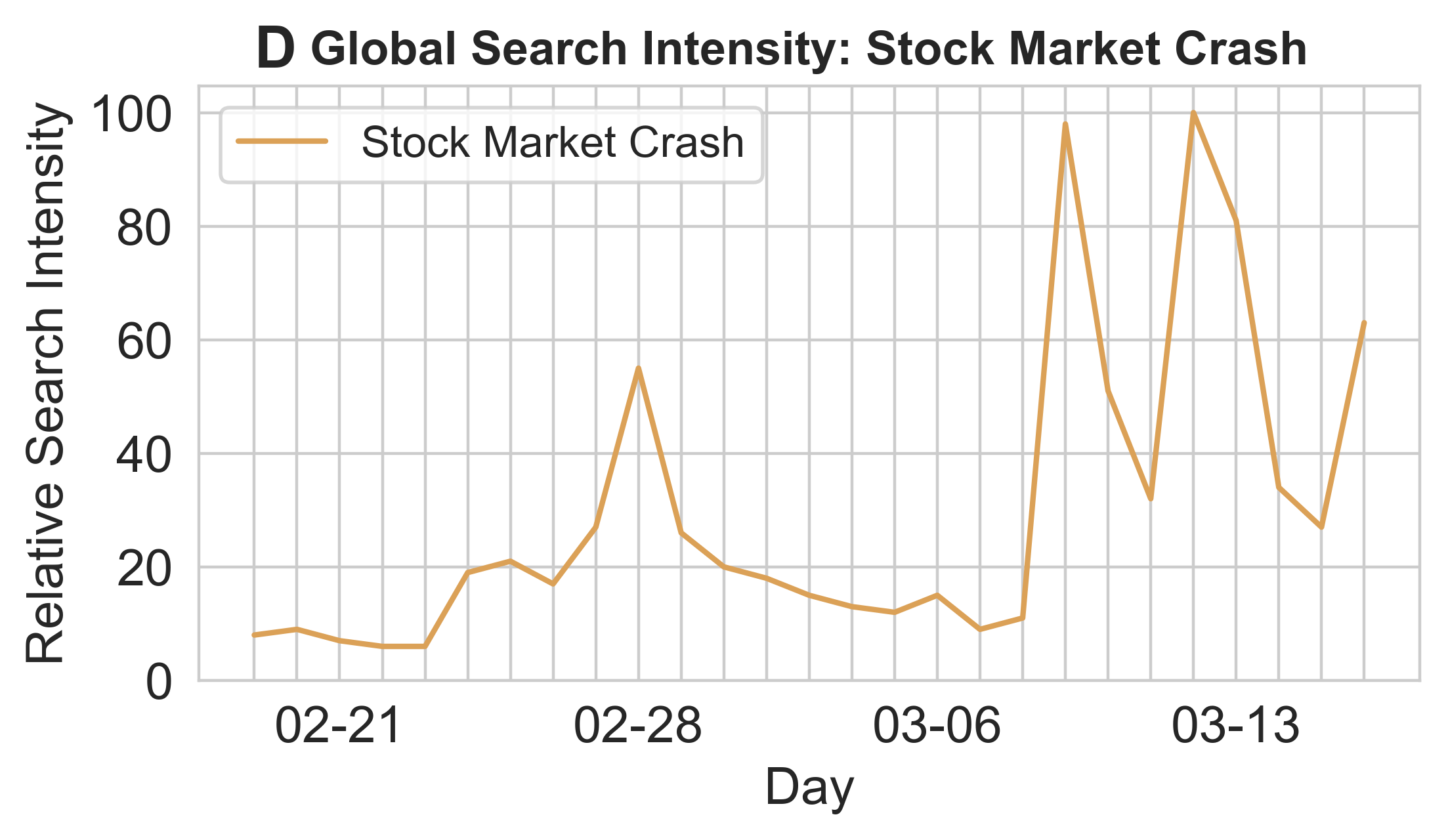}\\
\includegraphics[height=4cm]{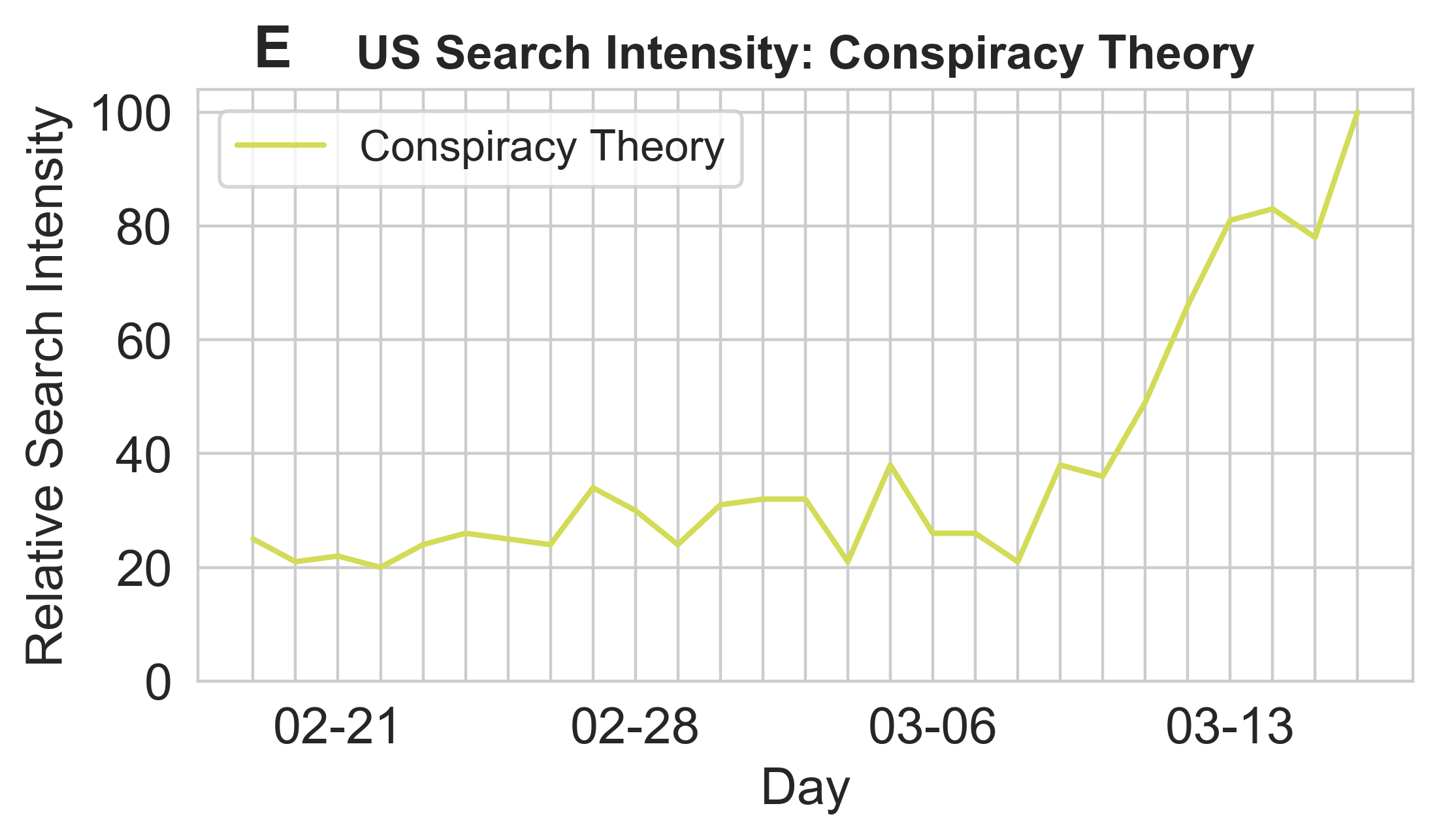} & \includegraphics[height=4cm]{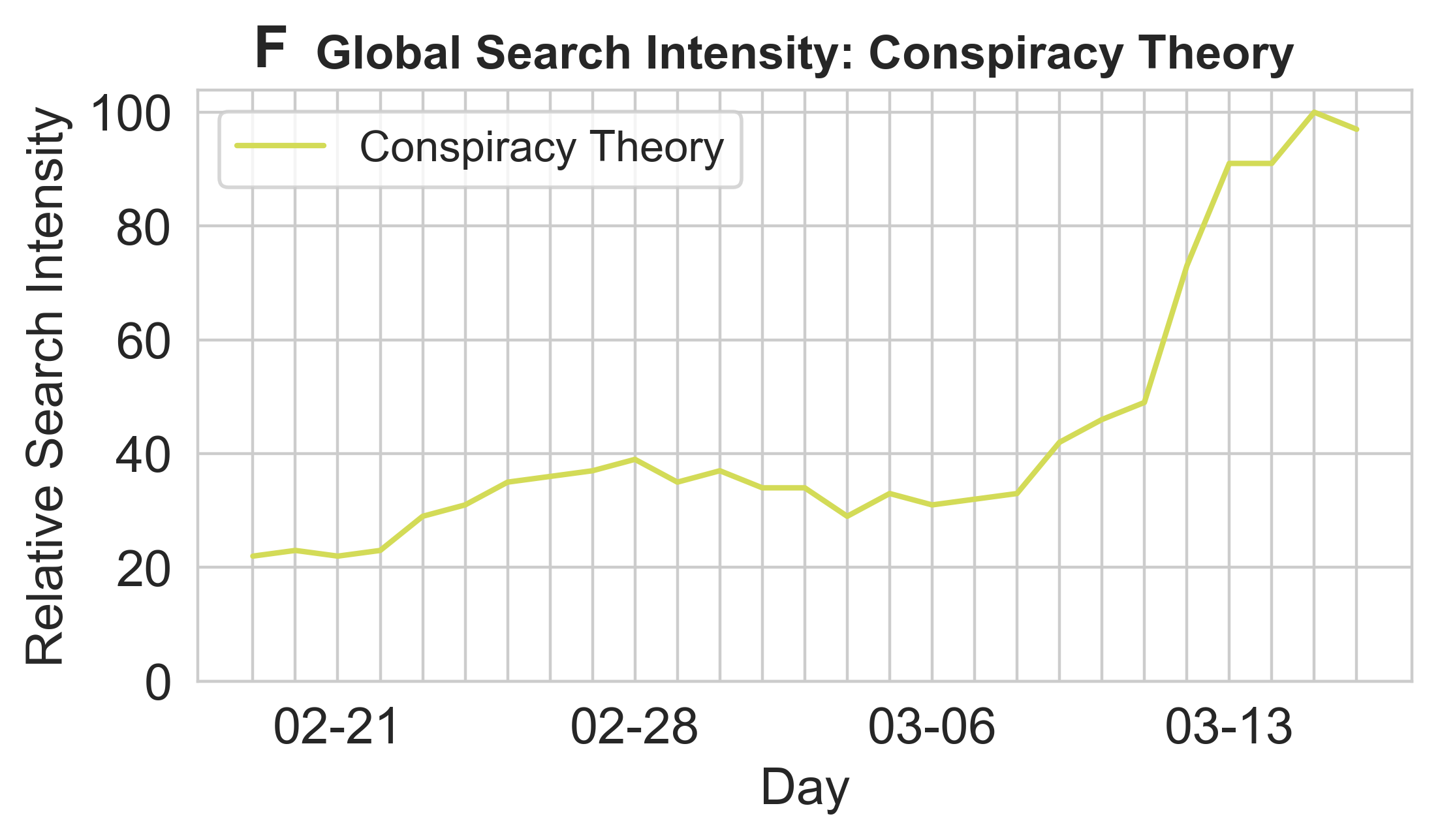}\\
\includegraphics[height=4cm]{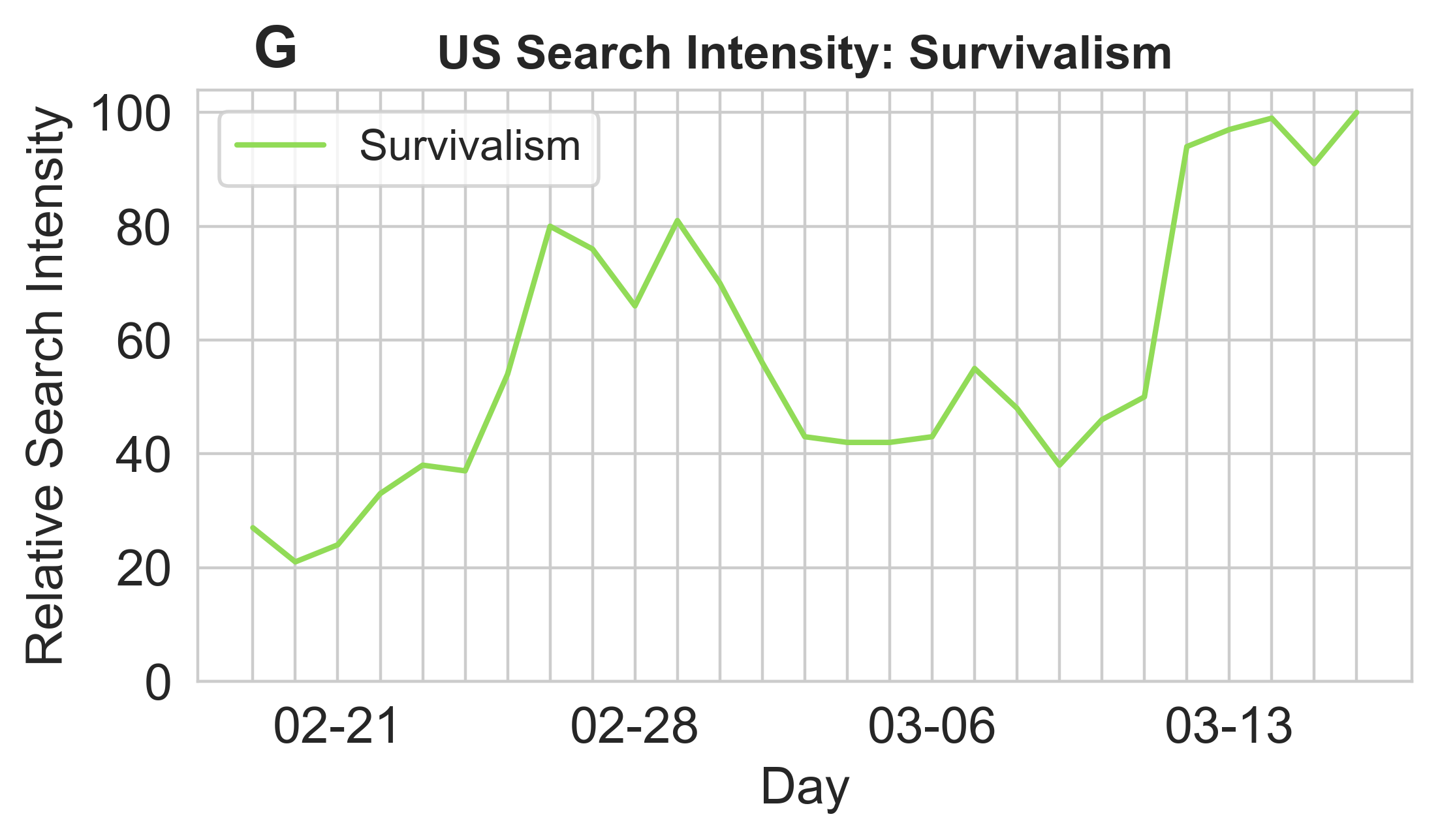} & \includegraphics[height=4cm]{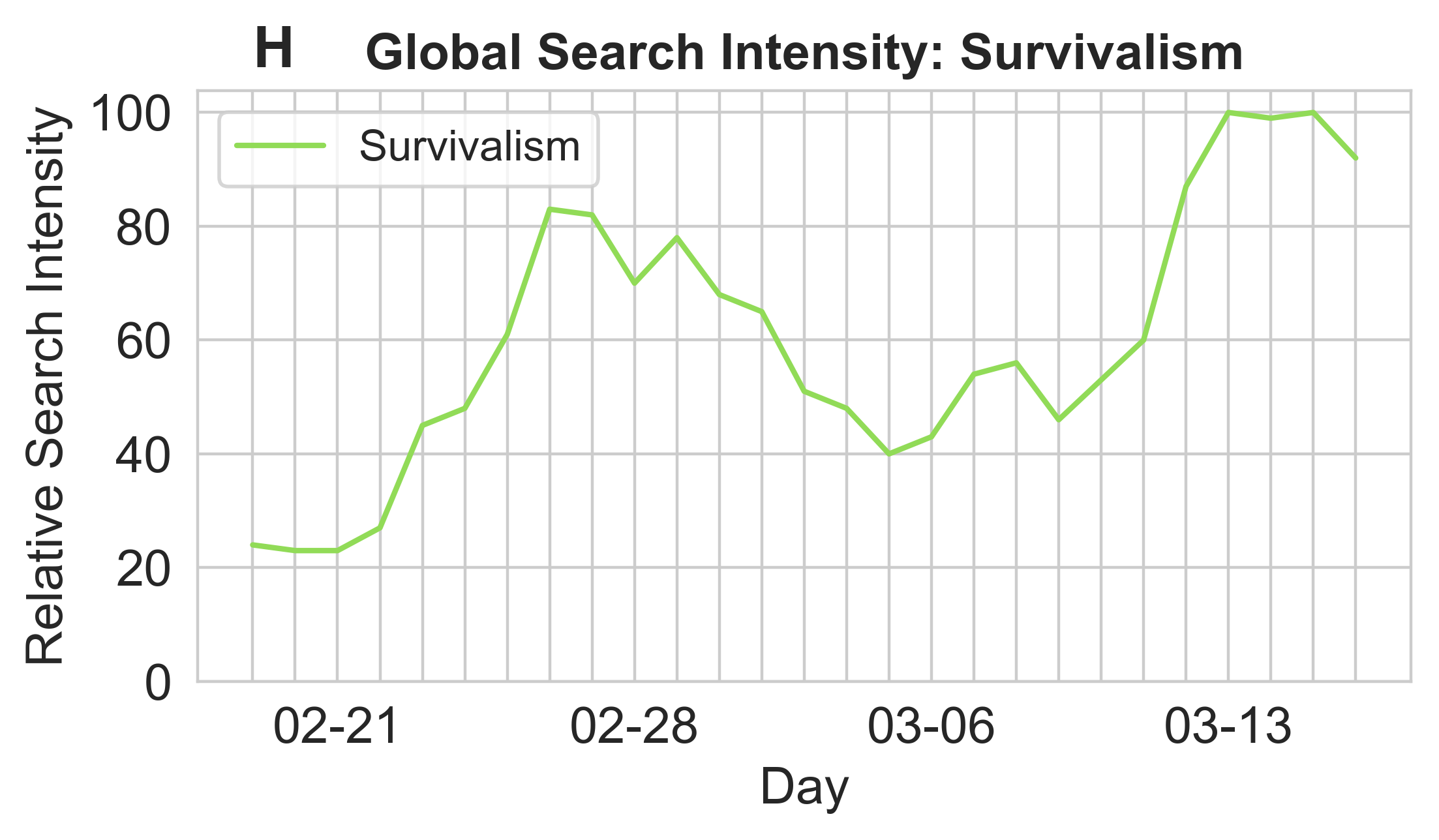}\\

\end{tabular}
\parbox{15cm}{\scriptsize  \textit{Notes:} Online Appendix Figure \ref{fig:trend_late} shows time series of the search intensity for Google topics "Recession", "Stock Market Crash", "Conspiracy Theory", and "Survivalism" from 
February 19th to March 16th, 2020 for the United States and worldwide.}
\end{figure}

\begin{figure}[!h]\centering
\caption{Change in economic outlook in opinion polls}
\label{fig:polldata}
\begin{tabular}{c}
\includegraphics[width=0.75\textwidth]{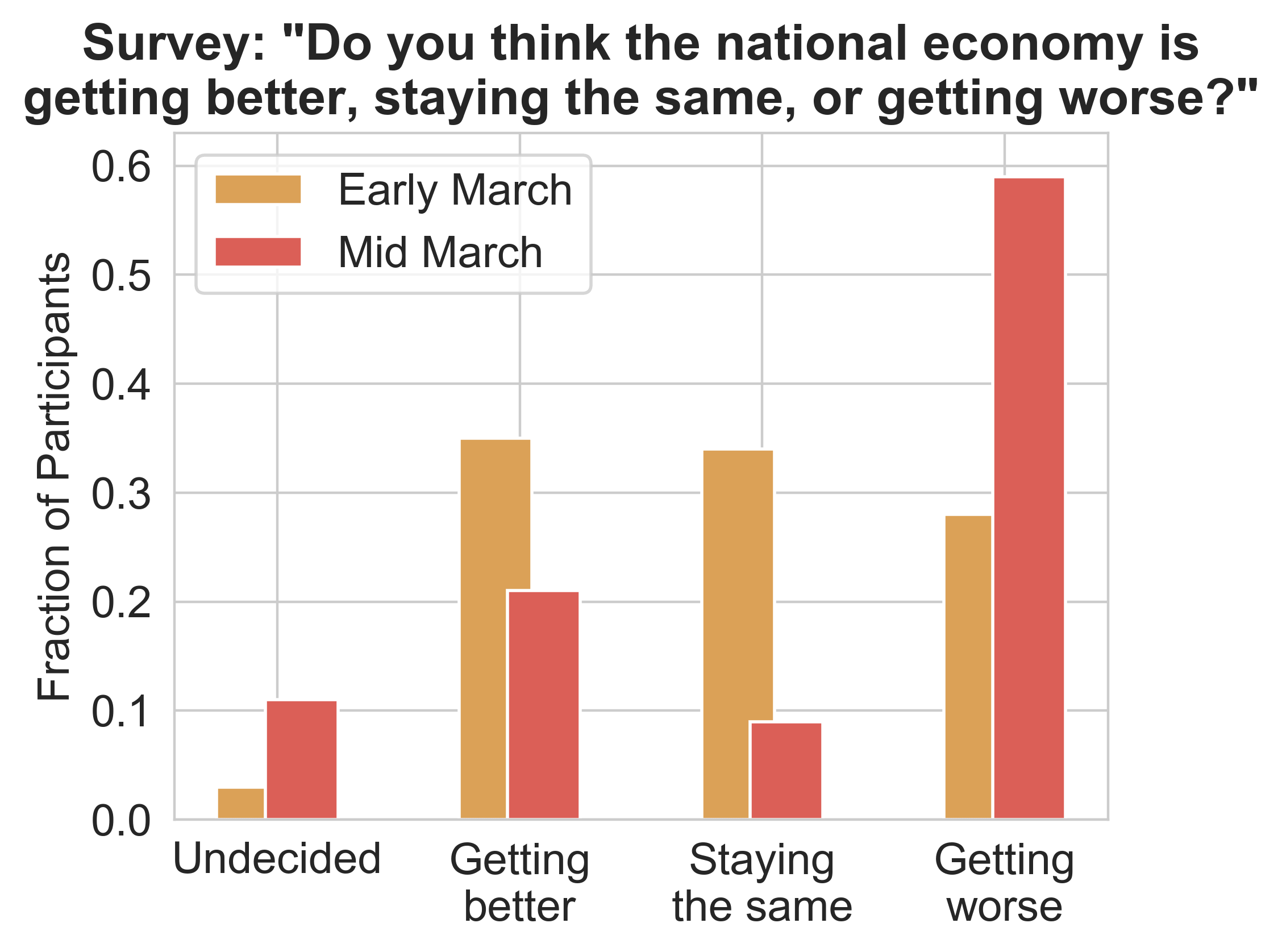} \\
\end{tabular}
\parbox{15cm}{\scriptsize  \textit{Notes:} Online Appendix Figure \ref{fig:polldata} compares polling results among representative samples of the US population from an opinion poll conducted between March 5 and 8 2020 (American Research Group Poll, Question 4, Roper Center Poll Identifier: 31117199.00003) with results from  an opinion poll conducted between March 16 and 19 2020 (Quinnipiac University Poll, Question 19, Roper Center Poll Identifier: 31 31117223.00023). The Figure displays answers to the following question in both opinion polls: ``Do you think the national economy is getting better, staying the same, or getting worse?''}
\end{figure}

\begin{figure}[!h]
\caption{Beliefs about Mortality and Contagiousness (R0) and Economic Anxieties}
\label{fig:nonpar}
\centering
\begin{tabular}{cc}
\multicolumn{1}{c}{\includegraphics[height=5cm]{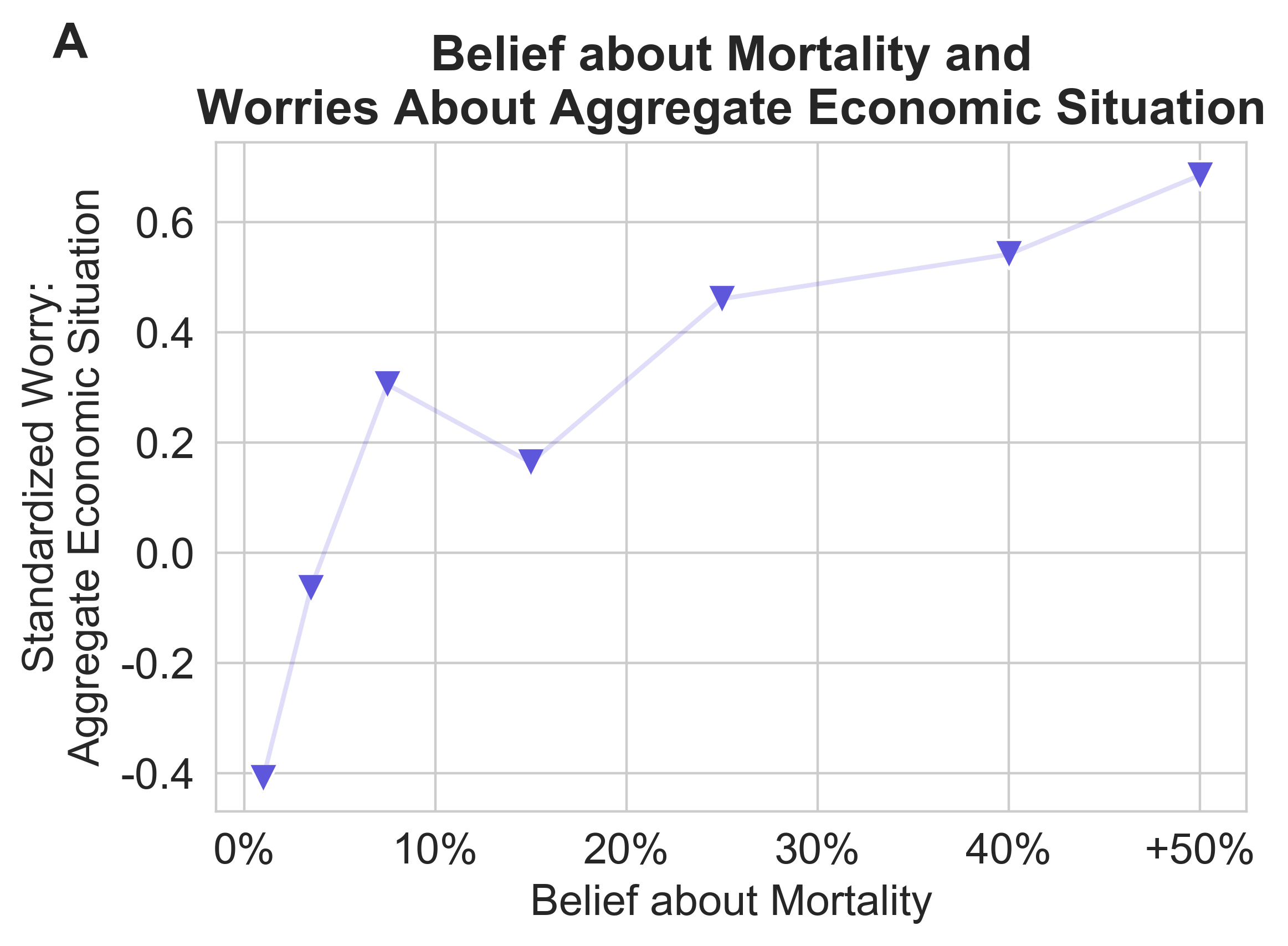}} &\multicolumn{1}{c}{\includegraphics[height=5cm]{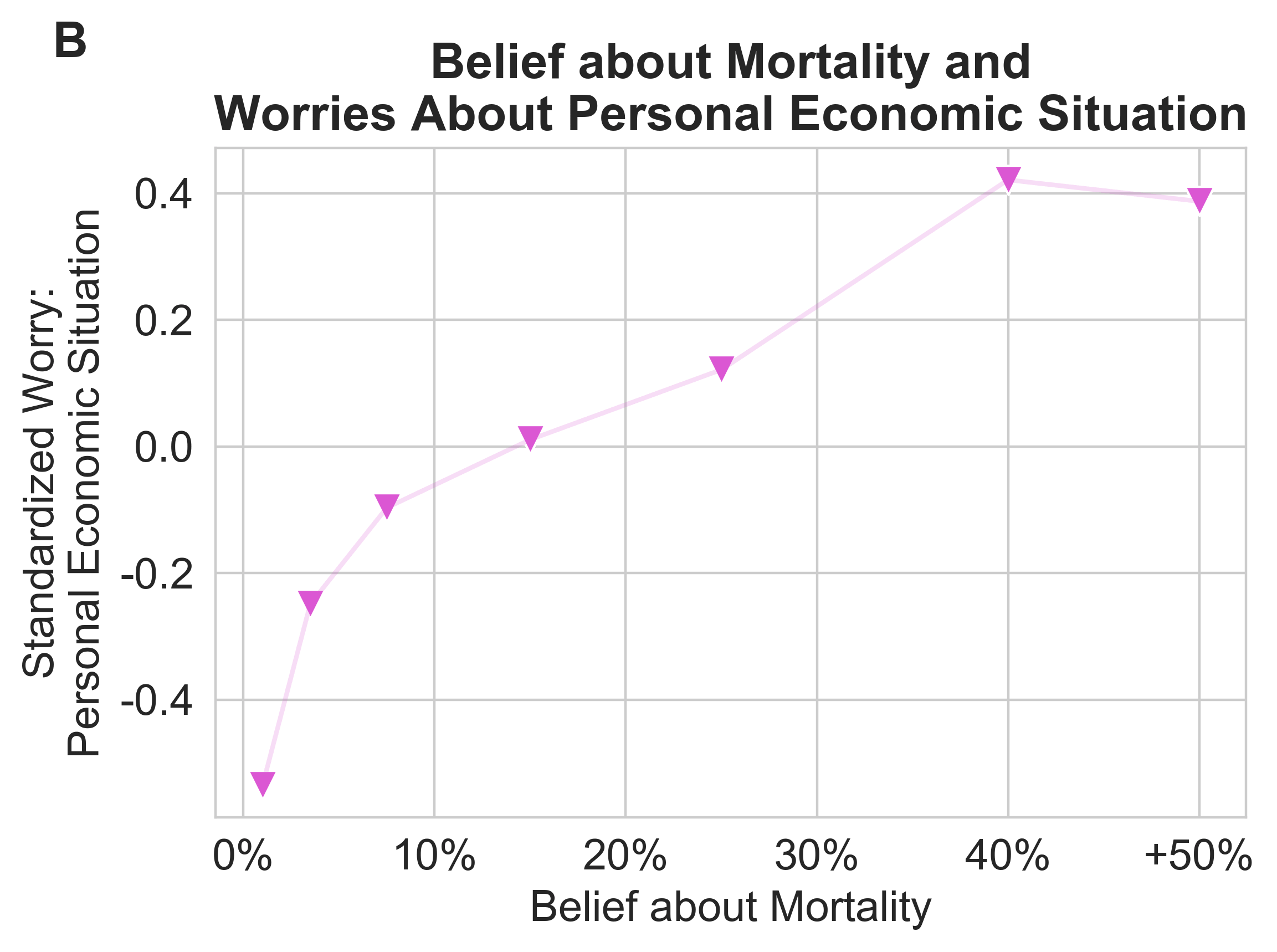}} \\
\multicolumn{1}{c}{\includegraphics[height=5cm]{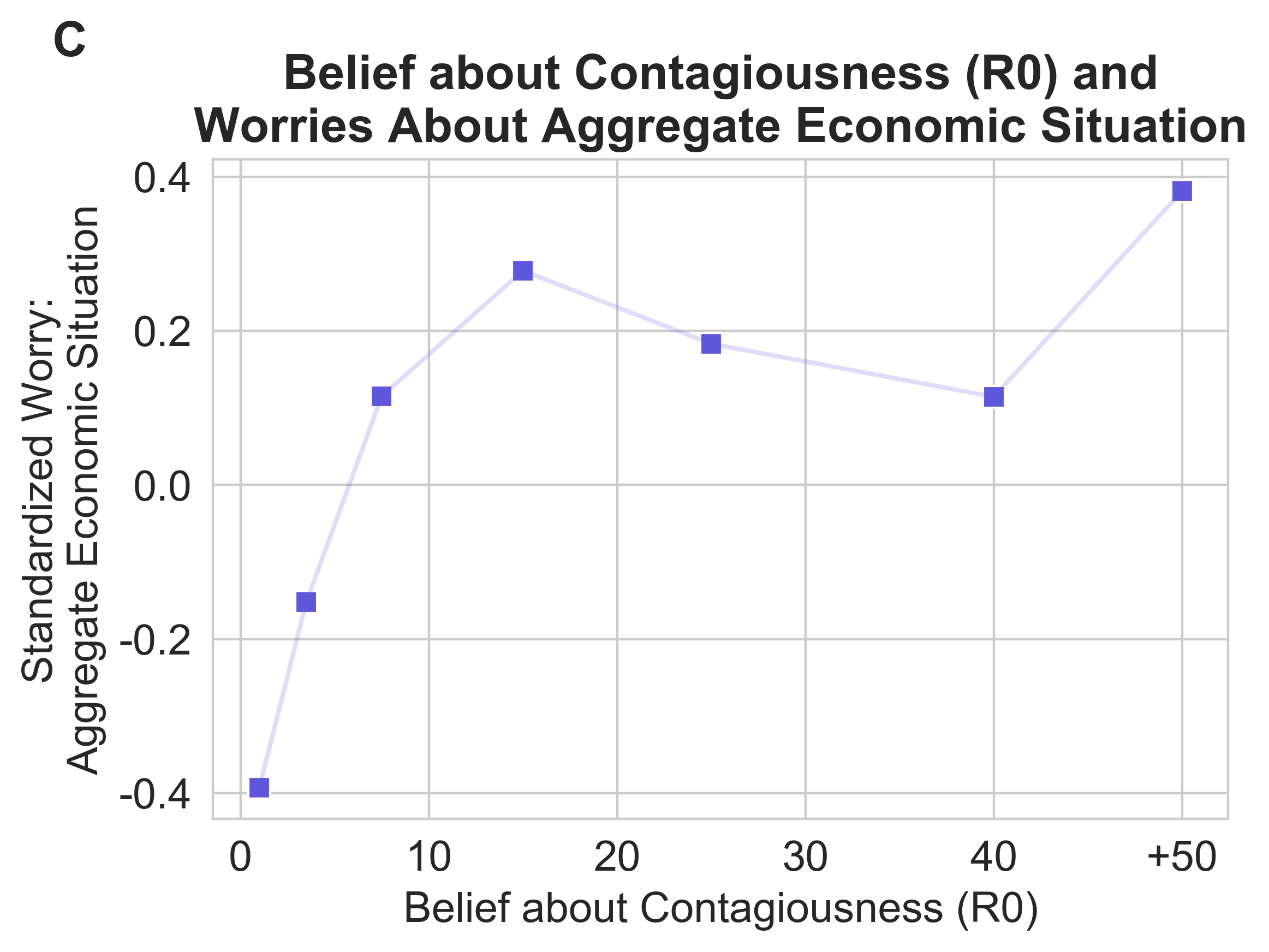}} &\multicolumn{1}{c}{\includegraphics[height=5cm]{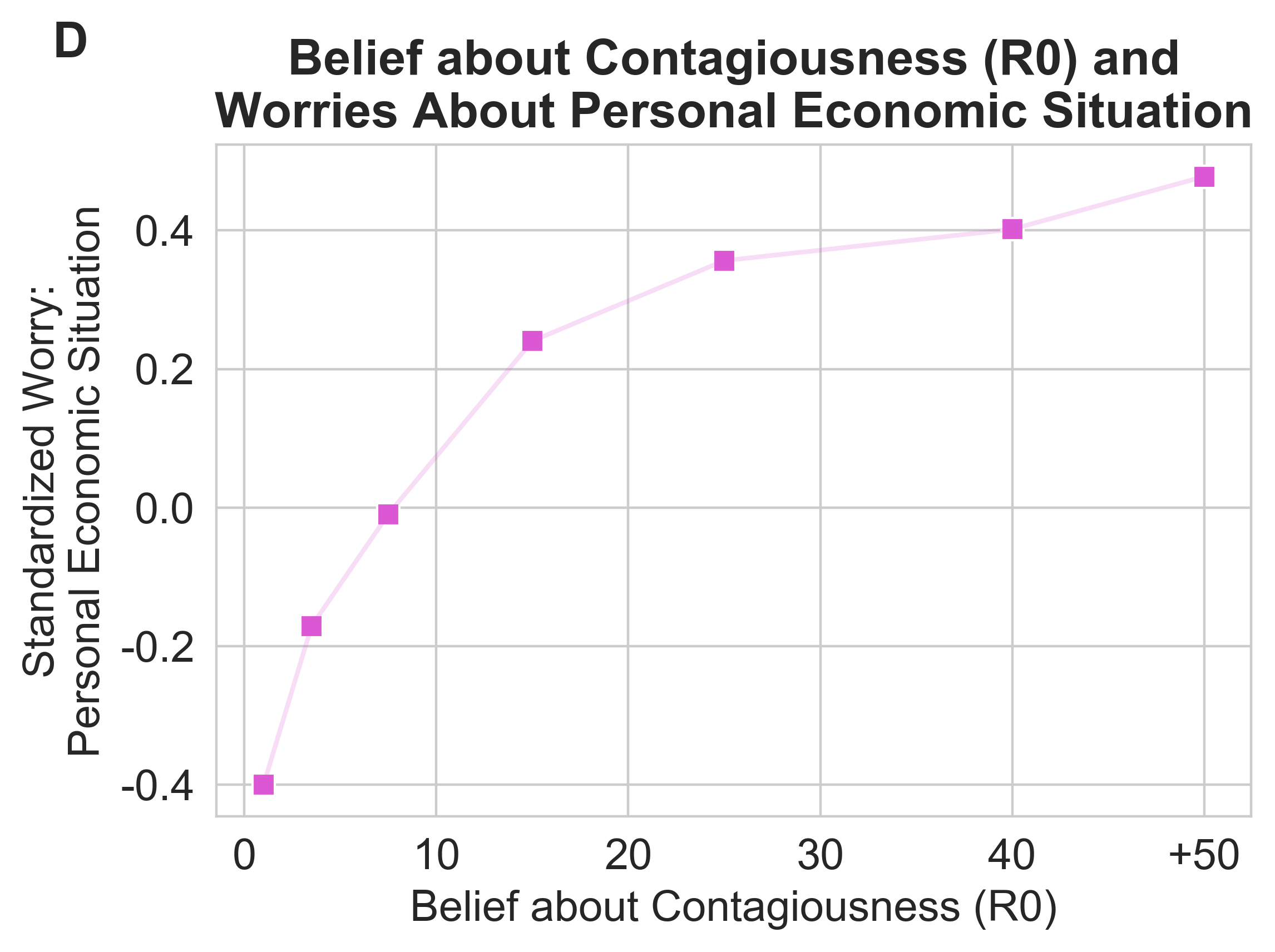}} \\
\end{tabular}
\parbox{15cm}{\scriptsize \textit{Notes:} Online Appendix Figure \ref{fig:nonpar} displays the non-parametric relationship between perceptions of the novel coronavirus and economic anxieties. The data were collected on March 5. Panel A and B show non-parametric relationships between respondents' worries about the aggregate and personal economic situation and their belief about coronavirus mortality. Panel C and D show non-parametric relationships between respondents' worries about the aggregate and their personal economic situation and their beliefs about coronavirus contagiousness (R0). Each plot shows a binscatter plot where x-values correspond to the midpoint of each bin. y-values indicate the mean of the outcome variable for the respective bin. }
\end{figure}

\begin{figure}[!h]
\caption{Classification of Mental Models of Infectious Disease Spread}
\label{fig:mentalclass}
\centering
\begin{tabular}{cc}
\\
\includegraphics[height=5.5cm]{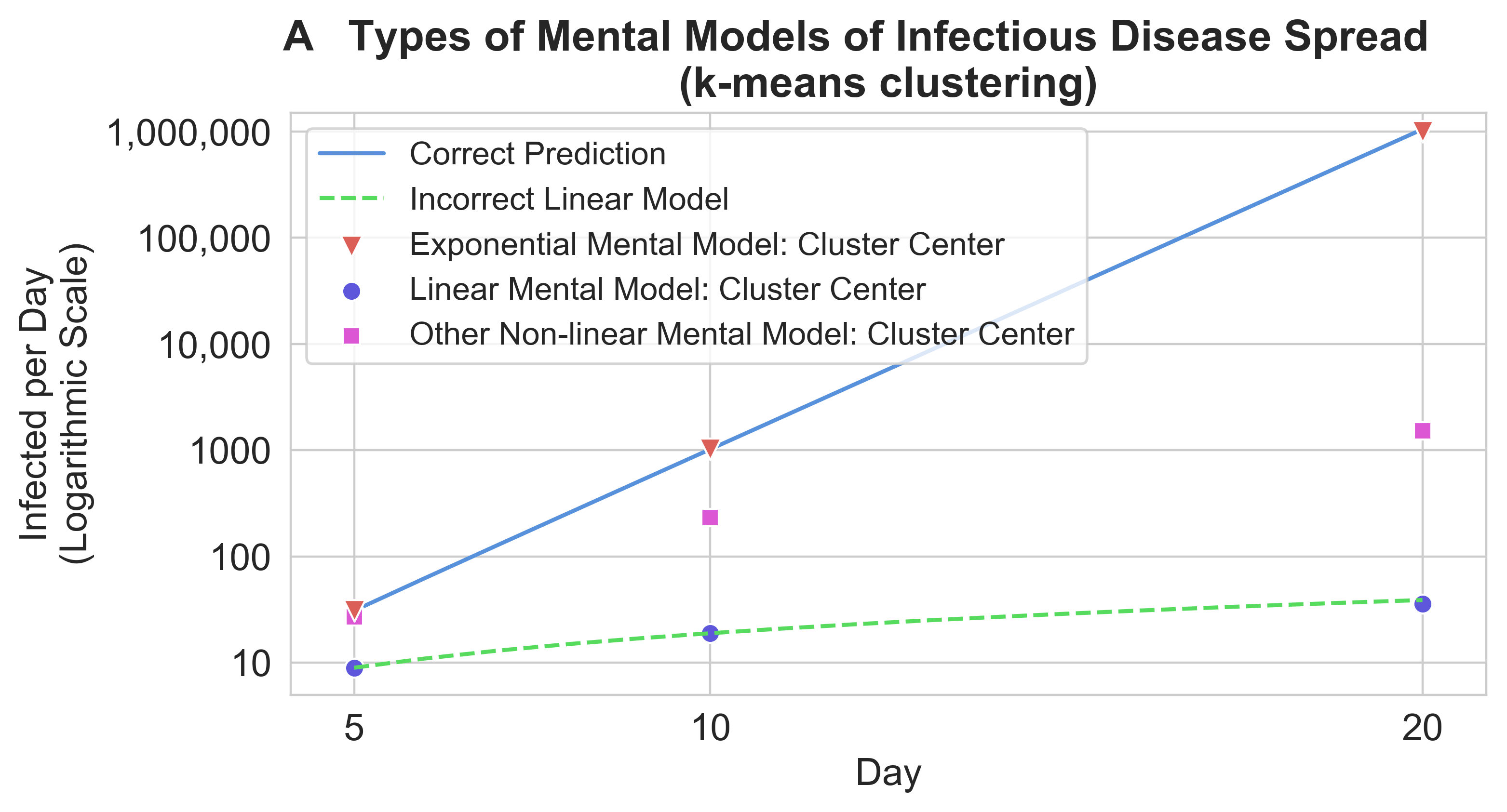} & \includegraphics[height=5.5cm]{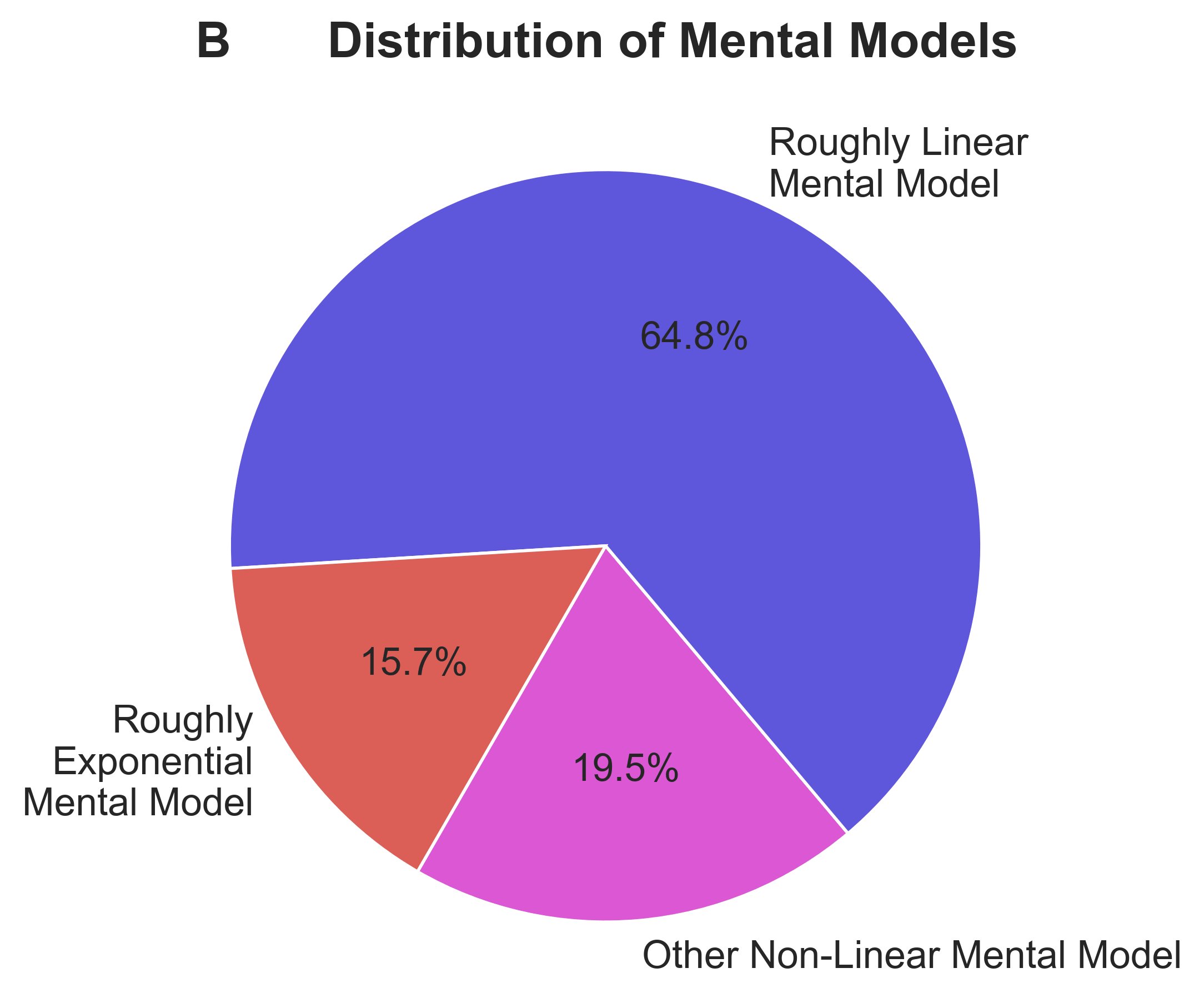} \\
\\
\multicolumn{2}{c}{\includegraphics[height=5.5cm]{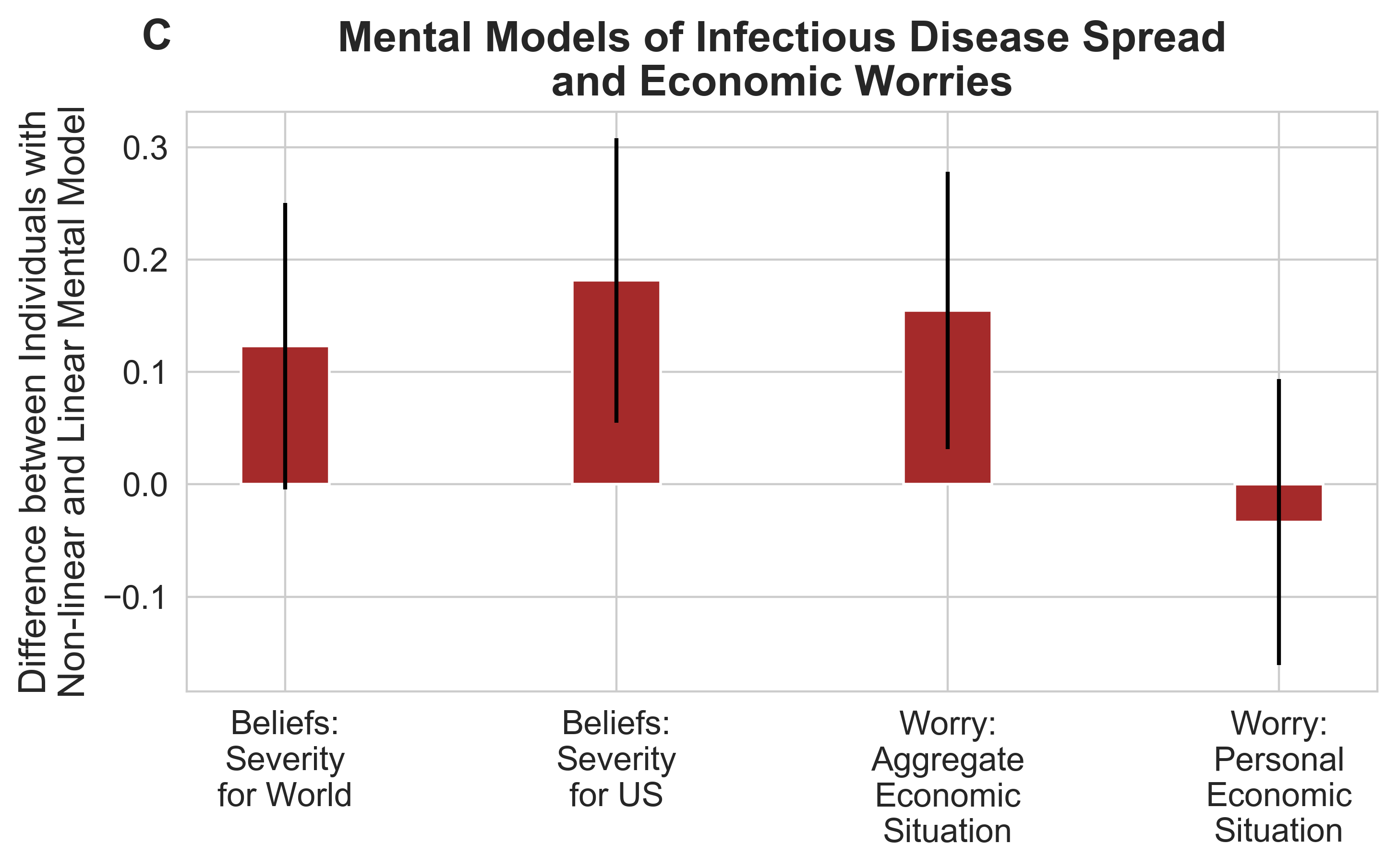}} \\
\end{tabular}
\parbox{15cm}{\scriptsize \textit{Notes:} Online Appendix Figure \ref{fig:mentalclass} shows a classification of mental models of infectious disease spread obtained from k-means clustering using 3 clusters in the space of predicted log-number of cases at day 5, 10, and 20. Panel A shows the median number of predicted cases in the three clusters (dots) as well as the cluster centers (shaded areas) that contain 50\% of the mass of each cluster. The blue and green lines represent the correct prediction values as well as an incorrect linear model with a slope of 2. The three types obtained from the cluster approach can be described as a correct 'exponential mental model', an incorrect 'linear mental model', or an incorrect 'other mental model'. Panel B visualizes the type distribution in the data. Panel C shows estimates of OLS regressions with an indicator for exhibiting a non-linear mental model as the independent variable and using as the dependent variables participants' beliefs about the crisis' severity for the world and the United States as well as their worries about the aggregate economy and their personal economic situation. Error bars indicate 95\% confidence intervals.

}
\end{figure}

This Online Appendix contains supplementary materials mentioned in the main text. Section \ref{sec:opinionpolling} provides an overview of the methodology underlying the analysis of the impact of state-level coronavirus arrival on outcomes in the US. In Appendix Section \ref{sec:appfig} we display additional figures. Online Appendix Figure \ref{fig:searchapp} displays the global trends in search intensity for our four main indicators. Online Appendix Figure \ref{fig:robdropping} displays the impact of dropping individual countries from the main search intensity analysis. Online Appendix Figure \ref{fig:robdroppingsubregions} displays the impact of dropping subregions from the main search intensity analysis. Online Appendix Figure \ref{fig:subgroupevol} displays how perceptions of severity and economic worries change over time for different subgroups. Online Appendix Figure \ref{fig:trend_late} displays search intensity trends in the US and globally for our four main measures. Online Appendix Figure \ref{fig:polldata} displays changes in economic expectations from early to mid-March using Roper Center polling data. Online Appendix Figure \ref{fig:nonpar} plots the non-parametric relationship between mortality and contagiousness perceptions and economic anxieties. Finally, Online Appendix Figure \ref{fig:mentalclass} displays the results of categorizing individuals in linear, exponential, and other mental models of disease spread. It also shows the correlation between mental models and economic anxieties. 

Online Appendix Section \ref{sec:apptab} displays additional tables. Online Appendix Table \ref{table:countrylist} contains all countries used for the Google search intensity analysis. Online Appendix Table \ref{table:recessionindicators} displays correlations of Google search intensity with GDP and its components. Online Appendix Table \ref{table:main_DD} contains the main Google search intensity results also displayed in Figure \ref{fig:search}. Online Appendix Table \ref{table:main_DD_nonormalisation} displays the main results without normalization of the outcome variables. Online Appendix Table \ref{table:placebo_DD} displays a series of placebo difference-in-differences regressions. Online Appendix Table \ref{table:sum_exp} displays summary statistics for the March 5 survey. Online Appendix Table \ref{table:sum_expgrowth} displays summary statistics for the March 16 survey. Online Appendix Table \ref{table:opinionpoll} shows the correlation between having any coronavirus case and coronavirus-related concerns in mid February. Online Appendix Table \ref{table:comparison_time} shows differences in coronavirus perceptions and economic anxieties over time. Online Appendix Table \ref{table:survey_DiD} show the results of the difference-in-differences regression described in Section \ref{sec:opinionpolling}. Online Appendix Tables \ref{table:experiment_correlations} and \ref{table:experiment_correlations_cont} display correlations of coronavirus perceptions and economic anxieties for binary and continuous variables, respectively. Online Appendix Table \ref{table:balance} shows balance tests for both experiments. Online Appendix Table \ref{table:experiment_mainpart1} shows the impact of information about relative mortality on the perceived severity of the crisis. Online Appendix Table \ref{table:experiment_main} shows the impact of coronavirus-related information on economic worries. Online Appendix Table \ref{table:experiment_main_crossrcheck} shows the impact of contagiousness information controlling for treatment assignment in the relative mortality experiment. Online Appendix Table \ref{table:experiment_main_int} explores interactions effects between the experiments. Online Appendix Table \ref{table:pred_expgrowth} displays predictors of mental models of disease spread. Finally, Online Appendix Table \ref{table:correlations_expgrowth} displays the correlations between predicted disease spread and severity perceptions as well as economic worries.

 \newpage

\section{The Impact of Coronavirus Arrival in the US}
\label{sec:opinionpolling}
In this section, we describe the methodology for the additional analysis linking confirmed coronavirus cases in the US to the perceived threat of the pandemic and economic anxieties. For this purpose, we leverage a public opinion poll and our own survey data.

\subsection{Cross-sectional Evidence from the US in mid-February}
The US reported its first case of coronavirus on January 22, 2020 \citep{dong2020}. For most of February, the case count within the US remained fairly flat, increasing from 8 cases on February 1st to 24 cases by February 29. We present  descriptive evidence documenting that there is an association between the spread of coronavirus within the US and increased anxieties using opinion polling data from individuals across US states. The Kaiser Family Foundation poll was conducted from February 13 to February 18, 2020 among a sample of 1207 US residents and included a few questions relating to coronavirus.\footnote{The underlying micro data are made available through the Roper Center ID 31117209.}

During that whole period, there were no reported new cases of coronavirus across the US with the total confirmed case count staying flat at 13 cases. 46 states reported no case. Three states (Washington, Massachusetts and Arizona) reported a single case each, Illinois reported two cases and California reported eight cases. We assess associations between residing in a state with at least any coronavirus case and responses to coronavirus-related questions. To do so, we estimate the following simple regression:

$$y_{i,s,t}  =  \gamma \times  anycase_{s} + \beta'X_i + \eta_t + \epsilon_{is}$$

where $anycase_{s}$ is a dummy indicating the presence of at least one coronavirus case. As indicated, there was no further spread recorded during the sample period according to data from \cite{dong2020}.\footnote{The fact that the US reported little intracommunity spread during much of February is likely not the result of no spread occurring, but rather due to the failure of the US to ramp up testing, and the use of a faulty test, see \url{nature.com/articles/d41586-020-01068-3}.}  

The dependent variable $y_{i,s,t}$ measures a survey respondent’s responses to a set of coronavirus related questions: whether individuals are (very) concerned or (not at all) concerned about: the family or oneself getting sick, a negative economic impact, or a widespread outbreak of coronavirus in the US. 

We study whether individuals living in states with any coronavirus cases during the time period give different responses. We control for interview date fixed effects, $\eta_t$, along with a set of individual-level controls. 

\paragraph{Results} The results are presented in Table \ref{table:opinionpoll} and suggest that respondents from states with any case of coronavirus during that sample period are more concerned about: themselves or family members getting sick;  the negative impact on the US economy; and about a widespread outbreak of coronavirus in the US.

\subsection{Difference-in-differences Analysis in March}

To go beyond a purely cross-sectional analysis of the relationship between the presence of coronavirus cases and economic anxieties we conduct a state-level difference-in-differences analysis. We use the fact that in the 11-day period between our surveys on March 5 and 16, 31 states recorded their first coronavirus case leaving only three states without a case on March 15. This allows us to estimate a regression of the following form:

\begin{equation}
  y_{ist}= \alpha \times anycase_{st}+\delta_s+\eta_t+\beta'X_i+\varepsilon_{ist}
\end{equation}

where $anycase_{st}$ is a dummy variable indicating whether a given state has a confirmed coronavirus case by the time of the survey. We also include state fixed effects to account for permanent differences in economic anxieties across states. Finally, time fixed effects account for any level differences across time that affect all states in the same way. We conduct this analysis for the main four outcomes measuring perceptions of the severity of the pandemic and worries about its impact on the economy.

\paragraph{Results}
Online Appendix Table \ref{table:survey_DiD} displays estimation results. We find that individuals in states with at least one coronavirus case exhibit a significantly higher level of perceived threat by the pandemic and worries about the US economy. The relationship impact on worries about personal economic circumstances is also positive but it is smaller and not statistically significant.

\newpage

\section{Online Appendix Figures}
\label{sec:appfig}


\newpage 

\FloatBarrier

\newpage

\FloatBarrier

\newpage

\FloatBarrier

\newpage

\newpage

\newpage

\newpage
\section{Online Appendix Tables}
\label{sec:apptab}


\begin{center}
\begin{table}[th!]\def\sym#1{\ifmmode^{#1}\else\(^{#1}\)\fi} 
\caption{Countries and territories included in the analysis of Google searches} \label{table:countrylist}

\scalebox{0.7}{
\begin{tabular}{llll}
\toprule
Andorra$^{}$&	Dominica$^{}$&	Korea, Republic of$^{abc}$&	Palestine, State of$^{}$\\
United Arab Emirates$^{b}$&	Dominican Republic$^{}$&	Kuwait$^{b}$&	Portugal$^{abc}$\\
Afghanistan$^{}$&	Algeria$^{b}$&	Cayman Islands$^{}$&	Paraguay$^{}$\\
Antigua and Barbuda$^{}$&	Ecuador$^{ac}$&	Kazakhstan$^{ab}$&	Qatar$^{ab}$\\
Albania$^{}$&	Estonia$^{ab}$&	Laos$^{}$&	Reunion$^{}$\\
Armenia$^{}$&	Egypt$^{ac}$&	Lebanon$^{}$&	Romania$^{abc}$\\
Angola$^{}$&	Spain$^{abc}$&	Saint Lucia$^{}$&	Serbia$^{abc}$\\
Argentina$^{abc}$&	Ethiopia$^{}$&	Sri Lanka$^{b}$&	Russian Federation$^{abc}$\\
Austria$^{abc}$&	Finland$^{abc}$&	Liberia$^{}$&	Rwanda$^{}$\\
Australia$^{abc}$&	Fiji$^{}$&	Lesotho$^{}$&	Saudi Arabia$^{a}$\\
Aruba$^{}$&	Faroe Islands$^{}$&	Lithuania$^{abc}$&	Sudan$^{}$\\
Azerbaijan$^{ab}$&	France$^{abc}$&	Luxembourg$^{}$&	Sweden$^{abc}$\\
Bosnia and Herzegovina$^{}$&	Gabon$^{}$&	Latvia$^{abc}$&	Singapore$^{abc}$\\
Barbados$^{}$&	United Kingdom$^{abc}$&	Libya$^{}$&	Saint Helena$^{}$\\
Bangladesh$^{b}$&	Grenada$^{}$&	Morocco$^{b}$&	Slovenia$^{abc}$\\
Belgium$^{abc}$&	Georgia$^{}$&	Moldova, Republic of$^{}$&	Slovakia$^{abc}$\\
Burkina Faso$^{}$&	French Guiana$^{}$&	Montenegro$^{}$&	Sierra Leone$^{}$\\
Bulgaria$^{abc}$&	Guernsey$^{}$&	Madagascar$^{}$&	Senegal$^{}$\\
Bahrain$^{b}$&	Ghana$^{}$&	North Macedonia$^{}$&	Somalia$^{}$\\
Burundi$^{}$&	Gibraltar$^{}$&	Mali$^{}$&	Suriname$^{}$\\
Benin$^{}$&	Greenland$^{}$&	Myanmar$^{}$&	El Salvador$^{ab}$\\
Bermuda$^{}$&	Guinea$^{}$&	Mongolia$^{}$&	Sint Maarten (Dutch part)$^{}$\\
Brunei Darussalam$^{}$&	Guadeloupe$^{}$&	Macao$^{}$&	Syrian Arab Republic$^{}$\\
Bolivia (Plurinational State of)$^{}$&	Greece$^{abc}$&	Martinique$^{}$&	Eswatini$^{}$\\
Bonaire, Sint Eustatius and Saba$^{}$&	Guatemala$^{}$&	Mauritania$^{}$&	Togo$^{}$\\
Brazil$^{abc}$&	Guam$^{}$&	Malta$^{}$&	Thailand$^{abc}$\\
Bahamas$^{}$&	Guyana$^{}$&	Mauritius$^{}$&	Tajikistan$^{}$\\
Bhutan$^{}$&	Hong Kong$^{abc}$&	Maldives$^{}$&	Turkmenistan$^{}$\\
Botswana$^{}$&	Honduras$^{}$&	Malawi$^{}$&	Tunisia$^{ab}$\\
Belarus$^{}$&	Croatia$^{ab}$&	Mexico$^{abc}$&	Tonga$^{}$\\
Belize$^{}$&	Haiti$^{}$&	Malaysia$^{abc}$&	Turkey$^{abc}$\\
Canada$^{abc}$&	Hungary$^{abc}$&	Mozambique$^{}$&	Trinidad and Tobago$^{}$\\
Congo, Democratic Republic of the$^{}$&	Indonesia$^{abc}$&	New Caledonia$^{}$&	Taiwan, Province of China$^{abc}$\\
Congo$^{}$&	Ireland$^{abc}$&	Niger$^{}$&	Tanzania, United Republic of$^{}$\\
Switzerland$^{abc}$&	Israel$^{abc}$&	Nigeria$^{abc}$&	Ukraine$^{abc}$\\
C\^{o}te d'Ivoire$^{}$&	Isle of Man$^{}$&	Nicaragua$^{}$&	Uganda$^{}$\\
Chile$^{abc}$&	India$^{abc}$&	Netherlands$^{abc}$&	United States of America$^{abc}$\\
Cameroon$^{}$&	Iraq$^{}$&	Norway$^{abc}$&	Uruguay$^{}$\\
China$^{ab}$&	Iran (Islamic Republic of)$^{abc}$&	Nepal$^{}$&	Uzbekistan$^{}$\\
Colombia$^{abc}$&	Iceland$^{}$&	New Zealand$^{abc}$&	Saint Vincent and the Grenadines$^{}$\\
Costa Rica$^{ac}$&	Italy$^{abc}$&	Oman$^{}$&	Venezuela (Bolivarian Republic of)$^{ac}$\\
Cuba$^{}$&	Jersey$^{}$&	Panama$^{}$&	Virgin Islands (U.S.)$^{}$\\
Cabo Verde$^{}$&	Jamaica$^{}$&	Peru$^{abc}$&	Viet Nam$^{ab}$\\
Curacao$^{}$&	Jordan$^{ab}$&	French Polynesia$^{}$&	Yemen$^{}$\\
Cyprus$^{abc}$&	Japan$^{abc}$&	Papua New Guinea$^{}$&	South Africa$^{abc}$\\
Czechia$^{abc}$&	Kenya$^{a}$&	Philippines$^{abc}$&	Zambia$^{}$\\
Germany$^{abc}$&	Kyrgyzstan$^{}$&	Pakistan$^{b}$&	Zimbabwe$^{}$\\
Djibouti$^{}$&	Cambodia$^{}$&	Poland$^{abc}$&	\\
Denmark$^{abc}$&	Saint Kitts and Nevis$^{}$&	Puerto Rico$^{}$&\\	\bottomrule
\end{tabular}}

\begin{minipage}{\textwidth} \scriptsize
\textit{Notes:} Appendix Table \ref{table:countrylist} lists all countries included in the analysis of the impact of the coronavirus on internet searches on Google. The superscripts \textit{a},\textit{b},\textit{c} refer to data availability regarding aggregate demand components for the analysis in Online Appendix Table \ref{table:recessionindicators}: \textit{a} data available on real GDP; \textit{b} data available on industrial production; \textit{c} data available on demand factors.
\end{minipage}
\end{table}
\end{center}

\vfill
\pagebreak

\begin{center}
\begin{table}[th!]\def\sym#1{\ifmmode^{#1}\else\(^{#1}\)\fi} 

  \caption{Increases in `Recession' topic Google searches are a leading indicator of subsequent aggregate demand contractions} \label{table:recessionindicators}
  \begin{center}
\scalebox{0.7} {\begin{tabular}{l*{8}{c}}
\toprule
 &\multicolumn{1}{c}{(1)}   &\multicolumn{1}{c}{(2)}   &\multicolumn{1}{c}{(3)}   &\multicolumn{1}{c}{(4)}   &\multicolumn{1}{c}{(5)}  &\multicolumn{1}{c}{(6)}    &\multicolumn{1}{c}{(7)}      \\
            
 \midrule
 &\multicolumn{2}{c}{}  &\multicolumn{5}{c}{Demand factors}  \\
  & \multicolumn{1}{c}{Real GDP}  &\multicolumn{1}{c}{Industrial production} &\multicolumn{1}{c}{C}  &\multicolumn{1}{c}{I} &\multicolumn{1}{c}{G}  &\multicolumn{1}{c}{X} &\multicolumn{1}{c}{M}\\
                               

\addlinespace
\input{eiu-recession_hits-leadingindicator.tex} \\
\addlinespace
\addlinespace

Country FE  & X& X & X & X& X  & X & X \\
Year x Quarter FE & X& X & X &X & X & X  & X  \\
\bottomrule
\end{tabular}}
\end{center}
\begin{minipage}{\textwidth} \scriptsize
\textit{Notes:} Appendix Table \ref{table:recessionindicators} displays the relationship between year-on-year growth rates in GDP, industrial production and demand components and `Recession' topic Google searches. Econometrically, we perform country-level regressions controlling for country and year-by-quarter fixed effects in all specifications. The results show that increases in Google search activity for recession-related topics are associated with lower growth rates in GDP, consumption spending and imports in the subsequent quarter. The level of analysis is country and quarter. Data were collected by the Economist Intelligence Unit from 2015 to 2019. The dependent variable in  column (1) measures GDP growth. The dependent variable in  column (2) measures growth of industrial production. Columns (3) to (6) measure different components of aggregate demand. Column  (3) shows the association with aggregate consumption.  Column  (4) shows the association with investments. Column  (5) shows the association with government spending.  Column  (6) shows the association with exports. Column  (7) shows the association with imports. The independent variable measures Google search intensity for the topic ``recession''. For countries included in each regression, see Online Appendix Table \ref{table:countrylist}. Standard errors clustered at the country level are presented in parentheses. \sym{*} \(p<0.10\), \sym{**} \(p<0.05\), \sym{***} \(p<0.01\) 
\end{minipage}
\end{table}
\end{center}

\vfill
\pagebreak

\begin{table}[!ht] \centering \def\sym#1{\ifmmode^{#1}\else\(^{#1}\)\fi}

\caption{The impact of coronavirus arrival on Google searches related to economic anxiety } 
\label{table:main_DD}
\scalebox{0.75}{
\input{Main_DD_results}}
\begin{minipage}{\textwidth} \scriptsize
\emph{Notes:} Online Appendix Table \ref{table:main_DD} displays the impact of coronavirus arrival on Google searches for search terms related to economic anxiety. The results show that coronavirus arrival is a predictor of Google searches related to economic anxiety. Column 1 shows results for Google searches related to recessions. Column 2 shows results for Google searches related to stock market crashes. Column 3 shows results for Google searches related to conspiracy topics. Column 4 shows results for Google searches related to survivalism.  The dependent variable measures Google search intensity for the indicated topics normalized by the average search intensity in a country prior to the coronavirus arrival. The data on Google searches were downloaded from the Google API on March 3$^{rd}$. In panel A, we show the impact of a dummy variable indicating at least one coronavirus case. In Panel B, we show the impact of having at least one human-to-human transmission of coronavirus. The data on first cases stem from \cite{dong2020}. The data on human-to-human transmissions are based on official reports by the WHO and national authorities. The level of analysis is country-day. Dates included range from January 1$^{st}$ 2020 to February 29$^{th}$ 2020. The table displays coefficients that are estimated using a linear regression model with country fixed effects and day fixed effects. Standard errors clustered at the country level are presented in parentheses. \sym{*} \(p<0.10\), \sym{**} \(p<0.05\), \sym{***} \(p<0.01\) 
\end{minipage}
\end{table}

\vfill
\pagebreak
\vfill

\begin{table}[tbp] \centering \def\sym#1{\ifmmode^{#1}\else\(^{#1}\)\fi} 
  \caption{Robustness of results to not normalizing -- Impact of coronavirus arrival on Google searches related to economic anxiety }
  \label{table:main_DD_nonormalisation}
 \scalebox{0.75}{
 \input{Main_DD_results_nonormalization.tex}
 }
\begin{minipage}{\textwidth} \scriptsize
\emph{Notes:} The dependent variable measure country-specific search intensity for a topic indicated in the column head from January 2020 to February 29 2020.  Column 1 shows results for Google searches related to recessions. Column 2 shows results for Google searches related to stock market crashes. Column 3 shows results for Google searches related to conspiracy topics. Column 4 shows results for Google searches related to survivalism.  The dependent variable measures Google search intensity for the indicated topics. The data on Google searches were downloaded from the Google API on March 3$^{rd}$. In panel A, we show the impact of a dummy variable indicating at least one coronavirus case. In Panel B, we show the impact of having at least one human-to-human transmission of coronavirus. The data on first cases stem from \cite{dong2020}. The data on human-to-human transmissions are based on official reports by the WHO and national authorities. The level of analysis is country-day. Dates included range from January 1$^{st}$ 2020 to February 29$^{th}$ 2020. The table displays coefficients that are estimated using a linear regression model with country fixed effects and day fixed effects. Standard errors are clustered at the country are presented in parentheses. \sym{*} \(p<0.10\), \sym{**} \(p<0.05\), \sym{***} \(p<0.01\) 
\end{minipage}
\end{table}

\vfill
\pagebreak

\vfill
\begin{landscape}
\hfill
 \begin{table}[!ht] \centering \def\sym#1{\ifmmode^{#1}\else\(^{#1}\)\fi} \caption{The impact of coronavirus arrival on placebo Google searches} 
\scalebox{0.65}{\input{Placebo_DD_results}}
\begin{minipage}{\textheight} 
\vspace{1 em}
\scriptsize  \emph{Notes:} Online Appendix Table \ref{table:placebo_DD} displays the impact of coronavirus arrival on placebo Google searches that should not be affected by the arrival of the coronavirus.
 The results document that coronavirus arrival does not systematically predict  Google searches unrelated to economic anxiety.
 The dependent variable measures Google search intensity for the indicated topics normalized by the average search intensity in a country prior to the coronavirus arrival. The data on Google searches were downloaded from the Google API on March 3$^{rd}$. In panel A, we show the impact of a dummy variable indicating at least one coronavirus case. In Panel B, we show the impact of having at least one human-to-human transmission of coronavirus. The data stem from \cite{dong2020}. The data on human-to-human transmissions are based on official reports by the WHO and national authorities. The level of analysis is country-day. Dates included range from January 1$^{st}$ 2020 to February 29$^{th}$ 2020. The table displays coefficients that are estimated using a linear regression model with country fixed effects and day fixed effects. Standard errors clustered at the country level are presented in parentheses. \sym{*} \(p<0.10\), \sym{**} \(p<0.05\), \sym{***} \(p<0.01\).
\label{table:placebo_DD}
\end{minipage}
 \end{table}
 \end{landscape}

\pagebreak

\begin{table}[!ht] \centering \def\sym#1{\ifmmode^{#1}\else\(^{#1}\)\fi} \caption{Summary statistics: Experimental sample March 5} 
\scalebox{0.9}{\input{sumstats_experiment_main}}
\begin{minipage}{\textwidth} 
\vspace{1 em}
\scriptsize \emph{Notes:} Online Appendix Table \ref{table:sum_exp} displays summary statistics for the experimental sample. These data were collected on March 5.  Panel A shows shares of respondents with indicated characteristics. Panel B shows shares of respondents with indicated beliefs about the severity of the crisis and economic anxieties. Panel C shows variables measuring perceptions of the coronavirus.   
\label{table:sum_exp}
\end{minipage}
 \end{table}

\pagebreak

\begin{table}[!ht] \centering \def\sym#1{\ifmmode^{#1}\else\(^{#1}\)\fi} \caption{Summary statistics: Exponential growth survey March 16} 
\scalebox{0.8}{\input{sum_expgrowth}}
\begin{minipage}{\textwidth} 
\vspace{1 em}
\scriptsize \emph{Notes:} Online Appendix Table \ref{table:sum_expgrowth} displays summary statistics for the experimental sample. These data were collected on March 16. Panel A shows shares of respondents with indicated characteristics. Panel B shows shares of respondents with indicated beliefs about the severity of the crisis and economic anxieties. Panel C shows variables measuring coronavirus perceptions and predictions of fictitious infectious disease spread. 
\label{table:sum_expgrowth}
\end{minipage}
 \end{table}

\pagebreak


\begin{table}[tbp] \centering \def\sym#1{\ifmmode^{#1}\else\(^{#1}\)\fi}
  \caption{US Opinion Polling Data on Reaction to Coronavirus From 13 - 18 Feb 2020}
\label{table:opinionpoll}
\scalebox{0.55}{\begin{tabular}{l*{3}{c}}
\toprule
 &\multicolumn{1}{c}{(1)}   &\multicolumn{1}{c}{(2)}   &\multicolumn{1}{c}{(3)}         \\
            
 \midrule
&  \multicolumn{3}{c}{Strength of concern that } \\
		                                        \cmidrule(lr){2-4}

	& \multicolumn{1}{c}{you/your family getting sick from coronavirus} &\multicolumn{1}{c}{coronavirus will have negative impact on US economy}  &\multicolumn{1}{c}{widespread outbreak of coronavirus in US}  \\

\addlinespace \\
\input{table-roper-poll-anycovid_confirmed.tex} \\
\addlinespace

Individual Controls & X& X & X  \\
Interview Date FE & X& X & X   \\

\bottomrule
\end{tabular}}

\begin{minipage}{\textwidth} \vspace{1 em} \scriptsize
 \emph{Notes:} Online Appendix Table \ref{table:opinionpoll} presents results from studying the impact of the coronavirus on perceptions and awareness during the early period of the spread. Data come from an opinion poll conducted for the Kaiser Family Foundation Poll (Roper Center ID 31117209) from February 13 - February 18. During this time there were only 13 reported coronavirus cases in all of the US which were concentrated in 5 states. Outcomes are measured on a 4-point scale (Not at all concerned;  Not too concerned; Somewhat concerned; Very concerned) and standardized to have mean zero and standard deviation one. Individual controls include age, gender, education, income and political party affiliation (5 point scale). Observation counts vary due to ``don't knows'' or non-response. Standard errors clustered at the state level are presented in parentheses with stars indicating \sym{*} \(p<0.10\), \sym{**} \(p<0.05\), \sym{***} \(p<0.01\) 
 \end{minipage}
\end{table}


\pagebreak

\begin{landscape}
\begin{table}[!ht] \centering \def\sym#1{\ifmmode^{#1}\else\(^{#1}\)\fi} \caption{Coronavirus perceptions and economic anxieties over time} 
\scalebox{0.75}{\input{comparison_time.tex}}
\begin{minipage}{\textwidth} 
\vspace{1 em}
\footnotesize  \emph{Notes:} Online Appendix Table \ref{table:comparison_time} displays summary statistics for economic anxieties (Panel A) and coronavirus perceptions (Panel B). Columns (1) - (3) display descriptives for Experiment 1 conducted on March 5, while Columns (4) to (6) display the descriptives for Experiment 2 conducted on March 16.   
\label{table:comparison_time}
\end{minipage}
 \end{table}
 \end{landscape}

\pagebreak

\begin{landscape}
\begin{table}[!ht] \centering \def\sym#1{\ifmmode^{#1}\else\(^{#1}\)\fi} \caption{Impact of coronavirus arrival in US states on economic anxieties} 
\scalebox{0.75}{\input{DiD_survey}}
\begin{minipage}{\textwidth} 
\vspace{1 em}
\footnotesize  \emph{Notes:} Online Appendix Table \ref{table:survey_DiD} displays the effect of having at least one confirmed coronavirus case on economic anxieties using a difference-in-differences estimation with survey data collected on March 5 and 16. All regressions include state and day fixed effects and control for gender, age bin dummies, log income, log income squared, dummies for having a high school degree and having some college education, dummies for being unemployed, currently working, a student and dummies for self-identifying as Democrat or Republican. The dependent variables in columns (1) to (2) are agreement on a five-point Likert-scale (from ``strongly disagree '' to ``strongly agree'') with the statements ``The world will be severely affected by the coronavirus.'' (column (1)), and ``The US will be severely affected by the coronavirus.'' (column (2)). The dependent variables in columns (3) and (4) are answers on a four-point Likert-scale (from ``not at all worried'' to ``very worried'') to the questions ``Are you worried about the effects of the coronavirus on the US economy?'' (column (3)) and ``Are you worried about the effects of the coronavirus on your household's economic situation?'' (column (4)).  All outcomes are standardized to have mean 0 and standard deviation 1 within each survey wave. Standard errors clustered at the state level are presented in parentheses.  \sym{*} \(p<0.10\), \sym{**} \(p<0.05\), \sym{***} \(p<0.01\) 
\label{table:survey_DiD}
\end{minipage}
 \end{table}
 \end{landscape}

\pagebreak


 \begin{table}[!ht] \centering \def\sym#1{\ifmmode^{#1}\else\(^{#1}\)\fi} \caption{The association of misperceptions and economic anxieties}  
\scalebox{0.7}{\input{experiment_correlations_combined.tex}}
\begin{minipage}{\textwidth} 
\vspace{1 em}
\scriptsize  \emph{Notes:} Online Appendix Table \ref{table:experiment_correlations} displays the raw effect of overestimating mortality and contagiousness of coronavirus (relative to official estimates) on the perceived severity of the effects of the coronavirus and economic worries. The data were collected on March 5. Panel A shows the results without control variables. Panel B shows the results controlling for gender, age bin dummies, log income, log income squared, dummies for having a high school degree and having some college education, dummies for being unemployed, currently working, a student and dummies for self-identifying as Democrat or Republican. The table shows coefficients estimated using linear regressions that include indicators for respondents whose beliefs about coronavirus mortality were higher relative to official estimates  and for respondents whose beliefs about coronavirus contagiousness were higher relative to scientific estimates. The dependent variables in columns (1) to (2) are agreement on a five-point Likert-scale (from ``strongly disagree '' to ``strongly agree'') with the statements ``The world will be severely affected by the coronavirus.'' (column (1)), and ``The US will be severely affected by the coronavirus.'' (column (2)). The dependent variables in columns (3) and (4) are answers on a four-point Likert-scale (from ``not at all worried'' to ``very worried'') to the questions ``Are you worried about the effects of the coronavirus on the US economy?'' (column (3)) and ``Are you worried about the effects of the coronavirus on your household's economic situation?'' (column (4)).  All outcomes are standardized to have mean 0 and standard deviation 1. Heteroskedasticity robust standard errors are presented in parentheses.  \sym{*} \(p<0.10\), \sym{**} \(p<0.05\), \sym{***} \(p<0.01\) 
\label{table:experiment_correlations}
\end{minipage}
\end{table}

 \begin{table}[!ht] \centering \def\sym#1{\ifmmode^{#1}\else\(^{#1}\)\fi} \caption{The association of misperceptions and economic anxieties - continuous measures}  
\scalebox{0.7}{\input{experiment_correlations_combined_continuous.tex}}
\begin{minipage}{\textwidth} 
\vspace{1 em}
\scriptsize  \emph{Notes:} Online Appendix Table \ref{table:experiment_correlations_cont} displays the correlation between mortality and contagiousness of coronavirus on the perceived severity of the effects of the coronavirus and economic worries. The data were collected on March 5. Panel A shows the results without control variables. Panel B shows the results controlling for gender, age bin dummies, log income, log income squared, dummies for having a high school degree and having some college education, dummies for being unemployed, currently working, a student and dummies for self-identifying as Democrat or Republican. The table shows coefficients for beliefs about  coronavirus mortality and about coronavirus contagiousness estimated using linear regressions. Beliefs about mortality are rescaled to range from 0 to 1. Beliefs about contagiousness are rescaled to display the association with believing that 100 extra people get infected. Beliefs about mortality and contagiousness are winsorized at the 95th percentile to account for outliers. The dependent variables in columns (1) to (2) are agreement on a five-point Likert-scale (from ``strongly disagree '' to ``strongly agree'') with the statements ``The world will be severely affected by the coronavirus.'' (column (1)), and ``The US will be severely affected by the coronavirus.'' (column (2)). The dependent variables in columns (3) and (4) are answers on a four-point Likert-scale (from ``not at all worried'' to ``very worried'') to the questions ``Are you worried about the effects of the coronavirus on the US economy?'' (column (3)) and ``Are you worried about the effects of the coronavirus on your household's economic situation?'' (column (4)).  All outcomes are standardized to have mean 0 and standard deviation 1. Heteroskedasticity robust standard errors are presented in parentheses.  \sym{*} \(p<0.10\), \sym{**} \(p<0.05\), \sym{***} \(p<0.01\) 
\label{table:experiment_correlations_cont}
\end{minipage}
 \end{table}

\pagebreak

 \begin{table}[!ht] \centering \def\sym#1{\ifmmode^{#1}\else\(^{#1}\)\fi} \caption{Experimental integrity: Balance table} 
\scalebox{0.5}{\input{balance_experiment_main}}
\begin{minipage}{\textwidth} 
\vspace{1 em}
\scriptsize \emph{Notes:} Online Appendix Table \ref{table:balance} displays balance tests for the experimental sample. The data were collected on March 5. Columns (1) to (3) show means for both experimental groups  in the mortality information experiment and the p-value for a test of equality of means  across samples. Columns (4) to (6) show means for both experimental groups in the contagiousness experiment and the p-value for a test of equality of means across samples. p-values are obtained using heteroskedasticity robust standard errors. p-values for the test of joint significance are based on the F-statistic obtained  by regressing all observables on the treatment indicators. 
\label{table:balance}
\end{minipage}
 \end{table}

\pagebreak

 \begin{table}[!ht] \centering \def\sym#1{\ifmmode^{#1}\else\(^{#1}\)\fi} \caption{The impact of coronavirus-related information on perceived severity of the crisis} 
\scalebox{1}{\input{experiment_mainpart1_combined}}
\begin{minipage}{\textwidth} 
\vspace{1 em}
\scriptsize \emph{Notes:} Online Appendix Table \ref{table:experiment_mainpart1} displays the impact of information about the coronavirus on the perceived severity of the impacts of the coronavirus. The data were collected on March 5. Panel A shows the results without control variables. Panel B shows the results controlling for gender, age bin dummies, log income, log income squared, dummies for having a high school degree and having some college education, dummies for being unemployed, currently working, a student and dummies for self-identifying as Democrat or Republican. The table shows coefficients estimated using linear regressions that compare respondents who were truthfully informed that the death rate of the coronavirus is either “20 times higher than for the flu” (high mortality treatment), or “5 times lower than for SARS” (low mortality treatment). The dependent variables in columns (1) to (2) are agreement on a five-point Likert-scale (from ``strongly disagree '' to ``strongly agree'') with the statements ``The world will be severely affected by the coronavirus.'' (column (1)), and ``The US will be severely affected by the coronavirus.'' (column (2)). All outcomes are standardized to have mean 0 and standard deviation 1. Heteroskedasticity robust standard errors are presented in parentheses.  \sym{*} \(p<0.10\), \sym{**} \(p<0.05\), \sym{***} \(p<0.01\) 
\label{table:experiment_mainpart1}
\end{minipage}
 \end{table}

\pagebreak

 \begin{table}[!ht] \centering \def\sym#1{\ifmmode^{#1}\else\(^{#1}\)\fi} \caption{The impact of coronavirus-related  information on economic worries } 
\scalebox{0.75}{\input{experiment_mainpart2_combined}}
\begin{minipage}{\textwidth} 
\vspace{1 em}
\scriptsize  \emph{Notes:} Online Appendix Table \ref{table:experiment_main} displays the impact of information about the coronavirus on economic anxiety. The data were collected on March 5. The results show that information about coronavirus causally affects economic anxiety. Panel A shows the results without control variables. Panel B shows the results controlling for gender, age bin dummies, log income, log income squared, dummies for having a high school degree and having some college education, dummies for being unemployed, currently working, a student and dummies for self-identifying as Democrat or Republican. \textit{High relative mortality} shows coefficients estimated using linear regressions that compare respondents who were truthfully informed that the death rate of the coronavirus is either “20 times higher than for the flu” (high mortality treatment), or “5 times lower than for SARS” (low mortality treatment).   \textit{Contagion information}  shows regression coefficients that compare respondents who were truthfully informed about the estimated contagiousness of coronavirus (R0$\approx$2) to respondents who were given no information. Estimates for  \textit{Contagion information} are obtained with an ANCOVA specification using baseline outcomes obtained in the same survey prior to the information treatment. The dependent variables in columns (1) and (2) are answers on a four-point Likert-scale (from ``not at all worried'' to ``very worried'') to the questions ``Are you worried about the effects of the coronavirus on the US economy?'' (column (1)) and ``Are you worried about the effects of the coronavirus on your household's economic situation?'' (column (2)). All outcomes are standardized to have mean 0 and standard deviation 1. Heteroskedasticity robust standard errors are presented in parentheses.  \sym{*} \(p<0.10\), \sym{**} \(p<0.05\), \sym{***} \(p<0.01\) 
\label{table:experiment_main}
\end{minipage}
 \end{table}

\pagebreak

 \begin{table}[!ht] \centering \def\sym#1{\ifmmode^{#1}\else\(^{#1}\)\fi} \caption{Controlling for treatment assignment in the relative mortality experiment} 
\scalebox{1}{\input{contagion_cross_rand_check}}
\begin{minipage}{\textwidth} 
\vspace{1 em}
\scriptsize  \emph{Notes:} Online Appendix Table \ref{table:experiment_main_crossrcheck} displays the impact of information on the contagiousness of coronavirus on economic anxiety. The data were collected on March 5. All specifications include a dummy for whether the respondent saw the high relative mortality treatment during the first experimental variation. Panel A shows the results without control variables. Panel B shows the results controlling for gender, age bin dummies, log income, log income squared, dummies for having a high school degree and having some college education, dummies for being unemployed, currently working, a student and dummies for self-identifying as Democrat or Republican.  \textit{Contagion information} shows regression coefficients that compare respondents who were truthfully informed about the estimated contagiousness of coronavirus (R0$\approx$2) to respondents who were given no information. Estimates are obtained with an ANCOVA specification using baseline outcomes obtained in the same survey prior to the information treatment. The dependent variables in columns (1) and (2) are answers on a four-point Likert-scale (from ``not at all worried'' to ``very worried'') to the questions ``Are you worried about the effects of the coronavirus on the US economy?'' (column (1)) and ``Are you worried about the effects of the coronavirus on your household's economic situation?'' (column (2)). All outcomes are standardized to have mean 0 and standard deviation 1. Heteroskedasticity robust standard errors are presented in parentheses.  \sym{*} \(p<0.10\), \sym{**} \(p<0.05\), \sym{***} \(p<0.01\) 
\label{table:experiment_main_crossrcheck}
\end{minipage}
 \end{table}

\pagebreak

 \begin{table}[!ht] \centering \def\sym#1{\ifmmode^{#1}\else\(^{#1}\)\fi} \caption{Interaction effects between contagion and mortality information} 
\scalebox{0.8}{\input{treatment_interaction_effects.tex}}
\begin{minipage}{\textwidth} 
\vspace{1 em}
\scriptsize \emph{Notes:} Online Appendix Table \ref{table:experiment_main_int} shows the interaction effect of our treatments on economic anxiety. The data were collected on March 5. Panel A shows the results without control variables. Panel B shows the results controlling for gender, age bin dummies, log income, log income squared, dummies for having a high school degree and having some college education, dummies for being unemployed, currently working, a student and dummies for self-identifying as Democrat or Republican. The dependent variables in columns (1) and (2) are answers on a four-point Likert-scale (from ``not at all worried'' to ``very worried'') to the questions ``Are you worried about the effects of the coronavirus on the US economy?'' (column (1)) and ``Are you worried about the effects of the coronavirus on your household's economic situation?'' (column (2)). The outcome variables are elicited after both treatments. Estimates are obtained with an ANCOVA specification using baseline outcomes obtained in the same survey after the mortality treatment and prior to the contagiousness information treatment. We do not include a baseline indicator for the mortality treatment as it would be collinear with the baseline measures of economic worries which are themselves sufficient statistics for the assignment probabilities regarding the mortality treatments. For interpretation, we use the interaction between the low mortality treatment (as opposed to the high mortality treatment), as it is complementary to the contagiousness information given that both treatments render the severity of the coronavirus less intense. All outcomes are standardized to have mean 0 and standard deviation 1. Heteroskedasticity robust standard errors are presented in parentheses.  \sym{*} \(p<0.10\), \sym{**} \(p<0.05\), \sym{***} \(p<0.01\) 
\label{table:experiment_main_int}
\end{minipage}
 \end{table}

\pagebreak

 \begin{table}[!ht] \centering \def\sym#1{\ifmmode^{#1}\else\(^{#1}\)\fi} \caption{Predictors of mental models of disease spread}  
 
\scalebox{0.6}{\input{growth_perc_pred}}
\begin{minipage}{\textwidth} 
\vspace{1 em}
\scriptsize \emph{Notes:} Online Appendix Table \ref{table:pred_expgrowth} displays the correlations of covariates with different classifications of mental models. These data were collected on March 16. Columns (1) and (2) show the correlation with log predicted of cases after 20 days from a fictitious disease. Columns (3) and (4) show the correlation with an indicator variable being classified as having an exponential model by k-means clustering. Columns (1) and (3) show the result for a simple OLS regression. Columns (2) and (4)  show results using OLS with only the variables selected by the LASSO algorithm using $\alpha$ and $\lambda$ parameters chosen by 10-fold cross-validation. Heteroskedasticity robust standard errors are presented in parentheses. \sym{*} \(p<0.10\), \sym{**} \(p<0.05\), \sym{***} \(p<0.01\) 
\label{table:pred_expgrowth}
\end{minipage}
 \end{table}
 
\pagebreak

 \begin{table}[!ht] \centering \def\sym#1{\ifmmode^{#1}\else\(^{#1}\)\fi} \caption{The association of mental models of infectious disease spread and economic anxieties}  

\scalebox{0.6}{\input{correlations_expgrowth_combined.tex}}
\begin{minipage}{\textwidth} 
\vspace{1 em}
\scriptsize  \emph{Notes:} Online Appendix Table \ref{table:correlations_expgrowth} displays the regression coefficients of perceived severity of the effects of the coronavirus with participants' standardized log estimate of the spread of a fictitious disease. The data were collected on March 16. The table shows coefficients estimated using linear regressions that regress perceived crisis severity and economic anxieties on the z-scored log of estimated infections from a fictitious disease. The dependent variables in columns (1) to (4) are agreement on a five-point Likert-scale (from ``strongly disagree '' to ``strongly agree'') with the statements ``The world will be severely affected by the coronavirus.'' (columns (1) and (2)), and ``The US will be severely affected by the coronavirus.'' (columns (3) and (4)). The dependent variables in columns (5) to (8) are answers on a four-point Likert-scale (from ``not at all worried'' to ``very worried'') to the questions ``Are you worried about the effects of the coronavirus on the US economy?'' (columns (5) and (6)) and ``Are you worried about the effects of the coronavirus on your household's economic situation?'' (columns (7) and (8)). The right-hand-side variables are the standardized log of participants' estimates for the number of people infected with the fictitious disease on day 5, day 10 and day 20, respectively.  All outcomes are standardized to have mean 0 and standard deviation 1. Even columns control for gender, age bin dummies, log income, log income squared, dummies for having a high school degree and having some college education, dummies for being unemployed, being currently working, being a student and dummies for self-identifying as Democrat and Republican. Heteroskedasticity robust standard errors are presented in parentheses.  \sym{*} \(p<0.10\), \sym{**} \(p<0.05\), \sym{***} \(p<0.01\) 
\label{table:correlations_expgrowth}
\end{minipage}
 \end{table}
 
\pagebreak

\pagebreak

%
%
%

\pagebreak

\end{document}

%% file: eiu-recession_hits-leadingindicator.tex
L.Recession topic Google searches&      -1.009***&      -1.231*  &      -1.564***&      -1.847   &      -1.345   &       1.109*  &      -5.063***\\
                    &     (0.311)   &     (0.661)   &     (0.506)   &     (1.798)   &     (0.888)   &     (0.657)   &     (1.352)   \\
                    &               &               &               &               &               &               &               \\
R$^2$               &        .716   &        .446   &        .627   &         .27   &        .282   &        .236   &        .314   \\
Countries           &          70   &          72   &          58   &          58   &          58   &          58   &          58   \\
Observations        &        1350   &        1218   &        1087   &        1087   &        1087   &        1087   &        1087   \\

%% file: Main_DD_results.tex
{
\def\sym#1{\ifmmode^{#1}\else\(^{#1}\)\fi}
\begin{tabular}{l*{4}{c}}
\toprule
                &\multicolumn{4}{c}{Impact on Goolge search trends}\\\cmidrule(lr){2-5}
                &\multicolumn{1}{c}{(1)}&\multicolumn{1}{c}{(2)}&\multicolumn{1}{c}{(3)}&\multicolumn{1}{c}{(4)}\\
                &\multicolumn{1}{c}{Recession}&\multicolumn{1}{c}{Stock Market Crash}&\multicolumn{1}{c}{Conspiracy Theory}&\multicolumn{1}{c}{Survivalism}\\
\midrule
{\bf Panel  A: Any Covid-19 case}&         &         &         &         \\
                &         &         &         &         \\
Post any Covid-19 case&0.178\sym{**}&0.580\sym{***}&0.447\sym{***}&0.204\sym{***}\\
                &  (0.073)&  (0.124)&  (0.091)&  (0.073)\\
                &         &         &         &         \\
\midrule R$^2$  &     0.04&     0.10&     0.06&     0.13\\
Number of Observations &    11640&    11640&    11640&    11640\\
\midrule {\bf Panel B: Any human-to-human transmission}&         &         &         &         \\
                &         &         &         &         \\
Post any human-to-human transmission&0.351\sym{**}&0.293\sym{*}&0.388\sym{**}&0.354\sym{**}\\
                &  (0.141)&  (0.163)&  (0.164)&  (0.140)\\
                &         &         &         &         \\
\midrule R$^2$  &     0.04&     0.10&     0.06&     0.13\\
Number of Observations&    11640&    11640&    11640&    11640\\
 \midrule Number of countries&      194&      194&      194&      194\\
Country FE      &        X&        X&        X&        X\\
\bottomrule
\end{tabular}
}

%% file: Main_DD_results_nonormalization.tex
{
\def\sym#1{\ifmmode^{#1}\else\(^{#1}\)\fi}
\begin{tabular}{l*{4}{c}}
\toprule
                &\multicolumn{4}{c}{Impact on Goolge search trends}\\\cmidrule(lr){2-5}
                &\multicolumn{1}{c}{(1)}&\multicolumn{1}{c}{(2)}&\multicolumn{1}{c}{(3)}&\multicolumn{1}{c}{(4)}\\
                &\multicolumn{1}{c}{Recession}&\multicolumn{1}{c}{Stock Market Crash}&\multicolumn{1}{c}{Conspiracy Theory}&\multicolumn{1}{c}{Survivalism}\\
\midrule
{\bf Panel  A: Any Covid-19 case}&         &         &         &         \\
                &         &         &         &         \\
Post any Covid-19 case&2.522\sym{***}&2.824\sym{***}&3.252\sym{***}&1.493\sym{***}\\
                &  (0.677)&  (0.538)&  (0.580)&  (0.443)\\
                &         &         &         &         \\
\midrule R$^2$  &     0.27&     0.26&     0.31&     0.36\\
Number of Observations &    11640&    11640&    11640&    11640\\
\midrule {\bf Panel B: Any human-to-human transmission}&         &         &         &         \\
                &         &         &         &         \\
Post any human-to-human transmission&5.644\sym{***}&1.792\sym{*}&3.622\sym{*}&2.281\sym{**}\\
                &  (1.927)&  (1.069)&  (1.999)&  (0.978)\\
                &         &         &         &         \\
\midrule R$^2$  &     0.27&     0.26&     0.31&     0.36\\
Number of Observations&    11640&    11640&    11640&    11640\\
 \midrule Number of countries&      194&      194&      194&      194\\
Country FE      &        X&        X&        X&        X\\
\bottomrule
\end{tabular}
}

%% file: Placebo_DD_results.tex
{
\def\sym#1{\ifmmode^{#1}\else\(^{#1}\)\fi}
\begin{tabular}{l*{10}{c}}
\toprule
                &\multicolumn{10}{c}{Impact on google searches for}                                                 \\\cmidrule(lr){2-11}
                &\multicolumn{1}{c}{(1)}&\multicolumn{1}{c}{(2)}&\multicolumn{1}{c}{(3)}&\multicolumn{1}{c}{(4)}&\multicolumn{1}{c}{(5)}&\multicolumn{1}{c}{(6)}&\multicolumn{1}{c}{(7)}&\multicolumn{1}{c}{(8)}&\multicolumn{1}{c}{(9)}&\multicolumn{1}{c}{(10)}\\
                &\multicolumn{1}{c}{Dog}&\multicolumn{1}{c}{Horse}&\multicolumn{1}{c}{Insect}&\multicolumn{1}{c}{DaVinci}&\multicolumn{1}{c}{Nelson Mandela}&\multicolumn{1}{c}{Rain}&\multicolumn{1}{c}{Rainbow}&\multicolumn{1}{c}{Stars}&\multicolumn{1}{c}{Mars (planet)}&\multicolumn{1}{c}{Menstrual Cycle}\\
\midrule
{\bf Panel  A: First confirmed case}&         &         &         &         &         &         &         &         &         &         \\
                &         &         &         &         &         &         &         &         &         &         \\
First confirmed case&   -0.018&   -0.036&   -0.007&   -0.044&   -0.033&    0.044&   -0.042&   -0.012&   -0.001&    0.033\\
                &  (0.018)&  (0.027)&  (0.042)&  (0.038)&  (0.058)&  (0.048)&  (0.045)&  (0.033)&  (0.103)&  (0.033)\\
                &         &         &         &         &         &         &         &         &         &         \\
\midrule R$^2$  &     0.05&     0.03&     0.02&     0.04&     0.03&     0.04&     0.03&     0.02&     0.06&     0.04\\
Number of Observations &    11640&    11640&    11639&    11249&    11483&    11640&    11640&    11640&    11507&    11477\\
\midrule {\bf Panel  B: First human-to-human  transmission}&         &         &         &         &         &         &         &         &         &         \\
                &         &         &         &         &         &         &         &         &         &         \\
First human-to-human  transmission&   -0.021&   -0.085&    0.035&   -0.030&    0.027&   -0.041&   -0.047&   -0.024&    0.102&    0.039\\
                &  (0.038)&  (0.064)&  (0.040)&  (0.039)&  (0.059)&  (0.078)&  (0.055)&  (0.048)&  (0.324)&  (0.028)\\
                &         &         &         &         &         &         &         &         &         &         \\
\midrule R$^2$  &     0.05&     0.03&     0.02&     0.04&     0.03&     0.04&     0.03&     0.02&     0.06&     0.04\\
Number of Observations&    11640&    11640&    11639&    11249&    11483&    11640&    11640&    11640&    11507&    11477\\
\midrule  Number of countries&      194&      194&      194&      193&      194&      194&      194&      194&      194&      193\\
Country FE      &        X&        X&        X&        X&        X&        X&        X&        X&        X&        X\\
Day FE          &        X&        X&        X&        X&        X&        X&        X&        X&        X&        X\\
\bottomrule
\end{tabular}
}

%% file: sumstats_experiment_main.tex
\begin{tabular}{l*{4}{c}} \toprule
          &\multicolumn{1}{c}{(1)}&\multicolumn{1}{c}{(2)}&\multicolumn{1}{c}{(3)}&\multicolumn{1}{c}{(4)}\\
          &\multicolumn{1}{c}{Mean}&\multicolumn{1}{c}{SD}&\multicolumn{1}{c}{Median}&\multicolumn{1}{c}{Obs.}\\
\midrule
\textbf{Panel A: Demographics}&         &         &         &         \\
          &         &         &         &         \\
\%  Male  &    49.02&    50.02&         &      914\\
\%  Age < 35&     0.23&     0.42&         &      914\\
\% Highschool education&    17.61&    38.12&         &      914\\
\%  College eductation&    80.53&    39.62&         &      914\\
\% Currently working&    55.03&    49.77&         &      914\\
\% Democrat&    40.04&    49.03&         &      914\\
\%  Republican&    33.15&    47.10&         &      914\\
\% High trust in science&     1.09&    10.41&         &      914\\
          &         &         &         &         \\
\textbf{Panel B: Economic Anxieties}&         &         &         &         \\
          &         &         &         &         \\
\% agree: world severely affected by coronavirus&    67.61&    46.82&         &      914\\
\% agree: US severely affected by coronavirus&    55.14&    49.76&         &      914\\
\% worried about US economy&    68.05&    46.65&         &      914\\
\% worried about personal econ. situation&    47.16&    49.95&         &      914\\
          &         &         &         &         \\
\textbf{Panel C: Coronavirus perceptions }&         &         &         &         \\
          &         &         &         &         \\
Infectiousness (R0)&    43.23&   146.17&       10&      914\\
Predicted  mortality rate&    13.70&    20.84&        5&      914\\
\bottomrule \end{tabular}

%% file: sum_expgrowth.tex
\begin{tabular}{l*{4}{c}} \toprule
          &\multicolumn{1}{c}{(1)}&\multicolumn{1}{c}{(2)}&\multicolumn{1}{c}{(3)}&\multicolumn{1}{c}{(4)}\\
          &\multicolumn{1}{c}{Mean}&\multicolumn{1}{c}{SD}&\multicolumn{1}{c}{Median}&\multicolumn{1}{c}{Obs.}\\
\midrule
\textbf{Panel A: Demographics}&         &         &         &         \\
          &         &         &         &         \\
\%  Male  &    52.09&    49.98&         &     1006\\
\%  Age < 35&    22.66&    41.89&         &     1006\\
\% Highschool education&    19.98&    40.00&         &     1006\\
\%  College eductation&    76.64&    42.33&         &     1006\\
\% Currently working&    52.19&    49.98&         &     1006\\
\% Democrat&    38.57&    48.70&         &     1006\\
\%  Republican&    32.11&    46.71&         &     1006\\
\% High trust in science&     1.79&    13.26&         &     1006\\
          &         &         &         &         \\
\textbf{Panel B: Economic Anxieties}&         &         &         &         \\
          &         &         &         &         \\
\% agree: world severely affected by coronavirus&    80.12&    39.93&         &     1006\\
\% agree: US severely affected by coronavirus&    77.83&    41.56&         &     1006\\
\% worried about US economy&    87.57&    33.00&         &     1006\\
\% worried about personal econ. situation&    73.76&    44.02&         &     1006\\
          &         &         &         &         \\
\textbf{Panel C: Coronavirus perceptions }&         &         &         &         \\
          &         &         &         &         \\
Infectiousness (R0)&    49.81&   175.13&        5&     1006\\
Number of cases after 5 days (w)&    20.02&    20.72&       11&     1006\\
Number of cases after 10 days  (w)&   340.29&   678.64&       30&     1006\\
Number of cases after 20 days  (w)&122218.17&311256.66&       60&     1006\\
Predicted  mortality rate&    15.60&    21.47&        5&     1006\\
\bottomrule \end{tabular}

%% file: table-roper-poll-anycovid_confirmed.tex
Any Covid-19 case   &       0.111*  &       0.124** &       0.123** \\
                    &     (0.063)   &     (0.057)   &     (0.056)   \\
                    &               &               &               \\
Mean of DV          &      .00278   &      .00213   &     .000396   \\
R$^2$               &        .179   &        .139   &        .186   \\
States              &          50   &          50   &          50   \\
Observations        &        1197   &        1192   &        1197   \\

%% file: comparison_time.tex
\begin{tabular}{l*{8}{c}} \toprule
          &\multicolumn{3}{c}{March 5}  &\multicolumn{3}{c}{March 16} &\multicolumn{2}{c}{Comparison of means}\\\cmidrule(lr){2-4}\cmidrule(lr){5-7}\cmidrule(lr){8-9}
          &\multicolumn{1}{c}{(1)}&\multicolumn{1}{c}{(2)}&\multicolumn{1}{c}{(3)}&\multicolumn{1}{c}{(4)}&\multicolumn{1}{c}{(5)}&\multicolumn{1}{c}{(6)}&\multicolumn{1}{c}{(7)}&\multicolumn{1}{c}{(8)}\\
          &\multicolumn{1}{c}{Mean}&\multicolumn{1}{c}{SD}&\multicolumn{1}{c}{Obs.}&\multicolumn{1}{c}{Mean}&\multicolumn{1}{c}{SD}&\multicolumn{1}{c}{Obs.}&\multicolumn{1}{c}{$\Delta$}&\multicolumn{1}{c}{p(early = late)}\\
\midrule
\textbf{Panel A: Economic Anxieties}&         &         &         &         &         &         &         &         \\
          &         &         &         &         &         &         &         &         \\
\% agree: world severely affected by coronavirus&    80.12&    39.93&     1006&    67.61&    46.82&      914&   -12.50&    0.000\\
\% agree: US severely affected by coronavirus&    77.83&    41.56&     1006&    55.14&    49.76&      914&   -22.69&    0.000\\
\% worried about US economy&    87.57&    33.00&     1006&    68.05&    46.65&      914&   -19.52&    0.000\\
\% worried about personal econ. situation&    73.76&    44.02&     1006&    47.16&    49.95&      914&   -26.60&    0.000\\
          &         &         &         &         &         &         &         &         \\
\bottomrule \end{tabular}

%% file: DiD_survey.tex
{
\def\sym#1{\ifmmode^{#1}\else\(^{#1}\)\fi}
\begin{tabular}{l*{4}{c}}
\toprule
                &\multicolumn{2}{c}{Predicted impact on (standardized)}&\multicolumn{2}{c}{Worry about (standardized)}\\\cmidrule(lr){2-3}\cmidrule(lr){4-5}
                &\multicolumn{1}{c}{(1)}&\multicolumn{1}{c}{(2)}&\multicolumn{1}{c}{(3)}&\multicolumn{1}{c}{(4)}\\
                &\multicolumn{1}{c}{World}&\multicolumn{1}{c}{US}&\multicolumn{1}{c}{US Economy}&\multicolumn{1}{c}{Pers. Economic Sit.}\\
\midrule
                &         &         &         &         \\
Any case        &0.2318\sym{**}&0.2579\sym{***}&0.2159\sym{*}&   0.0636\\
                & (0.0985)& (0.0946)& (0.1079)& (0.1258)\\
                &         &         &         &         \\
\midrule R$^2$  &     0.07&     0.07&     0.08&     0.09\\
Number of Observations&     1920&     1920&     1920&     1920\\
\midrule Including controls&        X&        X&        X&        X\\
State FE        &        X&        X&        X&        X\\
Survey FE       &        X&        X&        X&        X\\
\bottomrule
\end{tabular}
}

%% file: experiment_correlations_combined.tex
{
\def\sym#1{\ifmmode^{#1}\else\(^{#1}\)\fi}
\begin{tabular}{l*{4}{c}}
\toprule
                &\multicolumn{2}{c}{Predicted impact on (standardized)}&\multicolumn{2}{c}{Worry about (standardized)}\\\cmidrule(lr){2-3}\cmidrule(lr){4-5}
                &\multicolumn{1}{c}{(1)}&\multicolumn{1}{c}{(2)}&\multicolumn{1}{c}{(3)}&\multicolumn{1}{c}{(4)}\\
                &\multicolumn{1}{c}{World}&\multicolumn{1}{c}{US}&\multicolumn{1}{c}{US Economy}&\multicolumn{1}{c}{Pers. Economic Sit.}\\
\midrule
{\bf Panel A: No control variables}&         &         &         &         \\
                &         &         &         &         \\
Overestimate mortality&0.3655\sym{***}&0.4516\sym{***}&0.1892\sym{***}&0.4783\sym{***}\\
                & (0.0647)& (0.0637)& (0.0655)& (0.0635)\\
                &         &         &         &         \\
Overestimate contagiousness&0.5263\sym{***}&0.5722\sym{***}&0.4504\sym{***}&0.4096\sym{***}\\
                & (0.0899)& (0.0843)& (0.0879)& (0.0825)\\
                &         &         &         &         \\
\midrule R$^2$  &     0.08&     0.11&     0.04&     0.09\\
\midrule        &         &         &         &         \\
                &         &         &         &         \\
{\bf Panel B: Including control variables}&         &         &         &         \\
                &         &         &         &         \\
Overestimate mortality&0.3830\sym{***}&0.4390\sym{***}&0.1850\sym{***}&0.4016\sym{***}\\
                & (0.0654)& (0.0642)& (0.0668)& (0.0645)\\
                &         &         &         &         \\
Overestimate contagiousness&0.5077\sym{***}&0.5392\sym{***}&0.4252\sym{***}&0.3780\sym{***}\\
                & (0.0885)& (0.0832)& (0.0849)& (0.0810)\\
                &         &         &         &         \\
\midrule R$^2$  &     0.13&     0.16&     0.10&     0.16\\
 Number of Observations &      914&      914&      914&      914\\
\bottomrule
\end{tabular}
}

%% file: experiment_correlations_combined_continuous.tex
{
\def\sym#1{\ifmmode^{#1}\else\(^{#1}\)\fi}
\begin{tabular}{l*{4}{c}}
\toprule
                &\multicolumn{2}{c}{Predicted impact on (standardized)}&\multicolumn{2}{c}{Worry about (standardized)}\\\cmidrule(lr){2-3}\cmidrule(lr){4-5}
                &\multicolumn{1}{c}{(1)}&\multicolumn{1}{c}{(2)}&\multicolumn{1}{c}{(3)}&\multicolumn{1}{c}{(4)}\\
                &\multicolumn{1}{c}{World}&\multicolumn{1}{c}{US}&\multicolumn{1}{c}{US Economy}&\multicolumn{1}{c}{Pers. Economic Sit.}\\
\midrule
{\bf Panel A: No control variables}&         &         &         &         \\
                &         &         &         &         \\
Perceived mortality&0.8220\sym{***}&1.1333\sym{***}&0.6213\sym{***}&1.3450\sym{***}\\
                & (0.1887)& (0.1730)& (0.2013)& (0.1841)\\
                &         &         &         &         \\
Perceived contagiousness &0.2714\sym{**}&0.6121\sym{***}&0.4826\sym{***}&0.5845\sym{***}\\
                & (0.1243)& (0.1071)& (0.1200)& (0.1149)\\
                &         &         &         &         \\
\midrule R$^2$  &     0.03&     0.09&     0.04&     0.11\\
\midrule        &         &         &         &         \\
                &         &         &         &         \\
{\bf Panel B: Including control variables}&         &         &         &         \\
                &         &         &         &         \\
Perceived mortality&0.8841\sym{***}&1.1090\sym{***}&0.6747\sym{***}&1.1382\sym{***}\\
                & (0.1932)& (0.1804)& (0.2098)& (0.1919)\\
                &         &         &         &         \\
Perceived contagiousness&0.2728\sym{**}&0.6084\sym{***}&0.4810\sym{***}&0.5596\sym{***}\\
                & (0.1239)& (0.1070)& (0.1226)& (0.1187)\\
                &         &         &         &         \\
\midrule R$^2$  &     0.08&     0.14&     0.10&     0.17\\
 Number of Observations &      914&      914&      914&      914\\
\bottomrule
\end{tabular}
}

%% file: balance_experiment_main.tex
\begin{tabular}{l*{6}{c}} \toprule
          &\multicolumn{3}{c}{Mortality information experiment}&\multicolumn{3}{c}{Contagion information experiment}\\\cmidrule(lr){2-4}\cmidrule(lr){5-7}
          &\multicolumn{1}{c}{(1)}&\multicolumn{1}{c}{(2)}&\multicolumn{1}{c}{(3)}&\multicolumn{1}{c}{(4)}&\multicolumn{1}{c}{(5)}&\multicolumn{1}{c}{(6)}\\
          &\multicolumn{1}{c}{Mean low rel. mortality}&\multicolumn{1}{c}{Mean high rel. mortality}&\multicolumn{1}{c}{p(low rel. mort. = high rel. mort)}&\multicolumn{1}{c}{Mean no contagion info}&\multicolumn{1}{c}{Mean contagion info}&\multicolumn{1}{c}{p(no info = info)}\\
\midrule
\% Male   &    50.55&    47.49&     0.36&    50.11&    47.93&     0.51\\
\% Age < 35&    23.74&    24.62&     0.76&    23.96&    24.40&     0.88\\
\% Highschool education&    18.90&    16.34&     0.31&    16.70&    18.52&     0.47\\
\% College eductation&    78.90&    82.14&     0.22&    81.98&    79.08&     0.27\\
\% Currently working&    58.46&    51.63&     0.04&    55.38&    54.68&     0.83\\
\% Democrat&    38.90&    41.18&     0.48&    37.58&    42.48&     0.13\\
\% Republican&    33.41&    32.90&     0.87&    34.95&    31.37&     0.25\\
\% High trust in science&     1.98&     0.22&     0.01&     0.88&     1.31&     0.53\\
          &         &         &         &         &         &         \\
p-value of joint significance&         &         &     0.00&         &         &     0.74\\
\bottomrule \end{tabular}

%% file: experiment_mainpart1_combined.tex
{
\def\sym#1{\ifmmode^{#1}\else\(^{#1}\)\fi}
\begin{tabular}{l*{2}{c}}
\toprule
                &\multicolumn{2}{c}{Predicted impact on (standardized)}\\\cmidrule(lr){2-3}
                &\multicolumn{1}{c}{(1)}&\multicolumn{1}{c}{(2)}\\
                &\multicolumn{1}{c}{World}&\multicolumn{1}{c}{US}\\
\midrule
{\bf Panel A: No control variables} &         &         \\
                &         &         \\
High relative mortality&0.2833\sym{***}&0.2246\sym{***}\\
                & (0.0655)& (0.0658)\\
                &         &         \\
\midrule R$^2$  &     0.02&     0.01\\
 Number of Observations &      914&      914\\
\midrule        &         &         \\
                &         &         \\
{\bf Panel B: Including control variables} &         &         \\
                &         &         \\
High relative mortality&0.2705\sym{***}&0.2245\sym{***}\\
                & (0.0659)& (0.0651)\\
                &         &         \\
\midrule R$^2$  &     0.06&     0.07\\
 Number of Observations &      914&      914\\
\bottomrule
\end{tabular}
}

%% file: experiment_mainpart2_combined.tex
{
\def\sym#1{\ifmmode^{#1}\else\(^{#1}\)\fi}
\begin{tabular}{l*{2}{c}}
\toprule
                &\multicolumn{2}{c}{Worry about (standardized)}\\\cmidrule(lr){2-3}
                &\multicolumn{1}{c}{(1)}&\multicolumn{1}{c}{(2)}\\
                &\multicolumn{1}{c}{US Economy}&\multicolumn{1}{c}{Pers. Economic Sit.}\\
\midrule
{\bf Panel A: No control variables}&         &         \\
                &         &         \\
High relative mortality&0.1557\sym{**}&0.1560\sym{**}\\
                & (0.0660)& (0.0660)\\
                &         &         \\
\midrule R$^2$  &     0.01&     0.01\\
 Number of Observations &      914&      914\\
\midrule        &         &         \\
Contagion information&  -0.0116&-0.0866\sym{**}\\
                & (0.0434)& (0.0415)\\
                &         &         \\
\midrule R$^2$  &     0.57&     0.61\\
 Number of Observations &      914&      914\\
\midrule        &         &         \\
{\bf Panel B: Including control variables}&         &         \\
                &         &         \\
High relative mortality&0.1556\sym{**}&0.1834\sym{***}\\
                & (0.0660)& (0.0645)\\
                &         &         \\
\midrule R$^2$  &     0.07&     0.10\\
 Number of Observations &      914&      914\\
\midrule        &         &         \\
Contagion information&  -0.0064&-0.0826\sym{**}\\
                & (0.0435)& (0.0416)\\
                &         &         \\
\midrule R$^2$  &     0.58&     0.62\\
 Number of Observations &      914&      914\\
\bottomrule
\end{tabular}
}

%% file: contagion_cross_rand_check.tex
{
\def\sym#1{\ifmmode^{#1}\else\(^{#1}\)\fi}
\begin{tabular}{l*{2}{c}}
\toprule
                &\multicolumn{2}{c}{Worry about (standardized)}\\\cmidrule(lr){2-3}
                &\multicolumn{1}{c}{(1)}&\multicolumn{1}{c}{(2)}\\
                &\multicolumn{1}{c}{US Economy}&\multicolumn{1}{c}{Pers. Economic Sit.}\\
\midrule
{\bf Panel A: No control variables}&         &         \\
\midrule        &         &         \\
Contagion information&  -0.0107&-0.0876\sym{**}\\
                & (0.0435)& (0.0414)\\
                &         &         \\
\midrule R$^2$  &     0.57&     0.61\\
 Number of Observations &      914&      914\\
\midrule {\bf Panel B: Including control variables}&         &         \\
\midrule        &         &         \\
 Contagion information&  -0.0052&-0.0836\sym{**}\\
                & (0.0436)& (0.0415)\\
                &         &         \\
\midrule R$^2$  &     0.58&     0.62\\
 Number of Observations &      914&      914\\
\bottomrule
\end{tabular}
}

%% file: treatment_interaction_effects.tex
{
\def\sym#1{\ifmmode^{#1}\else\(^{#1}\)\fi}
\begin{tabular}{l*{2}{c}}
\toprule
                &\multicolumn{2}{c}{Worry about (standardized)}\\\cmidrule(lr){2-3}
                &\multicolumn{1}{c}{(1)}&\multicolumn{1}{c}{(2)}\\
                &\multicolumn{1}{c}{US Economy}&\multicolumn{1}{c}{Pers. Economic Sit.}\\
\midrule
{\bf Panel A: No control variables}&         &         \\
\midrule        &         &         \\
Contagion information&   0.0270&-0.1243\sym{**}\\
                & (0.0576)& (0.0493)\\
                &         &         \\
Contagion information $\times$ Low relative mortality&  -0.0759&   0.0743\\
                & (0.0675)& (0.0571)\\
                &         &         \\
Baseline value  &0.7534\sym{***}&0.7803\sym{***}\\
                & (0.0248)& (0.0228)\\
                &         &         \\
\midrule R$^2$  &     0.57&     0.61\\
 Number of Observations &      914&      914\\
\midrule {\bf Panel B: Including control variables}&         &         \\
\midrule        &         &         \\
Contagion information&   0.0427&-0.1188\sym{**}\\
                & (0.0576)& (0.0494)\\
                &         &         \\
  Contagion information $\times$ Low relative mortality&  -0.0965&   0.0709\\
                & (0.0667)& (0.0567)\\
                &         &         \\
Baseline value  &0.7467\sym{***}&0.7572\sym{***}\\
                & (0.0259)& (0.0250)\\
                &         &         \\
\midrule R$^2$  &     0.58&     0.62\\
 Number of Observations &      914&      914\\
\bottomrule
\end{tabular}
}

%% file: growth_perc_pred.tex
{
\def\sym#1{\ifmmode^{#1}\else\(^{#1}\)\fi}
\begin{tabular}{l*{4}{c}}
\toprule
                &\multicolumn{2}{c}{Log predicted cases after twenty days}&\multicolumn{2}{c}{Exponential mental model}\\\cmidrule(lr){2-3}\cmidrule(lr){4-5}
                &\multicolumn{1}{c}{(1)}&\multicolumn{1}{c}{(2)}&\multicolumn{1}{c}{(3)}&\multicolumn{1}{c}{(4)}\\
                &\multicolumn{1}{c}{OLS}&\multicolumn{1}{c}{Post LASSO }&\multicolumn{1}{c}{OLS}&\multicolumn{1}{c}{Post LASSO }\\
\midrule
Male            &    0.236         &    0.213         &   0.0163         &                  \\
                &  (0.287)         &  (0.285)         & (0.0240)         &                  \\
Aged 25 to 34   &  -0.0312         &   -0.244         & 0.000318         &  -0.0361         \\
                &  (0.490)         &  (0.381)         & (0.0415)         & (0.0300)         \\
Aged 35 to 44   &    0.153         &   -0.115         &   0.0539         &                  \\
                &  (0.531)         &  (0.407)         & (0.0455)         &                  \\
Aged 45 to 54   &    0.296         &                  &  -0.0193         &  -0.0603\sym{**} \\
                &  (0.495)         &                  & (0.0398)         & (0.0272)         \\
Aged 55 to 64   &    0.536         &    0.354         &   0.0400         &                  \\
                &  (0.534)         &  (0.421)         & (0.0431)         &                  \\
Aged 65 above   &    1.289\sym{**} &    1.097\sym{**} &    0.138\sym{***}&    0.106\sym{***}\\
                &  (0.560)         &  (0.450)         & (0.0472)         & (0.0353)         \\
Log income      &    2.792         &    0.370\sym{**} &    0.387         &   0.0342\sym{**} \\
                &  (3.450)         &  (0.182)         &  (0.271)         & (0.0139)         \\
Log income squared&   -0.113         &                  &  -0.0166         &                  \\
                &  (0.162)         &                  & (0.0127)         &                  \\
Highschool education&    1.265\sym{**} &                  &   0.0793\sym{**} &                  \\
                &  (0.499)         &                  & (0.0370)         &                  \\
College eductation&    2.257\sym{***}&    1.184\sym{***}&    0.144\sym{***}&   0.0781\sym{***}\\
                &  (0.477)         &  (0.314)         & (0.0354)         & (0.0255)         \\
Currently working&   -0.687\sym{*}  &   -0.667\sym{*}  &  -0.0428         &  -0.0366         \\
                &  (0.379)         &  (0.373)         & (0.0322)         & (0.0278)         \\
Student         &   -0.804         &   -0.780         &  -0.0317         &                  \\
                &  (0.602)         &  (0.601)         & (0.0472)         &                  \\
Democrat        &   -0.658\sym{**} &   -0.642\sym{*}  &  -0.0420         &  -0.0346         \\
                &  (0.330)         &  (0.329)         & (0.0278)         & (0.0230)         \\
Republican      &   -0.359         &   -0.336         &  -0.0165         &                  \\
                &  (0.376)         &  (0.376)         & (0.0302)         &                  \\
Constant        &   -12.79         &    1.335         &   -2.215         &   -0.250\sym{*}  \\
                &  (18.34)         &  (1.764)         &  (1.436)         &  (0.140)         \\
\midrule
R$^2$           &   0.0517         &   0.0485         &   0.0559         &   0.0510         \\
Number of observations&     1006         &     1006         &     1006         &     1006         \\
\bottomrule
\end{tabular}
}

%% file: correlations_expgrowth_combined.tex
{
\def\sym#1{\ifmmode^{#1}\else\(^{#1}\)\fi}
\begin{tabular}{l*{8}{c}}
\toprule
                &\multicolumn{4}{c}{Predicted impact on (standardized)}&\multicolumn{4}{c}{Worry about (standardized)}\\\cmidrule(lr){2-5}\cmidrule(lr){6-9}
                &\multicolumn{1}{c}{(1)}&\multicolumn{1}{c}{(2)}&\multicolumn{1}{c}{(3)}&\multicolumn{1}{c}{(4)}&\multicolumn{1}{c}{(5)}&\multicolumn{1}{c}{(6)}&\multicolumn{1}{c}{(7)}&\multicolumn{1}{c}{(8)}\\
                &\multicolumn{1}{c}{World}&\multicolumn{1}{c}{World}&\multicolumn{1}{c}{US}&\multicolumn{1}{c}{US}&\multicolumn{1}{c}{US Economy}&\multicolumn{1}{c}{US Economy}&\multicolumn{1}{c}{Pers. Economic Sit.}&\multicolumn{1}{c}{Pers. Economic Sit.}\\
\midrule
                &         &         &         &         &         &         &         &         \\
{\bf Panel A }  &         &         &         &         &         &         &         &         \\
                &         &         &         &         &         &         &         &         \\
Log(estimate day 5)- z-score&0.0971\sym{***}&0.0832\sym{***}&0.0942\sym{***}&0.0893\sym{***}&0.0747\sym{**}&0.0750\sym{**}&   0.0044&   0.0268\\
                & (0.0284)& (0.0282)& (0.0299)& (0.0305)& (0.0306)& (0.0306)& (0.0318)& (0.0325)\\
                &         &         &         &         &         &         &         &         \\
\midrule R$^2$  &     0.01&     0.07&     0.01&     0.06&     0.01&     0.06&     0.00&     0.09\\
Number of Observations &     1006&     1006&     1006&     1006&     1006&     1006&     1006&     1006\\
\midrule        &         &         &         &         &         &         &         &         \\
{\bf Panel B}   &         &         &         &         &         &         &         &         \\
                &         &         &         &         &         &         &         &         \\
Log(estimate day 10)- z-score&0.1044\sym{***}&0.0857\sym{***}&0.1137\sym{***}&0.1069\sym{***}&0.0714\sym{**}&0.0703\sym{**}&  -0.0229&   0.0086\\
                & (0.0291)& (0.0292)& (0.0295)& (0.0303)& (0.0313)& (0.0314)& (0.0327)& (0.0321)\\
                &         &         &         &         &         &         &         &         \\
\midrule R$^2$  &     0.01&     0.07&     0.01&     0.06&     0.01&     0.06&     0.00&     0.09\\
Number of Observations &     1006&     1006&     1006&     1006&     1006&     1006&     1006&     1006\\
\midrule        &         &         &         &         &         &         &         &         \\
{\bf Panel C}   &         &         &         &         &         &         &         &         \\
                &         &         &         &         &         &         &         &         \\
Log(estimate day 20)- z-score&0.0828\sym{***}&0.0608\sym{*}&0.0933\sym{***}&0.0841\sym{***}&   0.0502&   0.0489&  -0.0441&  -0.0079\\
                & (0.0313)& (0.0315)& (0.0314)& (0.0323)& (0.0324)& (0.0326)& (0.0332)& (0.0327)\\
                &         &         &         &         &         &         &         &         \\
\midrule R$^2$  &     0.01&     0.06&     0.01&     0.06&     0.00&     0.06&     0.00&     0.09\\
Number of Observations &     1006&     1006&     1006&     1006&     1006&     1006&     1006&     1006\\
\midrule Controls&         &        X&         &        X&         &        X&         &        X\\
\bottomrule
\end{tabular}
}

%% file: Final working paper/coronabib.bib
@article{Bartik2020,
author = {Bartik, Alexander W. and Bertrand, Marianne and Cullen, Zo{\"{e}} B. and Glaeser, Edward L. and Luca, Michael and Stanton, Christopher T.},
doi = {10.1017/CBO9781107415324.004},
isbn = {9788578110796},
issn = {1098-6596},
journal = {NBER Working Paper 26989},
pmid = {25246403},
title = {{How Are Small Businesses Adjusting to COVID-19? Early Evidence From a Survey}},
year = {2020}
}

@article{bursztyn2020misinformation,
	Author = {Bursztyn, Leonardo and Rao, Aakaash and Roth, Christopher and Yanagizawa-Drott, David},
	Journal = {Becker Friedman Institute for Economics Working Paper 2020-44},
	Title = {Misinformation During a Pandemic},
	Year = {2020}}

@article{hanspal2020income,
	Author = {Hanspal, Tobin and Weber, Annika and Wohlfart, Johannes},
	journal = {CEBI Working Paper No. 13/20},
	Title = {Income and Wealth Shocks and Expectations during the COVID-19 Pandemic},
	Year = {2020}}

@article{adams2020inequality,
	Author = {Adams-Prassl, Abi and Boneva, Teodora and Golin, Marta and Rauh, Christopher},
	Title = {Inequality in the Impact of the Coronavirus Shock: Evidence from Real Time Surveys},
	journal= {IZA Discussion Paper No. 13183},
	Year = {2020}}

@article{heimer2019yolo,
	Author = {Heimer, Rawley Z and Myrseth, Kristian Ove R and Schoenle, Raphael S},
	Journal = {The Journal of Finance},
	Number = {6},
	Pages = {2957--2996},
	Publisher = {Wiley Online Library},
	Title = {YOLO: Mortality Beliefs and Household Finance Puzzles},
	Volume = {74},
	Year = {2019}}

@article{levy2016exponential,
	Author = {Levy, Matthew and Tasoff, Joshua},
	Journal = {Journal of the European Economic Association},
	Number = {3},
	Pages = {545--583},
	Publisher = {Oxford University Press},
	Title = {Exponential-Growth Bias and Lifecycle Consumption},
	Volume = {14},
	Year = {2016}}

@article{coibion2020labor,
	Author = {Coibion, Olivier and Gorodnichenko, Yuriy and Weber, Michael},
	Journal = {Becker Friedman Institute for Economics Working Paper 2020-41},
	Title = {Labor Markets during the COVID-19 Crisis: A Preliminary View},
	Year = {2020}}

@article{kerwin2018scared,
	Author = {Kerwin, Jason},
	Journal = {Working Paper available at SSRN 2797493},
	Title = {Scared Straight or Scared to Death? The Effect of Risk Beliefs on Risky Behaviors},
	Year = {2018}}

@article{prat2013political,
	Author = {Prat, Andrea and Str{\"o}mberg, David},
	Journal = {Advances in Economics and Econometrics},
	Pages = {135},
	Publisher = {Cambridge University Press, Cambridge},
	Title = {The Political Economy of Mass Media},
	Volume = {2},
	Year = {2013}}

@article{chong2007framing,
	Author = {Chong, Dennis and Druckman, James N},
	Journal = {Annual Review of Political Science},
	Pages = {103--126},
	Publisher = {Annual Reviews},
	Title = {Framing theory},
	Volume = {10},
	Year = {2007}}

@article{massonnaud2020covid,
	Author = {Massonnaud, Cl{\'e}ment and Roux, Jonathan and Cr{\'e}pey, Pascal},
	Journal = {medRxiv},
	Publisher = {Cold Spring Harbor Laboratory Press},
	Title = {COVID-19: Forecasting Short Term Hospital Needs in France},
	Year = {2020}}

@article{wagenaar1975misperception,
	Author = {Wagenaar, William A and Sagaria, Sabato D},
	Journal = {Perception \& Psychophysics},
	Number = {6},
	Pages = {416--422},
	Publisher = {Springer},
	Title = {Misperception of Exponential Growth},
	Volume = {18},
	Year = {1975}}

@article{Fortson2011mortrisk,
	Author = {Fortson, Jane G},
	Doi = {10.1162/REST\_a\_00067},
	Eprint = {https://doi.org/10.1162/REST_a_00067},
	Journal = {The Review of Economics and Statistics},
	Number = {1},
	Pages = {1-15},
	Title = {Mortality Risk and Human Capital Investment: The Impact of HIV/AIDS in Sub-Saharan Africa},
	Url = {https://doi.org/10.1162/REST_a_00067},
	Volume = {93},
	Year = {2011}}

@article{Abba2016spread,
	Author = {Adda, J{\'e}r{\^o}me},
	Doi = {10.1093/qje/qjw005},
	Eprint = {https://academic.oup.com/qje/article-pdf/131/2/891/30636376/qjw005.pdf},
	Issn = {0033-5533},
	Journal = {The Quarterly Journal of Economics},
	Number = {2},
	Pages = {891-941},
	Title = {{Economic Activity and the Spread of Viral Diseases: Evidence from High Frequency Data}},
	Url = {https://doi.org/10.1093/qje/qjw005},
	Volume = {131},
	Year = {2016}}

@article{Oster2013testing,
	Author = {Oster, Emily and Shoulson, Ira and Dorsey, E. Ray},
	Doi = {10.1257/aer.103.2.804},
	Journal = {American Economic Review},
	Month = {April},
	Number = {2},
	Pages = {804-30},
	Title = {Optimal Expectations and Limited Medical Testing: Evidence from Huntington Disease},
	Url = {http://www.aeaweb.org/articles?id=10.1257/aer.103.2.804},
	Volume = {103},
	Year = {2013}}

@article{carbone2005smoking,
	Author = {Jared C. Carbone and Snorre Kverndokk and Ole J{\o}rgen R{\o}geberg},
	Doi = {https://doi.org/10.1016/j.jhealeco.2004.11.001},
	Issn = {0167-6296},
	Journal = {Journal of Health Economics},
	Number = {4},
	Pages = {631 - 653},
	Title = {Smoking, Health, Risk, and Perception},
	Url = {http://www.sciencedirect.com/science/article/pii/S0167629604001183},
	Volume = {24},
	Year = {2005}}

@article{winter2014obese,
	Author = {Winter, Joachim and Wuppermann, Amelie},
	Doi = {10.1002/hec.2933},
	Eprint = {https://onlinelibrary.wiley.com/doi/pdf/10.1002/hec.2933},
	Journal = {Health Economics},
	Number = {5},
	Pages = {564-585},
	Title = {Do They Know What is at Risk? Health Risk Perceptions Among The Obese?},
	Url = {https://onlinelibrary.wiley.com/doi/abs/10.1002/hec.2933},
	Volume = {23},
	Year = {2014}}

@article{liu1995risk,
	Author = {Liu, Jin-Tan and Hsieh, Chee-Ruey},
	Journal = {Journal of Risk and Uncertainty},
	Number = {2},
	Pages = {139--157},
	Publisher = {Springer},
	Title = {Risk Perception and Smoking Behavior: Empirical Evidence from Taiwan},
	Volume = {11},
	Year = {1995}}

@article{weinstein1989optimistic,
	Author = {Weinstein, Neil D},
	Journal = {Science},
	Number = {4935},
	Pages = {1232--1234},
	Publisher = {American Association for the Advancement of Science},
	Title = {Optimistic Biases about Personal Risks},
	Volume = {246},
	Year = {1989}}

@article{binder2018experiment,
	Author = {Binder, Carola and Rodrigue, Alex},
	Journal = {Southern Economic Journal},
	Number = {2},
	Pages = {580--598},
	Publisher = {Wiley Online Library},
	Title = {Household Informedness and Long-Run Inflation Expectations: Experimental Evidence},
	Volume = {85},
	Year = {2018}}

@article{binder2018stuck,
	Author = {Binder, Carola and Makridis, Christos},
	Journal = {Working Paper available at SSRN 3267721},
	Title = {Stuck in the Seventies: Gas Prices and Macroeconomic Expectations},
	Year = {2018}}

@article{coibion2015phillips,
	Author = {Coibion, Olivier and Gorodnichenko, Yuriy},
	Journal = {American Economic Journal: Macroeconomics},
	Number = {1},
	Pages = {197--232},
	Publisher = {American Economic Association},
	Title = {Is the Phillips Curve Alive and Well after All? Inflation Expectations and the Missing Disinflation},
	Volume = {7},
	Year = {2015}}

@article{coibion2015information,
	Author = {Coibion, Olivier and Gorodnichenko, Yuriy},
	Journal = {The American Economic Review},
	Number = {8},
	Pages = {2644--2678},
	Publisher = {American Economic Association},
	Title = {Information Rigidity and the Expectations Formation Process: A Simple Framework and New Facts},
	Volume = {105},
	Year = {2015}}

@article{coibion2012can,
	Author = {Coibion, Olivier and Gorodnichenko, Yuriy},
	Journal = {Journal of Political Economy},
	Number = {1},
	Pages = {116--159},
	Publisher = {University of Chicago Press Chicago, IL},
	Title = {What Can Survey Forecasts Tell us about Information Rigidities?},
	Volume = {120},
	Year = {2012}}

@article{fuster2012natural,
	Author = {Fuster, Andreas and Hebert, Benjamin and Laibson, David},
	Journal = {NBER Macroeconomics Annual},
	Number = {1},
	Pages = {1--48},
	Publisher = {University of Chicago Press Chicago, IL},
	Title = {Natural Expectations, Macroeconomic Dynamics, and Asset Pricing},
	Volume = {26},
	Year = {2012}}

@article{fuster2010natural,
	Author = {Fuster, Andreas and Laibson, David and Mendel, Brock},
	Journal = {Journal of Economic Perspectives},
	Number = {4},
	Pages = {67--84},
	Title = {Natural Expectations and Macroeconomic Fluctuations},
	Volume = {24},
	Year = {2010}}

@article{kan2004obesity,
	Author = {Kan, Kamhon and Tsai, Wei-Der},
	Journal = {Journal of Health Economics},
	Number = {5},
	Pages = {907--934},
	Publisher = {Elsevier},
	Title = {Obesity and Risk Knowledge},
	Volume = {23},
	Year = {2004}}

@article{binder2020coronavirus,
	Author = {Binder, Carola},
	Date-Added = {2020-03-25 05:57:53 +0000},
	Date-Modified = {2020-03-25 05:57:53 +0000},
	Journal = {Review of Economics and Statistics},
	Title = {Coronavirus Fears and Macroeconomic Expectations},
	Number={forthcoming},
	Year = {2020}

@article{stango2009exponential,
	Author = {Stango, Victor and Zinman, Jonathan},
	Journal = {The Journal of Finance},
	Number = {6},
	Pages = {2807--2849},
	Publisher = {Wiley Online Library},
	Title = {Exponential Growth Bias and Household Finance},
	Volume = {64},
	Year = {2009}}

@article{mackowiak2018lack,
	Author = {Ma{\'c}kowiak, Bartosz and Wiederholt, Mirko},
	Journal = {Journal of Monetary Economics},
	Pages = {35--47},
	Publisher = {Elsevier},
	Title = {Lack of Preparation for Rare Events},
	Volume = {100},
	Year = {2018}}

@article{tversky1973availability,
	Author = {Tversky, Amos and Kahneman, Daniel},
	Journal = {Cognitive Psychology},
	Number = {2},
	Pages = {207--232},
	Publisher = {Elsevier},
	Title = {Availability: A Heuristic for Judging Frequency and Probability},
	Volume = {5},
	Year = {1973}}

@article{gallagher2014learning,
	Author = {Gallagher, Justin},
	Journal = {American Economic Journal: Applied Economics},
	Pages = {206--233},
	Publisher = {JSTOR},
	Title = {Learning About an Infrequent Event: Evidence from Flood Insurance Take-Up in the United States},
	Year = {2014}}

@article{rabin2002inference,
	Author = {Rabin, Matthew},
	Journal = {The Quarterly Journal of Economics},
	Number = {3},
	Pages = {775--816},
	Publisher = {MIT Press},
	Title = {Inference by Believers in the Law of Small Numbers},
	Volume = {117},
	Year = {2002}}

@article{bordalo2013salience,
	Author = {Bordalo, Pedro and Gennaioli, Nicola and Shleifer, Andrei},
	Journal = {Journal of Political Economy},
	Number = {5},
	Pages = {803--843},
	Publisher = {University of Chicago Press Chicago, IL},
	Title = {Salience and Consumer Choice},
	Volume = {121},
	Year = {2013}}

@article{andre2019subjective,
	Author = {Andre, Peter and Pizzinelli, Carlo and Roth, Christopher and Wohlfart, Johannes},
	Journal = {CESifo Working Paper No. 7850},
	Title = {Subjective Models of the Macroeconomy: Evidence from Experts and a Representative Sample},
	Year = {2019}}

@article{kuchler2015personal,
	Author = {Kuchler, Theresa and Zafar, Basit},
	Journal = {The Journal of Finance},
	Number = {5},
	Pages = {2491--2542},
	Publisher = {Wiley Online Library},
	Title = {Personal Experiences and Expectations about Aggregate Outcomes},
	Volume = {74},
	Year = {2019}}

@article{coibion2019does,
	Author = {Coibion, Olivier and Georgarakos, Dimitris and Gorodnichenko, Yuriy and van Rooij, Maarten},
	Journal = {NBER Working Paper No. 26106},
	Title = {How Does Consumption Respond to News about Inflation? Field Evidence from a Randomized Control Trial},
	Year = {2019}}

@article{bailey2017beliefsleverage,
	Author = {Bailey, Michael and D{\'a}vila, Eduardo and Kuchler, Theresa and Stroebel, Johannes},
	Journal = {Review of Economic Studies},
	Title = {House Price Beliefs and Mortgage Leverage Choice},
	Year = {2019}}

@article{bailey2016social,
	Author = {Bailey, Michael and Cao, Ruiqing and Kuchler, Theresa and Stroebel, Johannes},
	Journal = {Journal of Political Economy},
	Number = {6},
	Pages = {2224--2276},
	Publisher = {University of Chicago Press Chicago, IL},
	Title = {The Economic Effects of Social Networks: Evidence from the Housing Market},
	Volume = {126},
	Year = {2018}}

@article{armona2016home,
	Author = {Armona, Luis and Fuster, Andreas and Zafar, Basit},
	Journal = {The Review of Economic Studies},
	Number = {4},
	Pages = {1371--1410},
	Publisher = {Oxford University Press},
	Title = {Home Price Expectations and Behaviour: Evidence from a Randomized Information Experiment},
	Volume = {86},
	Year = {2018}}

@article{dacunto2016effect,
	Author = {D'Acunto, Francesco and Hoang, Daniel and Weber, Michael},
	Journal = {CESifo Working Paper No. 7793},
	Title = {Managing Households' Expectations with Simple Economic Policies},
	Year = {2019}}

@article{d2019iq,
	Author = {D'Acunto, Francesco and Hoang, Daniel and Paloviita, Maritta and Weber, Michael},
	Journal = {NBER Working Paper No. 25496},
	Title = {IQ, Expectations, and Choice},
	Year = {2019}}

@article{coibion2019monetary,
	Author = {Coibion, Olivier and Gorodnichenko, Yuriy and Weber, Michael},
	Journal = {NBER Working Paper No. 25482},
	Title = {Monetary Policy Communications and their Effects on Household Inflation Expectations},
	Year = {2019}}

@article{coibion2018firms,
	Author = {Coibion, Olivier and Gorodnichenko, Yuriy and Kumar, Saten},
	Journal = {American Economic Review},
	Number = {9},
	Pages = {2671--2713},
	Title = {How Do Firms Form Their Expectations? New Survey Evidence},
	Volume = {108},
	Year = {2018}}

@article{kermack1927contribution,
	Author = {Kermack, William Ogilvy and McKendrick, Anderson G},
	Journal = {Proceedings of the Royal Society of London},
	Number = {772},
	Pages = {700--721},
	Publisher = {The Royal Society London},
	Title = {A Contribution to the Mathematical Theory of Epidemics},
	Volume = {115},
	Year = {1927}}

@article{bord2000sense,
	Author = {Bord, Richard J and O'connor, Robert E and Fisher, Ann},
	Journal = {Public Understanding of Science},
	Number = {3},
	Pages = {205--218},
	Publisher = {Bristol,[England]: Published by the Institute of Physics in association with~{\ldots}},
	Title = {In What Sense Does the Public Need to Understand Global Climate Change?},
	Volume = {9},
	Year = {2000}}

@book{keeling2011modeling,
	Author = {Keeling, Matt J and Rohani, Pejman},
	Publisher = {Princeton University Press},
	Title = {Modeling Infectious Diseases in Humans and Animals},
	Year = {2011}}

@article{dong2020,
	Author = {Dong, Ensheng and Du, Hongru and Gardner, Lauren},
	Booktitle = {The Lancet Infectious Diseases},
	Doi = {10.1016/S1473-3099(20)30120-1},
	Isbn = {1473-3099},
	Journal = {The Lancet Infectious Diseases},
	Publisher = {Elsevier},
	Title = {An Interactive Web-Based Dashboard to Track COVID-19 in Real Time},
	Ty = {JOUR},
	Url = {https://doi.org/10.1016/S1473-3099(20)30120-1},
	Year = {2020}}

@article{ginsberg2009,
	Author = {Ginsberg, Jeremy and Mohebbi, Matthew H. and Patel, Rajan S. and Brammer, Lynnette and Smolinski, Mark S. and Brilliant, Larry},
	Da = {2009/02/01},
	Doi = {10.1038/nature07634},
	Id = {Ginsberg2009},
	Isbn = {1476-4687},
	Journal = {Nature},
	Number = {7232},
	Pages = {1012--1014},
	Title = {Detecting Influenza Epidemics Using Search Engine 	Query Data},
	Ty = {JOUR},
	Url = {https://doi.org/10.1038/nature07634},
	Volume = {457},
	Year = {2009}}

@article{roth2019expectations,
	Author = {Roth, Christopher and Wohlfart, Johannes},
	Journal = {Review of Economics and Statistics},
	Publisher = {MIT Press},
	Title = {How Do Expectations about the Macroeconomy Affect Personal Expectations and Behavior?},
	Number = {forthcoming},
	Year = {2020}

@article{coibion2020inflation,
	Author = {Coibion, Olivier and Gorodnichenko, Yuriy and Ropele, Tiziano},
	Journal = {The Quarterly Journal of Economics},
	Number = {1},
	Pages = {165--219},
	Publisher = {Oxford University Press},
	Title = {Inflation Expectations and Firm Decisions: New Causal Evidence},
	Volume = {135},
	Year = {2020}}

@article{malmendier2011depression,
	Author = {Malmendier, Ulrike and Nagel, Stefan},
	Journal = {The Quarterly Journal of Economics},
	Number = {1},
	Pages = {373--416},
	Publisher = {MIT Press},
	Title = {Depression Babies: Do Macroeconomic Experiences Affect Risk Taking?},
	Volume = {126},
	Year = {2011}}

@article{vosen2011forecasting,
	Author = {Vosen, Simeon and Schmidt, Torsten},
	Journal = {Journal of Forecasting},
	Number = {6},
	Pages = {565--578},
	Publisher = {Wiley Online Library},
	Title = {Forecasting Private Consumption: Survey-Based Indicators vs. Google Trends},
	Volume = {30},
	Year = {2011}}

@article{choi2012predicting,
	Author = {Choi, Hyunyoung and Varian, Hal},
	Journal = {Economic Record},
	Pages = {2--9},
	Publisher = {Wiley Online Library},
	Title = {Predicting the present with Google Trends},
	Volume = {88},
	Year = {2012}}

@article{person2004fear,
	Author = {Person, Bobbie and Sy, Francisco and Holton, Kelly and Govert, Barbara and Liang, Arthur and others},
	Journal = {Emerging Infectious Diseases},
	Number = {2},
	Pages = {358},
	Publisher = {Centers for Disease Control and Prevention},
	Title = {Fear and Stigma: the Epidemic Within the SARS Outbreak},
	Volume = {10},
	Year = {2004}}

@article{razum2003sars,
	Author = {Razum, Oliver and Becher, Heiko and Kapaun, Annette and Junghanss, Thomas},
	Journal = {The Lancet},
	Number = {9370},
	Pages = {1739--1740},
	Publisher = {Elsevier},
	Title = {SARS, Lay Epidemiology, and Fear},
	Volume = {361},
	Year = {2003}}

@article{zhu2020novel,
	Author = {Zhu, Na and Zhang, Dingyu and Wang, Wenling and Li, Xingwang and Yang, Bo and Song, Jingdong and Zhao, Xiang and Huang, Baoying and Shi, Weifeng and Lu, Roujian and others},
	Journal = {New England Journal of Medicine},
	Publisher = {Mass Medical Soc},
	Title = {A Novel Coronavirus from Patients with Pneumonia in China, 2019},
	Year = {2020}}

@article{li2020early,
	Author = {Li, Qun and Guan, Xuhua and Wu, Peng and Wang, Xiaoye and Zhou, Lei and Tong, Yeqing and Ren, Ruiqi and Leung, Kathy SM and Lau, Eric HY and Wong, Jessica Y and others},
	Journal = {New England Journal of Medicine},
	Publisher = {Mass Medical Soc},
	Title = {Early Transmission Dynamics in Wuhan, China, of Novel Coronavirus--Infected Pneumonia},
	Year = {2020}}

@article{wu2020nowcasting,
	Author = {Wu, Joseph T and Leung, Kathy and Leung, Gabriel M},
	Journal = {The Lancet},
	Publisher = {Elsevier},
	Title = {Nowcasting and Forecasting the Potential Domestic and International Spread of the 2019-nCoV Outbreak Originating in Wuhan, China: a Modelling Study},
	Year = {2020}}
